%
%
%
\def\unredoffs{} \def\redoffs{\voffset=-.31truein\hoffset=-.48truein}
\def\speclscape{}
%
%
%
%
%
\newbox\leftpage \newdimen\fullhsize \newdimen\hstitle \newdimen\hsbody
\tolerance=1000\hfuzz=2pt
\catcode`\@=11 
\ifx\hyperdef\UNd@FiNeD\def\hyperdef#1#2#3#4{#4}\def\hyperref#1#2#3#4{#4}\fi
\def\bigans{b }
\def\answ{b }
%
\ifx\answ\bigans\message{(This will come out unreduced.}
\magnification=1200\unredoffs\baselineskip=16pt plus 2pt minus 1pt
\hsbody=\hsize \hstitle=\hsize 
\else\message{(This will be reduced.} \let\l@r=L
\magnification=1000\baselineskip=16pt plus 2pt minus 1pt \vsize=7truein
\redoffs \hstitle=8truein\hsbody=4.75truein\fullhsize=10truein\hsize=\hsbody
\output={\ifnum\pageno=0 
  \shipout\vbox{\speclscape{\hsize\fullhsize\makeheadline}
    \hbox to \fullhsize{\hfill\pagebody\hfill}}\advancepageno
  \else
  \almostshipout{\leftline{\vbox{\pagebody\makefootline}}}\advancepageno
  \fi}
\def\almostshipout#1{\if L\l@r \count1=1 \message{[\the\count0.\the\count1]}
      \global\setbox\leftpage=#1 \global\let\l@r=R
 \else \count1=2
  \shipout\vbox{\speclscape{\hsize\fullhsize\makeheadline}
      \hbox to\fullhsize{\box\leftpage\hfil#1}}  \global\let\l@r=L\fi}
\fi
%
\newcount\yearltd\yearltd=\year\advance\yearltd by -2000

\def\Title#1#2{\nopagenumbers\abstractfont\hsize=\hstitle\rightline{#1}%
\vskip 1in\centerline{\titlefont #2}\abstractfont\vskip .5in\pageno=0}
\def\Date#1{\vfill\leftline{#1}\tenpoint\supereject\global\hsize=\hsbody%
\footline={\hss\tenrm\hyperdef\hypernoname{page}\folio\folio\hss}}%
%

\def\draftmode{\message{ DRAFTMODE }\def\draftdate{{\rm preliminary draft:
\number\month/\number\day/\number\yearltd\ \ \hourmin}}%
\headline={\hfil\draftdate}\writelabels\baselineskip=20pt plus 2pt minus 2pt
 {\count255=\time\divide\count255 by 60 \xdef\hourmin{\number\count255}
  \multiply\count255 by-60\advance\count255 by\time
  \xdef\hourmin{\hourmin:\ifnum\count255<10 0\fi\the\count255}}}
\def\nolabels{\def\wrlabeL##1{}\def\eqlabeL##1{}\def\reflabeL##1{}}
\def\writelabels{\def\wrlabeL##1{\leavevmode\vadjust{\rlap{\smash%
{\line{{\escapechar=` \hfill\rlap{\sevenrm\hskip.03in\string##1}}}}}}}%
\def\eqlabeL##1{{\escapechar-1\rlap{\sevenrm\hskip.05in\string##1}}}%
\def\reflabeL##1{\noexpand\llap{\noexpand\sevenrm\string\string\string##1}}}
\nolabels
%
\global\newcount\secno \global\secno=0
\global\newcount\meqno \global\meqno=1
\def\s@csym{}
\def\newsec#1{\global\advance\secno by1%
{\toks0{#1}\message{(\the\secno. \the\toks0)}}%
\global\subsecno=0\eqnres@t\let\s@csym\secsym\xdef\secn@m{\the\secno}\noindent
{\bf\hyperdef\hypernoname{section}{\the\secno}{\the\secno.} #1}%
\writetoca{{\string\hyperref{}{section}{\the\secno}{\the\secno.}} {#1}}%
\par\nobreak\medskip\nobreak}
\def\eqnres@t{\xdef\secsym{\the\secno.}\global\meqno=1\bigbreak\bigskip}
\def\sequentialequations{\def\eqnres@t{\bigbreak}}\xdef\secsym{}
\global\newcount\subsecno \global\subsecno=0
\def\subsec#1{\global\advance\subsecno by1%
{\toks0{#1}\message{(\s@csym\the\subsecno. \the\toks0)}}%
\ifnum\lastpenalty>9000\else\bigbreak\fi
\noindent{\it\hyperdef\hypernoname{subsection}{\secn@m.\the\subsecno}%
{\secn@m.\the\subsecno.} #1}\writetoca{\string\quad
{\string\hyperref{}{subsection}{\secn@m.\the\subsecno}{\secn@m.\the\subsecno.}}
{#1}}\par\nobreak\medskip\nobreak}
\def\appendix#1#2{\global\meqno=1\global\subsecno=0\xdef\secsym{\hbox{#1.}}%
\bigbreak\bigskip\noindent{\bf Appendix \hyperdef\hypernoname{appendix}{#1}%
{#1.} #2}{\toks0{(#1. #2)}\message{\the\toks0}}%
\xdef\s@csym{#1.}\xdef\secn@m{#1}%
\writetoca{\string\hyperref{}{appendix}{#1}{Appendix {#1.}} {#2}}%
\par\nobreak\medskip\nobreak}
%
%
\def\checkm@de#1#2{\ifmmode{\def\f@rst##1{##1}\hyperdef\hypernoname{equation}%
{#1}{#2}}\else\hyperref{}{equation}{#1}{#2}\fi}
\def\eqnn#1{\DefWarn#1\xdef #1{(\noexpand\relax\noexpand\checkm@de%
{\s@csym\the\meqno}{\secsym\the\meqno})}%
\wrlabeL#1\writedef{#1\leftbracket#1}\global\advance\meqno by1}
\def\f@rst#1{\c@t#1a\em@ark}\def\c@t#1#2\em@ark{#1}
\def\eqna#1{\DefWarn#1\wrlabeL{#1$\{\}$}%
\xdef #1##1{(\noexpand\relax\noexpand\checkm@de%
{\s@csym\the\meqno\noexpand\f@rst{##1}}{\hbox{$\secsym\the\meqno##1$}})}
\writedef{#1\numbersign1\leftbracket#1{\numbersign1}}\global\advance\meqno by1}
\def\eqn#1#2{\DefWarn#1%
\xdef #1{(\noexpand\hyperref{}{equation}{\s@csym\the\meqno}%
{\secsym\the\meqno})}$$#2\eqno(\hyperdef\hypernoname{equation}%
{\s@csym\the\meqno}{\secsym\the\meqno})\eqlabeL#1$$%
\writedef{#1\leftbracket#1}\global\advance\meqno by1}
\def\xeqn{\expandafter\xe@n}\def\xe@n(#1){#1}
\def\xeqna#1{\expandafter\xe@n#1}
\def\eqns#1{(\e@ns #1{\hbox{}})}
\def\e@ns#1{\ifx\UNd@FiNeD#1\message{eqnlabel \string#1 is undefined.}%
\xdef#1{(?.?)}\fi{\let\hyperref=\relax\xdef\next{#1}}%
\ifx\next\em@rk\def\next{}\else%
\ifx\next#1\xeqn#1\else\def\n@xt{#1}\ifx\n@xt\next#1\else\xeqna#1\fi
\fi\let\next=\e@ns\fi\next}

\def\DefWarn#1{\ifx\UNd@FiNeD#1\else
\immediate\write16{*** WARNING: the label \string#1 is already defined ***}\fi}
%
\newskip\footskip\footskip14pt plus 1pt minus 1pt 
\def\footnotefont{\ninepoint}\def\f@t#1{\footnotefont #1\@foot}
\def\f@@t{\baselineskip\footskip\bgroup\footnotefont\aftergroup\@foot\let\next}
\setbox\strutbox=\hbox{\vrule height9.5pt depth4.5pt width0pt}
\global\newcount\ftno \global\ftno=0
\def\foot{\global\advance\ftno by1\def\foot@rg{\hyperref{}{footnote}%
{\the\ftno}{\the\ftno}\xdef\foot@rg{\noexpand\hyperdef\noexpand\hypernoname%
{footnote}{\the\ftno}{\the\ftno}}}\footnote{$^{\foot@rg}$}}
%
\newwrite\ftfile
\def\footend{\def\foot{\global\advance\ftno by1\chardef\wfile=\ftfile
\hyperref{}{footnote}{\the\ftno}{$^{\the\ftno}$}%
\ifnum\ftno=1\immediate\openout\ftfile=\jobname.fts\fi%
\immediate\write\ftfile{\noexpand\smallskip%
\noexpand\item{\noexpand\hyperdef\noexpand\hypernoname{footnote}
{\the\ftno}{f\the\ftno}:\ }\pctsign}\findarg}%
\def\footatend{\vfill\eject\immediate\closeout\ftfile{\parindent=20pt
\centerline{\bf Footnotes}\nobreak\bigskip\input \jobname.fts }}}
\def\footatend{}
%
%
\global\newcount\refno \global\refno=1
\newwrite\rfile
\def\ref{[\hyperref{}{reference}{\the\refno}{\the\refno}]\nref}
\def\nref#1{\DefWarn#1%
\xdef#1{[\noexpand\hyperref{}{reference}{\the\refno}{\the\refno}]}%
\writedef{#1\leftbracket#1}%
\ifnum\refno=1\immediate\openout\rfile=\jobname.refs\fi
\chardef\wfile=\rfile\immediate\write\rfile{\noexpand\item{[\noexpand\hyperdef%
\noexpand\hypernoname{reference}{\the\refno}{\the\refno}]\ }%
\reflabeL{#1\hskip.31in}\pctsign}\global\advance\refno by1\findarg}
\def\findarg#1#{\begingroup\obeylines\newlinechar=`\^^M\pass@rg}
{\obeylines\gdef\pass@rg#1{\writ@line\relax #1^^M\hbox{}^^M}%
\gdef\writ@line#1^^M{\expandafter\toks0\expandafter{\striprel@x #1}%
\edef\next{\the\toks0}\ifx\next\em@rk\let\next=\endgroup\else\ifx\next\empty%
\else\immediate\write\wfile{\the\toks0}\fi\let\next=\writ@line\fi\next\relax}}
\def\striprel@x#1{} \def\em@rk{\hbox{}}
\def\lref{\begingroup\obeylines\lr@f}
\def\lr@f#1#2{\DefWarn#1\gdef#1{\let#1=\UNd@FiNeD\ref#1{#2}}\endgroup\unskip}

\def\addref#1{\immediate\write\rfile{\noexpand\item{}#1}} 
\def\listrefs{\footatend\vfill\supereject\immediate\closeout\rfile\writestoppt
\baselineskip=\footskip\centerline{{\bf References}}\bigskip{\parindent=20pt%
\frenchspacing\escapechar=` \input \jobname.refs\vfill\eject}\nonfrenchspacing}
\def\startrefs#1{\immediate\openout\rfile=\jobname.refs\refno=#1}
\def\xref{\expandafter\xr@f}\def\xr@f[#1]{#1}
\def\refs#1{\count255=1[\r@fs #1{\hbox{}}]}
\def\r@fs#1{\ifx\UNd@FiNeD#1\message{reflabel \string#1 is undefined.}%
\nref#1{need to supply reference \string#1.}\fi%
\vphantom{\hphantom{#1}}{\let\hyperref=\relax\xdef\next{#1}}%
\ifx\next\em@rk\def\next{}%
\else\ifx\next#1\ifodd\count255\relax\xref#1\count255=0\fi%
\else#1\count255=1\fi\let\next=\r@fs\fi\next}
%

%
\newwrite\ffile\global\newcount\figno \global\figno=1
\def\fig{fig.~\hyperref{}{figure}{\the\figno}{\the\figno}\nfig}
\def\nfig#1{\DefWarn#1%
\xdef#1{fig.~\noexpand\hyperref{}{figure}{\the\figno}{\the\figno}}%
\writedef{#1\leftbracket fig.\noexpand~\xfig#1}%
\ifnum\figno=1\immediate\openout\ffile=\jobname.figs\fi\chardef\wfile=\ffile%
{\let\hyperref=\relax
\immediate\write\ffile{\noexpand\medskip\noexpand\item{Fig.\ %
\noexpand\hyperdef\noexpand\hypernoname{figure}{\the\figno}{\the\figno}. }
\reflabeL{#1\hskip.55in}\pctsign}}\global\advance\figno by1\findarg}
\def\listfigs{\vfill\eject\immediate\closeout\ffile{\parindent40pt
\baselineskip14pt\centerline{{\bf Figure Captions}}\nobreak\medskip
\escapechar=` \input \jobname.figs\vfill\eject}}
\def\xfig{\expandafter\xf@g}\def\xf@g fig.\penalty\@M\ {}
\def\figs#1{figs.~\f@gs #1{\hbox{}}}
\def\f@gs#1{{\let\hyperref=\relax\xdef\next{#1}}\ifx\next\em@rk\def\next{}\else
\ifx\next#1\xfig #1\else#1\fi\let\next=\f@gs\fi\next}
\def\figin{\epsfcheck\figin}\def\figins{\epsfcheck\figins}
\def\epsfcheck{\ifx\epsfbox\UNd@FiNeD
\message{(NO epsf.tex, FIGURES WILL BE IGNORED)}
\gdef\figin##1{\vskip2in}\gdef\figins##1{\hskip.5in}
\else\message{(FIGURES WILL BE INCLUDED)}%
\gdef\figin##1{##1}\gdef\figins##1{##1}\fi}
\def\DefWarn#1{}
\def\figinsert{\goodbreak\midinsert}
\def\ifig#1#2#3{\DefWarn#1\xdef#1{fig.~\noexpand\hyperref{}{figure}%
{\the\figno}{\the\figno}}\writedef{#1\leftbracket fig.\noexpand~\xfig#1}%
\figinsert\figin{\centerline{#3}}\medskip\centerline{\vbox{\baselineskip12pt
\advance\hsize by -1truein\noindent\wrlabeL{#1=#1}\footnotefont%
{\bf Fig.~\hyperdef\hypernoname{figure}{\the\figno}{\the\figno}:} #2}}
\bigskip\endinsert\global\advance\figno by1}
\newwrite\lfile
{\escapechar-1\xdef\pctsign{\string\%}\xdef\leftbracket{\string\{}
\xdef\rightbracket{\string\}}\xdef\numbersign{\string\#}}
\def\writedefs{\immediate\openout\lfile=\jobname.defs \def\writedef##1{%
{\let\hyperref=\relax\let\hyperdef=\relax\let\hypernoname=\relax
 \immediate\write\lfile{\string\def\string##1\rightbracket}}}}%
\def\writestop{\def\writestoppt{\immediate\write\lfile{\string\pageno
 \the\pageno\string\startrefs\leftbracket\the\refno\rightbracket
 \string\def\string\secsym\leftbracket\secsym\rightbracket
 \string\secno\the\secno\string\meqno\the\meqno}\immediate\closeout\lfile}}
\def\writestoppt{}\def\writedef#1{}
\def\seclab#1{\DefWarn#1%
\xdef #1{\noexpand\hyperref{}{section}{\the\secno}{\the\secno}}%
\writedef{#1\leftbracket#1}\wrlabeL{#1=#1}}
\def\subseclab#1{\DefWarn#1%
\xdef #1{\noexpand\hyperref{}{subsection}{\secn@m.\the\subsecno}%
{\secn@m.\the\subsecno}}\writedef{#1\leftbracket#1}\wrlabeL{#1=#1}}
\def\applab#1{\DefWarn#1%
\xdef #1{\noexpand\hyperref{}{appendix}{\secn@m}{\secn@m}}%
\writedef{#1\leftbracket#1}\wrlabeL{#1=#1}}
\newwrite\tfile \def\writetoca#1{}
\def\leaderfill{\leaders\hbox to 1em{\hss.\hss}\hfill}
\def\writetoc{\immediate\openout\tfile=\jobname.toc
   \def\writetoca##1{{\edef\next{\write\tfile{\noindent ##1
   \string\leaderfill {\string\hyperref{}{page}{\noexpand\number\pageno}%
                       {\noexpand\number\pageno}} \par}}\next}}}
\newread\ch@ckfile
\def\listtoc{\immediate\closeout\tfile\immediate\openin\ch@ckfile=\jobname.toc
\ifeof\ch@ckfile\message{no file \jobname.toc, no table of contents this pass}%
\else\closein\ch@ckfile\centerline{\bf Contents}\nobreak\medskip%
{\baselineskip=12pt\footnotefont\parskip=0pt\catcode`\@=11\input\jobname.toc
\catcode`\@=12\bigbreak\bigskip}\fi}
\catcode`\@=12 
%
\edef\tfontsize{\ifx\answ\bigans scaled\magstep3\else scaled\magstep4\fi}
\font\titlerm=cmr10 \tfontsize \font\titlerms=cmr7 \tfontsize
\font\titlermss=cmr5 \tfontsize \font\titlei=cmmi10 \tfontsize
\font\titleis=cmmi7 \tfontsize \font\titleiss=cmmi5 \tfontsize
\font\titlesy=cmsy10 \tfontsize \font\titlesys=cmsy7 \tfontsize
\font\titlesyss=cmsy5 \tfontsize \font\titleit=cmti10 \tfontsize
\skewchar\titlei='177 \skewchar\titleis='177 \skewchar\titleiss='177
\skewchar\titlesy='60 \skewchar\titlesys='60 \skewchar\titlesyss='60
\def\titlefont{\def\rm{\fam0\titlerm}
\textfont0=\titlerm \scriptfont0=\titlerms \scriptscriptfont0=\titlermss
\textfont1=\titlei \scriptfont1=\titleis \scriptscriptfont1=\titleiss
\textfont2=\titlesy \scriptfont2=\titlesys \scriptscriptfont2=\titlesyss
\textfont\itfam=\titleit \def\it{\fam\itfam\titleit}\rm}
 \ifx\answ\bigans\else scaled\magstep1\fi
\ifx\answ\bigans\def\abstractfont{\tenpoint}\else
\font\absit=cmti10 scaled \magstep1
\font\abssl=cmsl10 scaled \magstep1
\font\absrm=cmr10 scaled\magstep1 \font\absrms=cmr7 scaled\magstep1
\font\absrmss=cmr5 scaled\magstep1 \font\absi=cmmi10 scaled\magstep1
\font\absis=cmmi7 scaled\magstep1 \font\absiss=cmmi5 scaled\magstep1
\font\abssy=cmsy10 scaled\magstep1 \font\abssys=cmsy7 scaled\magstep1
\font\abssyss=cmsy5 scaled\magstep1 \font\absbf=cmbx10 scaled\magstep1
\skewchar\absi='177 \skewchar\absis='177 \skewchar\absiss='177
\skewchar\abssy='60 \skewchar\abssys='60 \skewchar\abssyss='60
\def\abstractfont{\def\rm{\fam0\absrm}
\textfont0=\absrm \scriptfont0=\absrms \scriptscriptfont0=\absrmss
\textfont1=\absi \scriptfont1=\absis \scriptscriptfont1=\absiss
\textfont2=\abssy \scriptfont2=\abssys \scriptscriptfont2=\abssyss
\textfont\itfam=\absit \def\it{\fam\itfam\absit}\def\footnotefont{\tenpoint}%
\textfont\slfam=\abssl \def\sl{\fam\slfam\abssl}%
\textfont\bffam=\absbf \def\bf
{\fam\bffam\absbf}\rm}\fi
\def\tenpoint{\def\rm{\fam0\tenrm}
\textfont0=\tenrm \scriptfont0=\sevenrm \scriptscriptfont0=\fiverm
\textfont1=\teni  \scriptfont1=\seveni  \scriptscriptfont1=\fivei
\textfont2=\tensy \scriptfont2=\sevensy \scriptscriptfont2=\fivesy
\textfont\itfam=\tenit \def\it{\fam\itfam\tenit}\def\footnotefont{\ninepoint}%
\textfont\bffam=\tenbf \def\bf{\fam\bffam\tenbf}\def\sl{\fam\slfam\tensl}\rm}
\font\ninerm=cmr9 \font\sixrm=cmr6 \font\ninei=cmmi9 \font\sixi=cmmi6
\font\ninesy=cmsy9 \font\sixsy=cmsy6 \font\ninebf=cmbx9
\font\nineit=cmti9 \font\ninesl=cmsl9 \skewchar\ninei='177
\skewchar\sixi='177 \skewchar\ninesy='60 \skewchar\sixsy='60
\def\ninepoint{\def\rm{\fam0\ninerm}
\textfont0=\ninerm \scriptfont0=\sixrm \scriptscriptfont0=\fiverm
\textfont1=\ninei \scriptfont1=\sixi \scriptscriptfont1=\fivei
\textfont2=\ninesy \scriptfont2=\sixsy \scriptscriptfont2=\fivesy
\textfont\itfam=\ninei \def\it{\fam\itfam\nineit}\def\sl{\fam\slfam\ninesl}%
\textfont\bffam=\ninebf \def\bf{\fam\bffam\ninebf}\rm}
%
%
\def\noblackbox{\overfullrule=0pt}
\hyphenation{anom-aly anom-alies coun-ter-term coun-ter-terms}
\def\inv{^{\raise.15ex\hbox{${\scriptscriptstyle -}$}\kern-.05em 1}}

\def\Dsl{\,\raise.15ex\hbox{/}\mkern-13.5mu D} 
\def\dsl{\raise.15ex\hbox{/}\kern-.57em\partial}
\def\del{\partial}

\def\lspace{\ifx\answ\bigans{}\else\qquad\fi}
\def\lbspace{\ifx\answ\bigans{}\else\hskip-.2in\fi} 

\def\boxeqn#1{\vcenter{\vbox{\hrule\hbox{\vrule\kern3pt\vbox{\kern3pt
	\hbox{${\displaystyle #1}$}\kern3pt}\kern3pt\vrule}\hrule}}}
\def\mbox#1#2{\vcenter{\hrule \hbox{\vrule height#2in
		\kern#1in \vrule} \hrule}}  

\def\grad#1{\,\nabla\!_{{#1}}\,}

\def\darr#1{\raise1.5ex\hbox{$\leftrightarrow$}\mkern-16.5mu #1}

\def\roughly#1{\raise.3ex\hbox{$#1$\kern-.75em\lower1ex\hbox{$\sim$}}}

\input amssym
\input epsf

\def\IZ{\relax\ifmmode\mathchoice
{\hbox{\cmss Z\kern-.4em Z}}{\hbox{\cmss Z\kern-.4em Z}} {\lower.9pt\hbox{\cmsss Z\kern-.4em Z}}
{\lower1.2pt\hbox{\cmsss Z\kern-.4em Z}}\else{\cmss Z\kern-.4em Z}\fi}

\newif\ifdraft\draftfalse
\newif\ifinter\interfalse
\ifdraft\draftmode\else\interfalse\fi
\def\journal#1&#2(#3){\unskip, \sl #1\ \bf #2 \rm(19#3) }
\def\andjournal#1&#2(#3){\sl #1~\bf #2 \rm (19#3) }

\def\frac#1#2{{#1\over#2}}

\def\ds{\displaystyle}

\def\inbar{\,\vrule height1.5ex width.4pt depth0pt}
\def\IC{\relax\hbox{$\inbar\kern-.3em{\rm C}$}}
\def\IR{\relax{\rm I\kern-.18em R}}
\def\IP{\relax{\rm I\kern-.18em P}}

%
%


%
\catcode`\@=11
\def\slash#1{\mathord{\mathpalette\c@ncel{#1}}}
\overfullrule=0pt

\def\Z{\hbox{$\bb Z$}}
\def\R{\hbox{$\bb R$}}

\def\underrel#1\over#2{\mathrel{\mathop{\kern\z@#1}\limits_{#2}}}

\catcode`\@=12


%

\def\mod{{\rm mod}}

\def\exp{{\rm exp}}


\def\[{[}
\def\]{]}

\def\comment#1{ }

%
\def\draftnote#1{\ifdraft{\baselineskip2ex
                 \vbox{\kern1em\hrule\hbox{\vrule\kern1em\vbox{\kern1ex
                 \noindent \underbar{NOTE}: #1
             \vskip1ex}\kern1em\vrule}\hrule}}\fi}
\def\internote#1{\ifinter{\baselineskip2ex
                 \vbox{\kern1em\hrule\hbox{\vrule\kern1em\vbox{\kern1ex
                 \noindent \underbar{Internal Note}: #1
             \vskip1ex}\kern1em\vrule}\hrule}}\fi}

%

%
%

%

\def\inv{^{-1}}



\def\b{\beta}


\def\bb{
\font\tenmsb=msbm10
\font\sevenmsb=msbm7
\font\fivemsb=msbm5
\textfont1=\tenmsb
\scriptfont1=\sevenmsb
\scriptscriptfont1=\fivemsb
}





\def\bar{\overline}
\def\b{\bar}
\def\bsq#1{{{\b{#1}}^{\lower 2.5pt\hbox{$\scriptstyle 2$}}}}
\def\bexp#1#2{{{\b{#1}}^{\lower 2.5pt\hbox{$\scriptstyle #2$}}}}
\def\dotexp#1#2{{{#1}^{\lower 2.5pt\hbox{$\scriptstyle #2$}}}}


\def\det{\mathop{\rm det}}

\def\Tr{\mathop{\rm Tr}}

\def\rt2{\sqrt{2}}

\def\grad{\nabla}
\def\mod{{\rm mod}}



\def\CF{{\cal F}}


\def\1{{\ds 1}}
\def\R{\hbox{$\bb R$}}
\def\C{\hbox{$\bb C$}}

\def\Z{\hbox{$\bb Z$}}

\def\P{\hbox{$\bb P$}}

\def\T{\hbox{$\bb T$}}


\noblackbox

\def\unit{\relax{\rm 1\kern-.26em I}}
\def\nada{\relax{\rm 0\kern-.30em l}}

\def\mod{{\rm \ mod \ }}

\noblackbox
\def\IL{\relax{\rm I\kern-.18em L}}
\def\IH{\relax{\rm I\kern-.18em H}}
\def\IR{\relax{\rm I\kern-.18em R}}
\def\IC{\relax\hbox{$\inbar\kern-.3em{\rm C}$}}
\def\IZ{\relax\ifmmode\mathchoice
{\hbox{\cmss Z\kern-.4em Z}}{\hbox{\cmss Z\kern-.4em Z}} {\lower.9pt\hbox{\cmsss Z\kern-.4em Z}}
{\lower1.2pt\hbox{\cmsss Z\kern-.4em Z}}\else{\cmss Z\kern-.4em Z}\fi}

\def\partialslash{\not{\hbox{\kern-2pt $\partial$}}}

\font\manual=manfnt \def\dbend{\lower3.5pt\hbox{\manual\char127}}

\def\IZ{\relax\ifmmode\mathchoice
{\hbox{\cmss Z\kern-.4em Z}}{\hbox{\cmss Z\kern-.4em Z}} {\lower.9pt\hbox{\cmsss Z\kern-.4em Z}}
{\lower1.2pt\hbox{\cmsss Z\kern-.4em Z}}\else{\cmss Z\kern-.4em Z}\fi}
\def\half{{1\over 2}}

\def\bar{\overline}

\def\rt2{\sqrt{2}}
\def\irt2{{1\over\sqrt{2}}}

\def\slashchar#1{\setbox0=\hbox{$#1$}           
   \dimen0=\wd0                                 
   \setbox1=\hbox{/} \dimen1=\wd1               
   \ifdim\dimen0>\dimen1                        
      \rlap{\hbox to \dimen0{\hfil/\hfil}}      
      #1                                        
   \else                                        
      \rlap{\hbox to \dimen1{\hfil$#1$\hfil}}   
      /                                         
   \fi}

\def\gcd{{\rm gcd}}

\def\figcaption#1#2{\DefWarn#1\xdef#1{Figure~\noexpand\hyperref{}{figure}%
{\the\figno}{\the\figno}}\writedef{#1\leftbracket Figure\noexpand~\xfig#1}%
\medskip\centerline{{\footnotefont\bf Figure~\hyperdef\hypernoname{figure}{\the\figno}{\the\figno}:}  #2 \wrlabeL{#1=#1}}%
\global\advance\figno by1}

\lref\VenezianoYB{
  G.~Veneziano,
  ``Construction of a crossing - symmetric, Regge behaved amplitude for linearly rising trajectories,''
Nuovo Cim.\ A {\bf 57}, 190 (1968).
}

\lref\GrossBR{
  D.~J.~Gross, R.~D.~Pisarski and L.~G.~Yaffe,
  ``QCD and Instantons at Finite Temperature,''
Rev.\ Mod.\ Phys.\  {\bf 53}, 43 (1981).
}

\lref\SvetitskyGS{
  B.~Svetitsky and L.~G.~Yaffe,
  ``Critical Behavior at Finite Temperature Confinement Transitions,''
Nucl.\ Phys.\ B {\bf 210}, 423 (1982).
}

\lref\SvetitskyYE{
  B.~Svetitsky,
  ``Symmetry Aspects of Finite Temperature Confinement Transitions,''
Phys.\ Rept.\  {\bf 132}, 1 (1986).
}

\lref\SiversIG{
  D.~Sivers and J.~Yellin,
  ``Review of recent work on narrow resonance models,''
Rev.\ Mod.\ Phys.\  {\bf 43}, 125 (1971).
}

\lref\WittenEY{
  E.~Witten,
  ``Dyons of Charge e theta/2 pi,''
Phys.\ Lett.\ B {\bf 86}, 283 (1979).
}

\lref\GreenSG{
  M.~B.~Green and J.~H.~Schwarz,
  ``Anomaly Cancellation in Supersymmetric D=10 Gauge Theory and Superstring Theory,''
Phys.\ Lett.\  {\bf 149B}, 117 (1984).
}

\lref\AcharyaDZ{
  B.~S.~Acharya and C.~Vafa,
  ``On domain walls of N=1 supersymmetric Yang-Mills in four-dimensions,''
[hep-th/0103011].
}

\lref\AffleckCH{
  I.~Affleck and F.~D.~M.~Haldane,
  ``Critical Theory of Quantum Spin Chains,''
Phys.\ Rev.\ B {\bf 36}, 5291 (1987).
}

\lref\BilloJDA{
  M.~Billó, M.~Caselle, D.~Gaiotto, F.~Gliozzi, M.~Meineri and R.~Pellegrini,
  ``Line defects in the 3d Ising model,''
JHEP {\bf 1307}, 055 (2013).
[arXiv:1304.4110 [hep-th]].
}

\lref\WittenABA{
  E.~Witten,
  ``Fermion Path Integrals And Topological Phases,''
Rev.\ Mod.\ Phys.\  {\bf 88}, no. 3, 035001 (2016).
[arXiv:1508.04715 [cond-mat.mes-hall]].
}

\lref\GaiottoYUP{
  D.~Gaiotto, A.~Kapustin, Z.~Komargodski and N.~Seiberg,
  ``Theta, Time Reversal, and Temperature,''
JHEP {\bf 1705}, 091 (2017).
[arXiv:1703.00501 [hep-th]].
}

\lref\GaiottoNVA{
  D.~Gaiotto, D.~Mazac and M.~F.~Paulos,
  min
  ``Bootstrapping the 3d Ising twist defect,''
JHEP {\bf 1403}, 100 (2014).
[arXiv:1310.5078 [hep-th]].
}

\lref\BrowerEA{
  R.~C.~Brower, J.~Polchinski, M.~J.~Strassler and C.~I.~Tan,
  ``The Pomeron and gauge/string duality,''
JHEP {\bf 0712}, 005 (2007).
[hep-th/0603115].
}

\lref\DieriglXTA{
  M.~Dierigl and A.~Pritzel,
  ``Topological Model for Domain Walls in (Super-)Yang-Mills Theories,''
Phys.\ Rev.\ D {\bf 90}, no. 10, 105008 (2014).
[arXiv:1405.4291 [hep-th]].
}

\lref\mandelstam{
S.~Mandelstam, ``Dual-resonance models." Physics Reports 13.6 (1974): 259-353.
}

\lref\FreundHW{
  P.~G.~O.~Freund,
  ``Finite energy sum rules and bootstraps,''
Phys.\ Rev.\ Lett.\  {\bf 20}, 235 (1968).
}

\lref\MeyerJC{
  H.~B.~Meyer and M.~J.~Teper,
  ``Glueball Regge trajectories and the pomeron: A Lattice study,''
Phys.\ Lett.\ B {\bf 605}, 344 (2005).
[hep-ph/0409183].
}

\lref\CoonYW{
  D.~D.~Coon,
Phys.\ Lett.\ B {\bf 29}, 669 (1969).
}

\lref\FairlieAD{
  D.~B.~Fairlie and J.~Nuyts,
Nucl.\ Phys.\ B {\bf 433}, 26 (1995).
[hep-th/9406043].
}

\lref\RedlichDV{
  A.~N.~Redlich,
  ``Parity Violation and Gauge Noninvariance of the Effective Gauge Field Action in Three-Dimensions,''
Phys.\ Rev.\ D {\bf 29}, 2366 (1984).
}

\lref\RedlichKN{
  A.~N.~Redlich,
  ``Gauge Noninvariance and Parity Violation of Three-Dimensional Fermions,''
Phys.\ Rev.\ Lett.\  {\bf 52}, 18 (1984).
}

\lref\PonomarevJQK{
  D.~Ponomarev and A.~A.~Tseytlin,
[arXiv:1603.06273 [hep-th]].
}

\lref\StromingerTalk{
  A.~Strominger, Talk at Strings 2014, Princeton.
}

\lref\CostaMG{
  M.~S.~Costa, J.~Penedones, D.~Poland and S.~Rychkov,
JHEP {\bf 1111}, 071 (2011).
[arXiv:1107.3554 [hep-th]].
}

\lref\AbanovQZ{
  A.~G.~Abanov and P.~B.~Wiegmann,
  ``Theta terms in nonlinear sigma models,''
Nucl.\ Phys.\ B {\bf 570}, 685 (2000).
[hep-th/9911025].
}

\lref\AffleckAS{
  I.~Affleck, J.~A.~Harvey and E.~Witten,
  ``Instantons and (Super)Symmetry Breaking in (2+1)-Dimensions,''
Nucl.\ Phys.\ B {\bf 206}, 413 (1982).
}

\lref\AharonyBX{
  O.~Aharony, A.~Hanany, K.~A.~Intriligator, N.~Seiberg and M.~J.~Strassler,
  ``Aspects of $N=2$ supersymmetric gauge theories in three-dimensions,''
Nucl.\ Phys.\ B {\bf 499}, 67 (1997).
[hep-th/9703110].
}

\lref\AharonyGP{
  O.~Aharony,
  ``IR duality in $d = 3$ $N=2$ supersymmetric $USp(2N_c)$ and $U(N_c)$ gauge theories,''
Phys.\ Lett.\ B {\bf 404}, 71 (1997).
[hep-th/9703215].
}

\lref\AharonyCI{
  O.~Aharony and I.~Shamir,
  ``On $O(N_c)$ $d=3$ ${\cal N}{=}2$ supersymmetric QCD Theories,''
JHEP {\bf 1112}, 043 (2011).
[arXiv:1109.5081 [hep-th]].
}

\lref\AharonyJZ{
  O.~Aharony, G.~Gur-Ari and R.~Yacoby,
  ``$d=3$ Bosonic Vector Models Coupled to Chern-Simons Gauge Theories,''
JHEP {\bf 1203}, 037 (2012).
[arXiv:1110.4382 [hep-th]].
}

\lref\AharonyNH{
  O.~Aharony, G.~Gur-Ari and R.~Yacoby,
  ``Correlation Functions of Large $N$ Chern-Simons-Matter Theories and Bosonization in Three Dimensions,''
JHEP {\bf 1212}, 028 (2012).
[arXiv:1207.4593 [hep-th]].
}

\lref\AharonyNS{
  O.~Aharony, S.~Giombi, G.~Gur-Ari, J.~Maldacena and R.~Yacoby,
  ``The Thermal Free Energy in Large $N$ Chern-Simons-Matter Theories,''
JHEP {\bf 1303}, 121 (2013).
[arXiv:1211.4843 [hep-th]].
}

\lref\AharonyHDA{
  O.~Aharony, N.~Seiberg and Y.~Tachikawa,
  ``Reading between the lines of four-dimensional gauge theories,''
JHEP {\bf 1308}, 115 (2013).
[arXiv:1305.0318 [hep-th]].
}

\lref\AharonyDHA{
  O.~Aharony, S.~S.~Razamat, N.~Seiberg and B.~Willett,
  ``3d dualities from 4d dualities,''
JHEP {\bf 1307}, 149 (2013).
[arXiv:1305.3924 [hep-th]].
}

\lref\AharonyKMA{
  O.~Aharony, S.~S.~Razamat, N.~Seiberg and B.~Willett,
  ``3d dualities from 4d dualities for orthogonal groups,''
JHEP {\bf 1308}, 099 (2013)
[arXiv:1307.0511 [hep-th]].
}

\lref\AharonyMJS{
  O.~Aharony,
  ``Baryons, monopoles and dualities in Chern-Simons-matter theories,''
JHEP {\bf 1602}, 093 (2016).
[arXiv:1512.00161 [hep-th]].
}

\lref\AharonyJVV{
  O.~Aharony, F.~Benini, P.~S.~Hsin and N.~Seiberg,
  ``Chern-Simons-matter dualities with $SO$ and $USp$ gauge groups,''
JHEP {\bf 1702}, 072 (2017).
[arXiv:1611.07874 [cond-mat.str-el]].
}

\lref\WangTXT{
  C.~Wang, A.~Nahum, M.~A.~Metlitski, C.~Xu and T.~Senthil,
  ``Deconfined quantum critical points: symmetries and dualities,''
[arXiv:1703.02426 [cond-mat.str-el]].
}

\lref\AlvarezGaumeNF{
  L.~Alvarez-Gaume, S.~Della Pietra and G.~W.~Moore,
  ``Anomalies and Odd Dimensions,''
Annals Phys.\  {\bf 163}, 288 (1985).
}

\lref\AnninosUI{
  D.~Anninos, T.~Hartman and A.~Strominger,
  ``Higher Spin Realization of the dS/CFT Correspondence,''
[arXiv:1108.5735 [hep-th]].
}

\lref\AnninosHIA{
  D.~Anninos, R.~Mahajan, D.~Radicevic and E.~Shaghoulian,
  ``Chern-Simons-Ghost Theories and de Sitter Space,''
JHEP {\bf 1501}, 074 (2015).
[arXiv:1405.1424 [hep-th]].
}

\lref\AtiyahJF{
  M.~F.~Atiyah, V.~K.~Patodi and I.~M.~Singer,
  ``Spectral Asymmetry in Riemannian Geometry, I,''
  Math.\ Proc.\ Camb.\ Phil.\ Soc.\ {\bf 77} (1975) 43--69.
}

\lref\BanksZN{
  T.~Banks and N.~Seiberg,
  ``Symmetries and Strings in Field Theory and Gravity,''
Phys.\ Rev.\ D {\bf 83}, 084019 (2011).
[arXiv:1011.5120 [hep-th]].
}

\lref\BarkeshliIDA{
  M.~Barkeshli and J.~McGreevy,
  ``Continuous transition between fractional quantum Hall and superfluid states,''
Phys.\ Rev.\ B {\bf 89}, 235116 (2014).
}

\lref\BeemMB{
  C.~Beem, T.~Dimofte and S.~Pasquetti,
  ``Holomorphic Blocks in Three Dimensions,''
[arXiv:1211.1986 [hep-th]].
}
\lref\VafaXG{
  C.~Vafa and E.~Witten,
  ``Parity Conservation in QCD,''
Phys.\ Rev.\ Lett.\  {\bf 53}, 535 (1984).
}

\lref\BeniniMF{
  F.~Benini, C.~Closset and S.~Cremonesi,
  ``Comments on 3d Seiberg-like dualities,''
JHEP {\bf 1110}, 075 (2011).
[arXiv:1108.5373 [hep-th]].
}

\lref\BernardXY{
  D.~Bernard,
  ``String Characters From {Kac-Moody} Automorphisms,''
  Nucl.\ Phys.\ B {\bf 288}, 628 (1987).
}

\lref\BhattacharyaZY{
  J.~Bhattacharya, S.~Bhattacharyya, S.~Minwalla and S.~Raju,
  ``Indices for Superconformal Field Theories in 3,5 and 6 Dimensions,''
JHEP {\bf 0802}, 064 (2008).
[arXiv:0801.1435 [hep-th]].
}

\lref\deBoerMP{
  J.~de Boer, K.~Hori, H.~Ooguri and Y.~Oz,
  ``Mirror symmetry in three-dimensional gauge theories, quivers and D-branes,''
Nucl.\ Phys.\ B {\bf 493}, 101 (1997).
[hep-th/9611063].
}

\lref\deBoerKA{
  J.~de Boer, K.~Hori, Y.~Oz and Z.~Yin,
  ``Branes and mirror symmetry in N=2 supersymmetric gauge theories in three-dimensions,''
Nucl.\ Phys.\ B {\bf 502}, 107 (1997).
[hep-th/9702154].
}

\lref\WilczekCY{
  F.~Wilczek and A.~Zee,
  ``Linking Numbers, Spin, and Statistics of Solitons,''
Phys.\ Rev.\ Lett.\  {\bf 51}, 2250 (1983).
}

\lref\BondersonPLA{
  P.~Bonderson, C.~Nayak and X.~L.~Qi,
  ``A time-reversal invariant topological phase at the surface of a 3D topological insulator,''
J.\ Stat.\ Mech.\  {\bf 2013}, P09016 (2013).
}

\lref\BorokhovIB{
  V.~Borokhov, A.~Kapustin and X.~k.~Wu,
  ``Topological disorder operators in three-dimensional conformal field theory,''
JHEP {\bf 0211}, 049 (2002).
[hep-th/0206054].
}

\lref\WittenTW{
  E.~Witten,
  ``Global Aspects of Current Algebra,''
Nucl.\ Phys.\ B {\bf 223}, 422 (1983).
}
\lref\WittenTX{
  E.~Witten,
  ``Current Algebra, Baryons, and Quark Confinement,''
Nucl.\ Phys.\ B {\bf 223}, 433 (1983).
}

\lref\GaiottoYUP{
  D.~Gaiotto, A.~Kapustin, Z.~Komargodski and N.~Seiberg,
  ``Theta, Time Reversal, and Temperature,''
[arXiv:1703.00501 [hep-th]].
}

\lref\BorokhovCG{
  V.~Borokhov, A.~Kapustin and X.~k.~Wu,
  ``Monopole operators and mirror symmetry in three-dimensions,''
JHEP {\bf 0212}, 044 (2002).
[hep-th/0207074].
}

\lref\KomargodskiDMC{
  Z.~Komargodski, A.~Sharon, R.~Thorngren and X.~Zhou,
  ``Comments on Abelian Higgs Models and Persistent Order,''
[arXiv:1705.04786 [hep-th]].
}

\lref\KomargodskiSMK{
  Z.~Komargodski, T.~Sulejmanpasic and M.~Unsal,
  ``Walls, Anomalies, and (De) Confinement in Quantum Anti-Ferromagnets,''
[arXiv:1706.05731 [cond-mat.str-el]].
}

\lref\Browder{
  W.~Browder and E.~Thomas,
  ``Axioms for the generalized Pontryagin cohomology operations,''
  Quart.\ J.\ Math.\ Oxford {\bf 13}, 55--60 (1962).
}

\lref\debult{
  F.~van~de~Bult,
  ``Hyperbolic Hypergeometric Functions,''
University of Amsterdam Ph.D. thesis
}

\lref\Camperi{
  M.~Camperi, F.~Levstein and G.~Zemba,
  ``The Large N Limit Of Chern-simons Gauge Theory,''
  Phys.\ Lett.\ B {\bf 247} (1990) 549.
}

\lref\ChenCD{
  W.~Chen, M.~P.~A.~Fisher and Y.~S.~Wu,
  ``Mott transition in an anyon gas,''
Phys.\ Rev.\ B {\bf 48}, 13749 (1993).
[cond-mat/9301037].
}

\lref\WittenDS{
  E.~Witten,
  ``Supersymmetric index of three-dimensional gauge theory,''
In *Shifman, M.A. (ed.): The many faces of the superworld* 156-184.
[hep-th/9903005].
}

\lref\ChenJHA{
  X.~Chen, L.~Fidkowski and A.~Vishwanath,
  ``Symmetry Enforced Non-Abelian Topological Order at the Surface of a Topological Insulator,''
Phys.\ Rev.\ B {\bf 89}, no. 16, 165132 (2014).
[arXiv:1306.3250 [cond-mat.str-el]].
}

\lref\IntriligatorLCA{
  K.~Intriligator and N.~Seiberg,
  ``Aspects of 3d N=2 Chern-Simons-Matter Theories,''
JHEP {\bf 1307}, 079 (2013).
[arXiv:1305.1633 [hep-th]].
}

\lref\ChengPDN{
  M.~Cheng and C.~Xu,
  ``Series of (2+1)-dimensional stable self-dual interacting conformal field theories,''
Phys.\ Rev.\ B {\bf 94}, 214415 (2016). 
[arXiv:1609.02560 [cond-mat.str-el]].
}

\lref\ClossetVG{
  C.~Closset, T.~T.~Dumitrescu, G.~Festuccia, Z.~Komargodski and N.~Seiberg,
  ``Contact Terms, Unitarity, and F-Maximization in Three-Dimensional Superconformal Theories,''
JHEP {\bf 1210}, 053 (2012).
[arXiv:1205.4142 [hep-th]].
}

\lref\ClossetVP{
  C.~Closset, T.~T.~Dumitrescu, G.~Festuccia, Z.~Komargodski and N.~Seiberg,
  ``Comments on Chern-Simons Contact Terms in Three Dimensions,''
JHEP {\bf 1209}, 091 (2012).
[arXiv:1206.5218 [hep-th]].
}

\lref\ClossetRU{
  C.~Closset, T.~T.~Dumitrescu, G.~Festuccia and Z.~Komargodski,
  ``Supersymmetric Field Theories on Three-Manifolds,''
JHEP {\bf 1305}, 017 (2013).
[arXiv:1212.3388 [hep-th]].
}

\lref\CveticXN{
  M.~Cvetic, T.~W.~Grimm and D.~Klevers,
  ``Anomaly Cancellation And Abelian Gauge Symmetries In F-theory,''
JHEP {\bf 1302}, 101 (2013).
[arXiv:1210.6034 [hep-th]].
}

\lref\DaiKQ{
  X.~z.~Dai and D.~S.~Freed,
  ``eta invariants and determinant lines,''
J.\ Math.\ Phys.\  {\bf 35}, 5155 (1994), Erratum: [J.\ Math.\ Phys.\  {\bf 42}, 2343 (2001)].
[hep-th/9405012].
}

\lref\DasguptaZZ{
  C.~Dasgupta and B.~I.~Halperin,
  ``Phase Transition in a Lattice Model of Superconductivity,''
Phys.\ Rev.\ Lett.\  {\bf 47}, 1556 (1981).
}

\lref\DaviesUW{
  N.~M.~Davies, T.~J.~Hollowood, V.~V.~Khoze and M.~P.~Mattis,
  ``Gluino condensate and magnetic monopoles in supersymmetric gluodynamics,''
Nucl.\ Phys.\ B {\bf 559}, 123 (1999).
[hep-th/9905015].
}

\lref\DaviesNW{
  N.~M.~Davies, T.~J.~Hollowood and V.~V.~Khoze,
  ``Monopoles, affine algebras and the gluino condensate,''
J.\ Math.\ Phys.\  {\bf 44}, 3640 (2003).
[hep-th/0006011].
}

\lref\DimoftePY{
  T.~Dimofte, D.~Gaiotto and S.~Gukov,
  ``3-Manifolds and 3d Indices,''
[arXiv:1112.5179 [hep-th]].
}

\lref\DolanQI{
  F.~A.~Dolan and H.~Osborn,
  ``Applications of the Superconformal Index for Protected Operators and q-Hypergeometric Identities to N=1 Dual Theories,''
Nucl.\ Phys.\ B {\bf 818}, 137 (2009).
[arXiv:0801.4947 [hep-th]].
}

\lref\DolanRP{
  F.~A.~H.~Dolan, V.~P.~Spiridonov and G.~S.~Vartanov,
  ``From 4d superconformal indices to 3d partition functions,''
Phys.\ Lett.\ B {\bf 704}, 234 (2011).
[arXiv:1104.1787 [hep-th]].
}

\lref\DouglasEX{
  M.~R.~Douglas,
  ``Chern-Simons-Witten theory as a topological Fermi liquid,''
[hep-th/9403119].
}

\lref\EagerHX{
  R.~Eager, J.~Schmude and Y.~Tachikawa,
  ``Superconformal Indices, Sasaki-Einstein Manifolds, and Cyclic Homologies,''
[arXiv:1207.0573 [hep-th]].
}

\lref\VenezianoYB{
  G.~Veneziano,
  ``Construction of a crossing - symmetric, Regge behaved amplitude for linearly rising trajectories,''
Nuovo Cim.\ A {\bf 57}, 190 (1968).
}

\lref\GrossBR{
  D.~J.~Gross, R.~D.~Pisarski and L.~G.~Yaffe,
  ``QCD and Instantons at Finite Temperature,''
Rev.\ Mod.\ Phys.\  {\bf 53}, 43 (1981).
}

\lref\SvetitskyGS{
  B.~Svetitsky and L.~G.~Yaffe,
  ``Critical Behavior at Finite Temperature Confinement Transitions,''
Nucl.\ Phys.\ B {\bf 210}, 423 (1982).
}

\lref\SvetitskyYE{
  B.~Svetitsky,
  ``Symmetry Aspects of Finite Temperature Confinement Transitions,''
Phys.\ Rept.\  {\bf 132}, 1 (1986).
}

\lref\WittenUKA{
  E.~Witten,
  ``Theta dependence in the large N limit of four-dimensional gauge theories,''
Phys.\ Rev.\ Lett.\  {\bf 81}, 2862 (1998).
[hep-th/9807109].
}

\lref\SiversIG{
  D.~Sivers and J.~Yellin,
  ``Review of recent work on narrow resonance models,''
Rev.\ Mod.\ Phys.\  {\bf 43}, 125 (1971).
}

\lref\WittenEY{
  E.~Witten,
  ``Dyons of Charge e theta/2 pi,''
Phys.\ Lett.\ B {\bf 86}, 283 (1979).
}

\lref\GreenSG{
  M.~B.~Green and J.~H.~Schwarz,
  ``Anomaly Cancellation in Supersymmetric D=10 Gauge Theory and Superstring Theory,''
Phys.\ Lett.\  {\bf 149B}, 117 (1984).
}

\lref\ColemanUZ{
  S.~R.~Coleman,
  ``More About the Massive Schwinger Model,''
Annals Phys.\  {\bf 101}, 239 (1976).
}

\lref\BaluniRF{
  V.~Baluni,
  ``CP Violating Effects in QCD,''
Phys.\ Rev.\ D {\bf 19}, 2227 (1979).
}

\lref\DashenET{
  R.~F.~Dashen,
  ``Some features of chiral symmetry breaking,''
Phys.\ Rev.\ D {\bf 3}, 1879 (1971).
}

\lref\WittenBC{
  E.~Witten,
  ``Instantons, the Quark Model, and the 1/n Expansion,''
Nucl.\ Phys.\ B {\bf 149}, 285 (1979).
}

\lref\WittenVV{
  E.~Witten,
  ``Current Algebra Theorems for the U(1) Goldstone Boson,''
Nucl.\ Phys.\ B {\bf 156}, 269 (1979).
}

\lref\WittenSP{
  E.~Witten,
  ``Large N Chiral Dynamics,''
Annals Phys.\  {\bf 128}, 363 (1980).
}

\lref\DAddaVBW{
  A.~D'Adda, M.~Luscher and P.~Di Vecchia,
  ``A 1/n Expandable Series of Nonlinear Sigma Models with Instantons,''
Nucl.\ Phys.\ B {\bf 146}, 63 (1978).
}

\lref\AcharyaDZ{
  B.~S.~Acharya and C.~Vafa,
  ``On domain walls of N=1 supersymmetric Yang-Mills in four-dimensions,''
[hep-th/0103011].
}

\lref\AffleckCH{
  I.~Affleck and F.~D.~M.~Haldane,
  ``Critical Theory of Quantum Spin Chains,''
Phys.\ Rev.\ B {\bf 36}, 5291 (1987).
}

\lref\BilloJDA{
  M.~Billó, M.~Caselle, D.~Gaiotto, F.~Gliozzi, M.~Meineri and R.~Pellegrini,
  ``Line defects in the 3d Ising model,''
JHEP {\bf 1307}, 055 (2013).
[arXiv:1304.4110 [hep-th]].
}

\lref\WittenABA{
  E.~Witten,
  ``Fermion Path Integrals And Topological Phases,''
Rev.\ Mod.\ Phys.\  {\bf 88}, no. 3, 035001 (2016).
[arXiv:1508.04715 [cond-mat.mes-hall]].
}

\lref\GaiottoNVA{
  D.~Gaiotto, D.~Mazac and M.~F.~Paulos,
  ``Bootstrapping the 3d Ising twist defect,''
JHEP {\bf 1403}, 100 (2014).
[arXiv:1310.5078 [hep-th]].
}

\lref\BrowerEA{
  R.~C.~Brower, J.~Polchinski, M.~J.~Strassler and C.~I.~Tan,
  ``The Pomeron and gauge/string duality,''
JHEP {\bf 0712}, 005 (2007).
[hep-th/0603115].
}

\lref\mandelstam{
S.~Mandelstam, ``Dual-resonance models." Physics Reports 13.6 (1974): 259-353.
}

\lref\FreundHW{
  P.~G.~O.~Freund,
  ``Finite energy sum rules and bootstraps,''
Phys.\ Rev.\ Lett.\  {\bf 20}, 235 (1968).
}

\lref\MeyerJC{
  H.~B.~Meyer and M.~J.~Teper,
  ``Glueball Regge trajectories and the pomeron: A Lattice study,''
Phys.\ Lett.\ B {\bf 605}, 344 (2005).
[hep-ph/0409183].
}

\lref\CoonYW{
  D.~D.~Coon,
  ``Uniqueness of the veneziano representation,''
Phys.\ Lett.\ B {\bf 29}, 669 (1969).
}

\lref\FairlieAD{
  D.~B.~Fairlie and J.~Nuyts,
  ``A fresh look at generalized Veneziano amplitudes,''
Nucl.\ Phys.\ B {\bf 433}, 26 (1995).
[hep-th/9406043].
}

\lref\CWpot{
  S.~R.~Coleman and E.~J.~Weinberg,
  ``Radiative Corrections as the Origin of Spontaneous Symmetry Breaking,''
Phys.\ Rev.\ D {\bf 7}, 1888 (1973).
}

\lref\RedlichDV{
  A.~N.~Redlich,
  ``Parity Violation and Gauge Noninvariance of the Effective Gauge Field Action in Three-Dimensions,''
Phys.\ Rev.\ D {\bf 29}, 2366 (1984).
}

\lref\RedlichKN{
  A.~N.~Redlich,
  ``Gauge Noninvariance and Parity Violation of Three-Dimensional Fermions,''
Phys.\ Rev.\ Lett.\  {\bf 52}, 18 (1984).
}

\lref\PonomarevJQK{
  D.~Ponomarev and A.~A.~Tseytlin,
[arXiv:1603.06273 [hep-th]].
}

\lref\StromingerTalk{
  A.~Strominger, Talk at Strings 2014, Princeton.
}

\lref\PoppitzNZ{
  E.~Poppitz, T.~Schäfer and M.~Ünsal,
  ``Universal mechanism of (semi-classical) deconfinement and theta-dependence for all simple groups,''
JHEP {\bf 1303}, 087 (2013).
[arXiv:1212.1238 [hep-th]].
}

\lref\UnsalZJ{
  M.~Unsal,
  ``Theta dependence, sign problems and topological interference,''
Phys.\ Rev.\ D {\bf 86}, 105012 (2012).
[arXiv:1201.6426 [hep-th]].
}

\lref\CostaMG{
  M.~S.~Costa, J.~Penedones, D.~Poland and S.~Rychkov,
JHEP {\bf 1111}, 071 (2011).
[arXiv:1107.3554 [hep-th]].
}

\lref\TachikawaXVS{
  Y.~Tachikawa and K.~Yonekura,
  ``Gauge interactions and topological phases of matter,''
PTEP {\bf 2016}, no. 9, 093B07 (2016).
[arXiv:1604.06184 [hep-th]].
}

\lref\CamanhoAPA{
  X.~O.~Camanho, J.~D.~Edelstein, J.~Maldacena and A.~Zhiboedov,
JHEP {\bf 1602}, 020 (2016).
[arXiv:1407.5597 [hep-th]].
}

\lref\LandsteinerCP{
  K.~Landsteiner, E.~Megias and F.~Pena-Benitez,
  ``Gravitational Anomaly and Transport,''
Phys.\ Rev.\ Lett.\  {\bf 107}, 021601 (2011).
[arXiv:1103.5006 [hep-ph]].
}

\lref\BanerjeeIZ{
  N.~Banerjee, J.~Bhattacharya, S.~Bhattacharyya, S.~Jain, S.~Minwalla and T.~Sharma,
  ``Constraints on Fluid Dynamics from Equilibrium Partition Functions,''
JHEP {\bf 1209}, 046 (2012).
[arXiv:1203.3544 [hep-th]].
}

\lref\JensenKJ{
  K.~Jensen, R.~Loganayagam and A.~Yarom,
  ``Thermodynamics, gravitational anomalies and cones,''
JHEP {\bf 1302}, 088 (2013).
[arXiv:1207.5824 [hep-th]].
}

\lref\BonettiELA{
  F.~Bonetti, T.~W.~Grimm and S.~Hohenegger,
  ``One-loop Chern-Simons terms in five dimensions,''
JHEP {\bf 1307}, 043 (2013).
[arXiv:1302.2918 [hep-th]].
}

\lref\DiPietroBCA{
  L.~Di Pietro and Z.~Komargodski,
  ``Cardy formulae for SUSY theories in $d =$ 4 and $d =$ 6,''
JHEP {\bf 1412}, 031 (2014).
[arXiv:1407.6061 [hep-th]].
}

\lref\ArdehaliHYA{
  A.~Arabi Ardehali, J.~T.~Liu and P.~Szepietowski,
  ``High-Temperature Expansion of Supersymmetric Partition Functions,''
JHEP {\bf 1507}, 113 (2015).
[arXiv:1502.07737 [hep-th]].
}

\lref\FeiOHA{
  L.~Fei, S.~Giombi, I.~R.~Klebanov and G.~Tarnopolsky,
  ``Generalized $F$-Theorem and the $\epsilon$ Expansion,''
JHEP {\bf 1512}, 155 (2015).
[arXiv:1507.01960 [hep-th]].
}

\lref\AcharyaDZ{
  B.~S.~Acharya and C.~Vafa,
  ``On domain walls of N=1 supersymmetric Yang-Mills in four-dimensions,''
[hep-th/0103011].
}

\lref\GiombiXXA{
  S.~Giombi and I.~R.~Klebanov,
  ``Interpolating between $a$ and $F$,''
JHEP {\bf 1503}, 117 (2015).
[arXiv:1409.1937 [hep-th]].
}

\lref\KapustinGUA{
  A.~Kapustin and N.~Seiberg,
  ``Coupling a QFT to a TQFT and Duality,''
JHEP {\bf 1404}, 001 (2014).
[arXiv:1401.0740 [hep-th]].
}

\lref\GaiottoKFA{
  D.~Gaiotto, A.~Kapustin, N.~Seiberg and B.~Willett,
  ``Generalized Global Symmetries,''
JHEP {\bf 1502}, 172 (2015).
[arXiv:1412.5148 [hep-th]].
}

\lref\DijkgraafPZ{
  R.~Dijkgraaf and E.~Witten,
  ``Topological Gauge Theories and Group Cohomology,''
Commun.\ Math.\ Phys.\  {\bf 129}, 393 (1990).
}

\lref\KadanoffKZ{
  L.~P.~Kadanoff and H.~Ceva,
  ``Determination of an opeator algebra for the two-dimensional Ising model,''
Phys.\ Rev.\ B {\bf 3}, 3918 (1971).
}

\lref\GrossKZA{
  D.~J.~Gross and P.~F.~Mende,
Phys.\ Lett.\ B {\bf 197}, 129 (1987).
}

\lref\KarlinerHD{
  M.~Karliner, I.~R.~Klebanov and L.~Susskind,
Int.\ J.\ Mod.\ Phys.\ A {\bf 3}, 1981 (1988).
}

\lref\Shankar{
  R.~Shankar and N.~Read,
  ``The $\theta = \pi$ Nonlinear $\sigma$ Model Is Massless,''
Nucl.\ Phys.\ B {\bf 336}, 457 (1990).
}

\lref\CachazoZK{
  F.~Cachazo, N.~Seiberg and E.~Witten,
  ``Phases of N=1 supersymmetric gauge theories and matrices,''
JHEP {\bf 0302}, 042 (2003).
[hep-th/0301006].
}

\lref\WeissRJ{
  N.~Weiss,
  ``The Effective Potential for the Order Parameter of Gauge Theories at Finite Temperature,''
Phys.\ Rev.\ D {\bf 24}, 475 (1981).
}

\lref\DashenET{
  R.~F.~Dashen,
  ``Some features of chiral symmetry breaking,''
Phys.\ Rev.\ D {\bf 3}, 1879 (1971).
}

\lref\Hitoshi{
http://hitoshi.berkeley.edu/221b/scattering3.pdf
}

\lref\GreensiteZZ{
  J.~Greensite,
  ``An introduction to the confinement problem,''
Lect.\ Notes Phys.\  {\bf 821}, 1 (2011).
}

\lref\SeibergRS{
  N.~Seiberg and E.~Witten,
  ``Electric - magnetic duality, monopole condensation, and confinement in N=2 supersymmetric Yang-Mills theory,''
Nucl.\ Phys.\ B {\bf 426}, 19 (1994), Erratum: [Nucl.\ Phys.\ B {\bf 430}, 485 (1994)].
[hep-th/9407087].
}

\lref\ColemanUZ{
  S.~R.~Coleman,
  ``More About the Massive Schwinger Model,''
Annals Phys.\  {\bf 101}, 239 (1976).
}
\lref\IntriligatorAU{
  K.~A.~Intriligator and N.~Seiberg,
  ``Lectures on supersymmetric gauge theories and electric-magnetic duality,''
Nucl.\ Phys.\ Proc.\ Suppl.\  {\bf 45BC}, 1 (1996), [Subnucl.\ Ser.\  {\bf 34}, 237 (1997)].
[hep-th/9509066].
}

\lref\SeibergAJ{
  N.~Seiberg and E.~Witten,
  ``Monopoles, duality and chiral symmetry breaking in N=2 supersymmetric QCD,''
Nucl.\ Phys.\ B {\bf 431}, 484 (1994).
[hep-th/9408099].
}

\lref\Deepak{
D.~Naidu, ``Categorical Morita equivalence for group-theoretical categories, " Communications in Algebra 35.11 (2007): 3544-3565.
APA	
}
\lref\AharonyDHA{
  O.~Aharony, S.~S.~Razamat, N.~Seiberg and B.~Willett,
  ``3d dualities from 4d dualities,''
JHEP {\bf 1307}, 149 (2013).
[arXiv:1305.3924 [hep-th]].
}

\lref\ArgyresJJ{
  P.~C.~Argyres and M.~R.~Douglas,
  ``New phenomena in SU(3) supersymmetric gauge theory,''
Nucl.\ Phys.\ B {\bf 448}, 93 (1995).
[hep-th/9505062].
}

\lref\SusskindAA{
  L.~Susskind,
Phys.\ Rev.\ D {\bf 49}, 6606 (1994).
[hep-th/9308139].
}

\lref\DubovskySH{
  S.~Dubovsky, R.~Flauger and V.~Gorbenko,
  ``Effective String Theory Revisited,''
JHEP {\bf 1209}, 044 (2012).
[arXiv:1203.1054 [hep-th]].
}

\lref\AharonyIPA{
  O.~Aharony and Z.~Komargodski,
  ``The Effective Theory of Long Strings,''
JHEP {\bf 1305}, 118 (2013).
[arXiv:1302.6257 [hep-th]].
}

\lref\HellermanCBA{
  S.~Hellerman, S.~Maeda, J.~Maltz and I.~Swanson,
  ``Effective String Theory Simplified,''
JHEP {\bf 1409}, 183 (2014).
[arXiv:1405.6197 [hep-th]].
}

\lref\AthenodorouCS{
  A.~Athenodorou, B.~Bringoltz and M.~Teper,
  ``Closed flux tubes and their string description in D=3+1 SU(N) gauge theories,''
JHEP {\bf 1102}, 030 (2011).
[arXiv:1007.4720 [hep-lat]].
}

\lref\AharonyHDA{
  O.~Aharony, N.~Seiberg and Y.~Tachikawa,
  ``Reading between the lines of four-dimensional gauge theories,''
JHEP {\bf 1308}, 115 (2013).
[arXiv:1305.0318 [hep-th]].
}

\lref\CallanSA{
  C.~G.~Callan, Jr. and J.~A.~Harvey,
  ``Anomalies and Fermion Zero Modes on Strings and Domain Walls,''
Nucl.\ Phys.\ B {\bf 250}, 427 (1985).
}

\lref\tHooftXSS{
  G.~'t Hooft et al.,
    ``Recent Developments in Gauge Theories. Proceedings, Nato Advanced Study Institute, Cargese, France, August 26 - September 8, 1979,''
NATO Sci.\ Ser.\ B {\bf 59}, pp.1 (1980).
}

\lref\AppelquistVG{
  T.~Appelquist and R.~D.~Pisarski,
  ``High-Temperature Yang-Mills Theories and Three-Dimensional Quantum Chromodynamics,''
Phys.\ Rev.\ D {\bf 23}, 2305 (1981).
}

\lref\HarveyIT{
  J.~A.~Harvey,
  ``TASI 2003 lectures on anomalies,''
[hep-th/0509097].
}

\lref\KonishiIZ{
  K.~Konishi,
  ``Confinement, supersymmetry breaking and theta parameter dependence in the Seiberg-Witten model,''
Phys.\ Lett.\ B {\bf 392}, 101 (1997).
[hep-th/9609021].
}

\lref\DineSGQ{
  M.~Dine, P.~Draper, L.~Stephenson-Haskins and D.~Xu,
  ``$\theta$ and the $\eta^\prime$ in Large $N$ Supersymmetric QCD,''
[arXiv:1612.05770 [hep-th]].
}

\lref\tHooftUJ{
  G.~'t Hooft,
  ``A Property of Electric and Magnetic Flux in Nonabelian Gauge Theories,''
Nucl.\ Phys.\ B {\bf 153}, 141 (1979).
}

\lref\Brower{
R.~C.~Brower and J.~Harte.
Physical Review 164.5 (1967): 1841.
}

\lref\ArgyresXN{
  P.~C.~Argyres, M.~R.~Plesser, N.~Seiberg and E.~Witten,
  ``New N=2 superconformal field theories in four-dimensions,''
Nucl.\ Phys.\ B {\bf 461}, 71 (1996).
[hep-th/9511154].
}

\lref\JakubskyKI{
  V.~Jakubsky, L.~M.~Nieto and M.~S.~Plyushchay,
  ``The origin of the hidden supersymmetry,''
Phys.\ Lett.\ B {\bf 692}, 51 (2010).
[arXiv:1004.5489 [hep-th]].
}

\lref\CorreaJE{
  F.~Correa and M.~S.~Plyushchay,
  ``Hidden supersymmetry in quantum bosonic systems,''
Annals Phys.\  {\bf 322}, 2493 (2007).
[hep-th/0605104].
}

\lref\HenningsonHP{
  M.~Henningson,
  ``Wilson-'t Hooft operators and the theta angle,''
JHEP {\bf 0605}, 065 (2006).
[hep-th/0603188].
}

\lref\CreutzXU{
  M.~Creutz,
  ``Spontaneous violation of CP symmetry in the strong interactions,''
Phys.\ Rev.\ Lett.\  {\bf 92}, 201601 (2004).
[hep-lat/0312018].
}

\lref\AffleckTJ{
  I.~Affleck,
  ``Nonlinear sigma model at Theta = pi: Euclidean lattice formulation and solid-on-solid models,''
Phys.\ Rev.\ Lett.\  {\bf 66}, 2429 (1991).
}

\lref\ChenPG{
  X.~Chen, Z.~C.~Gu, Z.~X.~Liu and X.~G.~Wen,
  ``Symmetry protected topological orders and the group cohomology of their symmetry group,''
Phys.\ Rev.\ B {\bf 87}, no. 15, 155114 (2013).
[arXiv:1106.4772 [cond-mat.str-el]].
}

\lref\PappadopuloJK{
  D.~Pappadopulo, S.~Rychkov, J.~Espin and R.~Rattazzi,
Phys.\ Rev.\ D {\bf 86}, 105043 (2012).
[arXiv:1208.6449 [hep-th]].
}

\lref\LegRed{
Askey, Richard. "Orthogonal expansions with positive coefficients." Proceedings of the American Mathematical Society 16.6 (1965): 1191-1194.
}

\lref\ClossetVP{
  C.~Closset, T.~T.~Dumitrescu, G.~Festuccia, Z.~Komargodski and N.~Seiberg,
  ``Comments on Chern-Simons Contact Terms in Three Dimensions,''
JHEP {\bf 1209}, 091 (2012).
[arXiv:1206.5218 [hep-th]].
}

\lref\DieriglXTA{
  M.~Dierigl and A.~Pritzel,
  ``Topological Model for Domain Walls in (Super-)Yang-Mills Theories,''
Phys.\ Rev.\ D {\bf 90}, no. 10, 105008 (2014).
[arXiv:1405.4291 [hep-th]].
}

\lref\WittenDF{
  E.~Witten,
  ``Constraints on Supersymmetry Breaking,''
Nucl.\ Phys.\ B {\bf 202}, 253 (1982).
}

\lref\SeibergRSG{
  N.~Seiberg and E.~Witten,
  ``Gapped Boundary Phases of Topological Insulators via Weak Coupling,''
PTEP {\bf 2016}, no. 12, 12C101 (2016).
[arXiv:1602.04251 [cond-mat.str-el]].
}

\lref\BeniniDUS{
  F.~Benini, P.~S.~Hsin and N.~Seiberg,
  ``Comments on Global Symmetries, Anomalies, and Duality in (2+1)d,''
[arXiv:1702.07035 [cond-mat.str-el]].
}

\lref\AmatiWQ{
  D.~Amati, M.~Ciafaloni and G.~Veneziano,
Phys.\ Lett.\ B {\bf 197}, 81 (1987).
}

\lref\HsinBLU{
  P.~S.~Hsin and N.~Seiberg,
  ``Level/rank Duality and Chern-Simons-Matter Theories,''
JHEP {\bf 1609}, 095 (2016).
[arXiv:1607.07457 [hep-th]].
}

\lref\KapustinLWA{
  A.~Kapustin and R.~Thorngren,
  ``Anomalies of discrete symmetries in three dimensions and group cohomology,''
Phys.\ Rev.\ Lett.\  {\bf 112}, no. 23, 231602 (2014).
[arXiv:1403.0617 [hep-th]].
}

\lref\ElitzurXJ{
  S.~Elitzur, Y.~Frishman, E.~Rabinovici and A.~Schwimmer,
  ``Origins of Global Anomalies in Quantum Mechanics,''
Nucl.\ Phys.\ B {\bf 273}, 93 (1986).
}

\lref\KapustinZVA{
  A.~Kapustin and R.~Thorngren,
  ``Anomalies of discrete symmetries in various dimensions and group cohomology,''
[arXiv:1404.3230 [hep-th]].
}

\lref\Hatcher{
A.~Hatcher, ``Algebraic topology,'' Cambridge University Press, 2002.
}

\lref\ZamolodchikovZR{
  A.~B.~Zamolodchikov and A.~B.~Zamolodchikov,
  ``Massless factorized scattering and sigma models with topological terms,''
Nucl.\ Phys.\ B {\bf 379}, 602 (1992).
}

\lref\ElitzurFH{
  S.~Elitzur, A.~Giveon and D.~Kutasov,
  ``Branes and N=1 duality in string theory,''
Phys.\ Lett.\ B {\bf 400}, 269 (1997).
[hep-th/9702014].
}

\lref\KapustinLWA{
  A.~Kapustin and R.~Thorngren,
  ``Anomalies of discrete symmetries in three dimensions and group cohomology,''
Phys.\ Rev.\ Lett.\  {\bf 112}, no. 23, 231602 (2014).
[arXiv:1403.0617 [hep-th]].
}
\lref\KapustinZVA{
  A.~Kapustin and R.~Thorngren,
  ``Anomalies of discrete symmetries in various dimensions and group cohomology,''
[arXiv:1404.3230 [hep-th]].
}

\lref\ElitzurHC{
  S.~Elitzur, A.~Giveon, D.~Kutasov, E.~Rabinovici and A.~Schwimmer,
  ``Brane dynamics and N=1 supersymmetric gauge theory,''
Nucl.\ Phys.\ B {\bf 505}, 202 (1997).
[hep-th/9704104].
}

\lref\EssinRQ{
  A.~M.~Essin, J.~E.~Moore and D.~Vanderbilt,
  ``Magnetoelectric polarizability and axion electrodynamics in crystalline insulators,''
Phys.\ Rev.\ Lett.\  {\bf 102}, 146805 (2009).
[arXiv:0810.2998 [cond-mat.mes-hall]].
}

\lref\slthreeZ{
  J.~Felder, A.~Varchenko,
  ``The elliptic gamma function and $SL(3,Z) \times Z^3$,'' $\;\;$
[arXiv:math/0001184].
}

\lref\FestucciaWS{
  G.~Festuccia and N.~Seiberg,
  ``Rigid Supersymmetric Theories in Curved Superspace,''
JHEP {\bf 1106}, 114 (2011).
[arXiv:1105.0689 [hep-th]].
}

\lref\FidkowskiJUA{
  L.~Fidkowski, X.~Chen and A.~Vishwanath,
  ``Non-Abelian Topological Order on the Surface of a 3D Topological Superconductor from an Exactly Solved Model,''
Phys.\ Rev.\ X {\bf 3}, no. 4, 041016 (2013).
[arXiv:1305.5851 [cond-mat.str-el]].
}

\lref\FradkinTT{
  E.~H.~Fradkin and F.~A.~Schaposnik,
  ``The Fermion - boson mapping in three-dimensional quantum field theory,''
Phys.\ Lett.\ B {\bf 338}, 253 (1994).
[hep-th/9407182].
}

\lref\GaddeEN{
  A.~Gadde, L.~Rastelli, S.~S.~Razamat and W.~Yan,
  ``On the Superconformal Index of N=1 IR Fixed Points: A Holographic Check,''
JHEP {\bf 1103}, 041 (2011).
[arXiv:1011.5278 [hep-th]].
}

\lref\GaddeIA{
  A.~Gadde and W.~Yan,
  ``Reducing the 4d Index to the $S^3$ Partition Function,''
JHEP {\bf 1212}, 003 (2012).
[arXiv:1104.2592 [hep-th]].
}

\lref\GaddeDDA{
  A.~Gadde and S.~Gukov,
  ``2d Index and Surface operators,''
[arXiv:1305.0266 [hep-th]].
}

\lref\VafaXH{
  C.~Vafa and E.~Witten,
  ``Eigenvalue Inequalities for Fermions in Gauge Theories,''
Commun.\ Math.\ Phys.\  {\bf 95}, 257 (1984).
}

\lref\GaiottoAK{
  D.~Gaiotto and E.~Witten,
  ``S-Duality of Boundary Conditions In N=4 Super Yang-Mills Theory,''
Adv.\ Theor.\ Math.\ Phys.\  {\bf 13}, no. 3, 721 (2009).
[arXiv:0807.3720 [hep-th]].
}

\lref\GaiottoBE{
  D.~Gaiotto, G.~W.~Moore and A.~Neitzke,
  ``Framed BPS States,''
[arXiv:1006.0146 [hep-th]].
}

\lref\GaiottoKFA{
  D.~Gaiotto, A.~Kapustin, N.~Seiberg and B.~Willett,
  ``Generalized Global Symmetries,''
JHEP {\bf 1502}, 172 (2015).
[arXiv:1412.5148 [hep-th]].
}

\lref\GeraedtsPVA{
  S.~D.~Geraedts, M.~P.~Zaletel, R.~S.~K.~Mong, M.~A.~Metlitski, A.~Vishwanath and O.~I.~Motrunich,
  ``The half-filled Landau level: the case for Dirac composite fermions,''
Science {\bf 352}, 197 (2016).
[arXiv:1508.04140 [cond-mat.str-el]].
}

\lref\GiombiYA{
  S.~Giombi and X.~Yin,
  ``On Higher Spin Gauge Theory and the Critical $O(N)$ Model,''
Phys.\ Rev.\ D {\bf 85}, 086005 (2012).
[arXiv:1105.4011 [hep-th]].
}

\lref\GiombiKC{
  S.~Giombi, S.~Minwalla, S.~Prakash, S.~P.~Trivedi, S.~R.~Wadia and X.~Yin,
  ``Chern-Simons Theory with Vector Fermion Matter,''
Eur.\ Phys.\ J.\ C {\bf 72}, 2112 (2012).
[arXiv:1110.4386 [hep-th]].
}

\lref\GiombiMS{
  S.~Giombi and X.~Yin,
  ``The Higher Spin/Vector Model Duality,''
J.\ Phys.\ A {\bf 46}, 214003 (2013).
[arXiv:1208.4036 [hep-th]].
}

\lref\GiombiZWA{
  S.~Giombi, V.~Gurucharan, V.~Kirilin, S.~Prakash and E.~Skvortsov,
  ``On the Higher-Spin Spectrum in Large $N$ Chern-Simons Vector Models,''
JHEP {\bf 1701}, 058 (2017).
[arXiv:1610.08472 [hep-th]].
}

\lref\GiveonSR{
  A.~Giveon and D.~Kutasov,
  ``Brane dynamics and gauge theory,''
Rev.\ Mod.\ Phys.\  {\bf 71}, 983 (1999).
[hep-th/9802067].
}

\lref\SmilgaDH{
  A.~V.~Smilga,
  ``QCD at theta similar to pi,''
Phys.\ Rev.\ D {\bf 59}, 114021 (1999).
[hep-ph/9805214].
}

\lref\GiveonZN{
  A.~Giveon and D.~Kutasov,
  ``Seiberg Duality in Chern-Simons Theory,''
Nucl.\ Phys.\ B {\bf 812}, 1 (2009).
[arXiv:0808.0360 [hep-th]].
}

\lref\GoddardQE{
  P.~Goddard, J.~Nuyts and D.~I.~Olive,
  ``Gauge Theories and Magnetic Charge,''
Nucl.\ Phys.\ B {\bf 125}, 1 (1977).
}

\lref\GoddardVK{
  P.~Goddard, A.~Kent and D.~I.~Olive,
  ``Virasoro Algebras and Coset Space Models,''
Phys.\ Lett.\ B {\bf 152}, 88 (1985).
}

\lref\GreenDA{
  D.~Green, Z.~Komargodski, N.~Seiberg, Y.~Tachikawa and B.~Wecht,
  ``Exactly Marginal Deformations and Global Symmetries,''
JHEP {\bf 1006}, 106 (2010).
[arXiv:1005.3546 [hep-th]].
}

\lref\GurPCA{
  G.~Gur-Ari and R.~Yacoby,
  ``Three Dimensional Bosonization From Supersymmetry,''
JHEP {\bf 1511}, 013 (2015).
[arXiv:1507.04378 [hep-th]].
}

\lref\GurAriXFF{
  G.~Gur-Ari, S.~A.~Hartnoll and R.~Mahajan,
  ``Transport in Chern-Simons-Matter Theories,''
JHEP {\bf 1607}, 090 (2016).
[arXiv:1605.01122 [hep-th]].
}

\lref\HalperinMH{
  B.~I.~Halperin, P.~A.~Lee and N.~Read,
  ``Theory of the half filled Landau level,''
Phys.\ Rev.\ B {\bf 47}, 7312 (1993).
}

\lref\DiVecchiaYFW{
  P.~Di Vecchia and G.~Veneziano,
  ``Chiral Dynamics in the Large n Limit,''
Nucl.\ Phys.\ B {\bf 171}, 253 (1980).
}

\lref\FreedRLK{
  D.~S.~Freed, Z.~Komargodski and N.~Seiberg,
  ``The Sum Over Topological Sectors and $\theta$ in the 2+1-Dimensional $\C\P^1$ $\sigma$-Model,''
[arXiv:1707.05448 [cond-mat.str-el]].
}

\lref\HamaEA{
  N.~Hama, K.~Hosomichi and S.~Lee,
  ``SUSY Gauge Theories on Squashed Three-Spheres,''
JHEP {\bf 1105}, 014 (2011).
[arXiv:1102.4716 [hep-th]].
}

\lref\Hasegawa{
K.~Hasegawa,
  ``Spin Module Versions of Weyl's Reciprocity Theorem for Classical Kac-Moody Lie Algebras - An Application to Branching Rule Duality,''
Publ.\ Res.\ Inst.\ Math.\ Sci.\ {\bf 25}, 741-828 (1989).
}

\lref\HoriDK{
  K.~Hori and D.~Tong,
  ``Aspects of Non-Abelian Gauge Dynamics in Two-Dimensional N=(2,2) Theories,''
JHEP {\bf 0705}, 079 (2007).
[hep-th/0609032].
}

\lref\HoriPD{
  K.~Hori,
  ``Duality In Two-Dimensional (2,2) Supersymmetric Non-Abelian Gauge Theories,''
[arXiv:1104.2853 [hep-th]].
}

\lref\HsinBLU{
  P.~S.~Hsin and N.~Seiberg,
  ``Level/rank Duality and Chern-Simons-Matter Theories,''
JHEP {\bf 1609}, 095 (2016).
[arXiv:1607.07457 [hep-th]].
}

\lref\HullMS{
  C.~M.~Hull and B.~J.~Spence,
  ``The Geometry of the gauged sigma model with Wess-Zumino term,''
Nucl.\ Phys.\ B {\bf 353}, 379 (1991).
}

\lref\HwangQT{
  C.~Hwang, H.~Kim, K.~-J.~Park and J.~Park,
  ``Index computation for 3d Chern-Simons matter theory: test of Seiberg-like duality,''
JHEP {\bf 1109}, 037 (2011).
[arXiv:1107.4942 [hep-th]].
}

\lref\HwangHT{
  C.~Hwang, K.~-J.~Park and J.~Park,
  ``Evidence for Aharony duality for orthogonal gauge groups,''
JHEP {\bf 1111}, 011 (2011).
[arXiv:1109.2828 [hep-th]].
}

\lref\HwangJH{
  C.~Hwang, H.~-C.~Kim and J.~Park,
  ``Factorization of the 3d superconformal index,''
[arXiv:1211.6023 [hep-th]].
}

\lref\ImamuraSU{
  Y.~Imamura and S.~Yokoyama,
  ``Index for three dimensional superconformal field theories with general R-charge assignments,''
JHEP {\bf 1104}, 007 (2011).
[arXiv:1101.0557 [hep-th]].
}

\lref\ImamuraUW{
  Y.~Imamura,
 ``Relation between the 4d superconformal index and the $S^3$ partition function,''
JHEP {\bf 1109}, 133 (2011).
[arXiv:1104.4482 [hep-th]].
}

\lref\ImamuraWG{
  Y.~Imamura and D.~Yokoyama,
 ``N=2 supersymmetric theories on squashed three-sphere,''
Phys.\ Rev.\ D {\bf 85}, 025015 (2012).
[arXiv:1109.4734 [hep-th]].
}

\lref\ImamuraRQ{
  Y.~Imamura and D.~Yokoyama,
 ``$S^3/Z_n$ partition function and dualities,''
JHEP {\bf 1211}, 122 (2012).
[arXiv:1208.1404 [hep-th]].
}

\lref\InbasekarTSA{
  K.~Inbasekar, S.~Jain, S.~Mazumdar, S.~Minwalla, V.~Umesh and S.~Yokoyama,
  ``Unitarity, crossing symmetry and duality in the scattering of ${\cal N}{=}1 $ SUSY matter Chern-Simons theories,''
JHEP {\bf 1510}, 176 (2015).
[arXiv:1505.06571 [hep-th]].
}

\lref\IntriligatorID{
  K.~A.~Intriligator and N.~Seiberg,
  ``Duality, monopoles, dyons, confinement and oblique confinement in supersymmetric SO(N(c)) gauge theories,''
Nucl.\ Phys.\ B {\bf 444}, 125 (1995).
[hep-th/9503179].
}

\lref\IntriligatorNE{
  K.~A.~Intriligator and P.~Pouliot,
  ``Exact superpotentials, quantum vacua and duality in supersymmetric SP(N(c)) gauge theories,''
Phys.\ Lett.\ B {\bf 353}, 471 (1995).
[hep-th/9505006].
}

\lref\IntriligatorER{
  K.~A.~Intriligator and N.~Seiberg,
  ``Phases of N=1 supersymmetric gauge theories and electric - magnetic triality,''
In *Los Angeles 1995, Future perspectives in string theory* 270-282.
[hep-th/9506084].
}

\lref\IntriligatorAU{
  K.~A.~Intriligator and N.~Seiberg,
  ``Lectures on supersymmetric gauge theories and electric - magnetic duality,''
Nucl.\ Phys.\ Proc.\ Suppl.\  {\bf 45BC}, 1 (1996).
[hep-th/9509066].
}

\lref\IntriligatorEX{
  K.~A.~Intriligator and N.~Seiberg,
  ``Mirror symmetry in three-dimensional gauge theories,''
Phys.\ Lett.\ B {\bf 387}, 513 (1996).
[hep-th/9607207].
}

\lref\IntriligatorLCA{
  K.~Intriligator and N.~Seiberg,
  ``Aspects of 3d ${\cal N}{=}2$ Chern-Simons-Matter Theories,''
JHEP {\bf 1307}, 079 (2013).
[arXiv:1305.1633 [hep-th]].
}

\lref\IvanovFN{
   E.~A.~Ivanov,
   ``Chern-Simons matter systems with manifest N=2 supersymmetry,''
Phys.\ Lett.\ B {\bf 268}, 203 (1991).
}

\lref\JafferisUN{
  D.~L.~Jafferis,
  ``The Exact Superconformal R-Symmetry Extremizes Z,''
JHEP {\bf 1205}, 159 (2012).
[arXiv:1012.3210 [hep-th]].
}

\lref\JafferisZI{
  D.~L.~Jafferis, I.~R.~Klebanov, S.~S.~Pufu and B.~R.~Safdi,
  ``Towards the F-Theorem: N=2 Field Theories on the Three-Sphere,''
JHEP {\bf 1106}, 102 (2011).
[arXiv:1103.1181 [hep-th]].
}

\lref\JainTX{
  J.~K.~Jain,
  ``Composite fermion approach for the fractional quantum Hall effect,''
Phys.\ Rev.\ Lett.\  {\bf 63}, 199 (1989).
}

\lref\JainPY{
  S.~Jain, S.~Minwalla, T.~Sharma, T.~Takimi, S.~R.~Wadia and S.~Yokoyama,
  ``Phases of large $N$ vector Chern-Simons theories on $S^2 {\times} S^1$,''
JHEP {\bf 1309}, 009 (2013).
[arXiv:1301.6169 [hep-th]].
}

\lref\JainGZA{
  S.~Jain, S.~Minwalla and S.~Yokoyama,
  ``Chern Simons duality with a fundamental boson and fermion,''
JHEP {\bf 1311}, 037 (2013).
[arXiv:1305.7235 [hep-th]].
}

\lref\JainNZA{
  S.~Jain, M.~Mandlik, S.~Minwalla, T.~Takimi, S.~R.~Wadia and S.~Yokoyama,
  ``Unitarity, Crossing Symmetry and Duality of the S-matrix in large $N$ Chern-Simons theories with fundamental matter,''
JHEP {\bf 1504}, 129 (2015).
[arXiv:1404.6373 [hep-th]].
}

\lref\KachruRMA{
  S.~Kachru, M.~Mulligan, G.~Torroba and H.~Wang,
  ``Mirror symmetry and the half-filled Landau level,''
Phys.\ Rev.\ B {\bf 92}, 235105 (2015).
[arXiv:1506.01376 [cond-mat.str-el]].
}

\lref\KachruRUI{
  S.~Kachru, M.~Mulligan, G.~Torroba and H.~Wang,
  ``Bosonization and Mirror Symmetry,''
Phys.\ Rev.\ D {\bf 94}, no. 8, 085009 (2016).
[arXiv:1608.05077 [hep-th]].
}

\lref\KachruAON{
  S.~Kachru, M.~Mulligan, G.~Torroba and H.~Wang,
  ``Nonsupersymmetric dualities from mirror symmetry,''
Phys.\ Rev.\ Lett.\  {\bf 118}, 011602 (2017).
[arXiv:1609.02149 [hep-th]].
}

\lref\KajantieVY{
  K.~Kajantie, M.~Laine, T.~Neuhaus, A.~Rajantie and K.~Rummukainen,
  ``Duality and scaling in three-dimensional scalar electrodynamics,''
Nucl.\ Phys.\ B {\bf 699}, 632 (2004).
[hep-lat/0402021].
}

\lref\KapustinHA{
  A.~Kapustin and M.~J.~Strassler,
  ``On mirror symmetry in three-dimensional Abelian gauge theories,''
JHEP {\bf 9904}, 021 (1999).
[hep-th/9902033].
}

\lref\KapustinPY{
  A.~Kapustin,
  ``Wilson-'t Hooft operators in four-dimensional gauge theories and S-duality,''
Phys.\ Rev.\ D {\bf 74}, 025005 (2006).
[hep-th/0501015].
}

\lref\KapustinKZ{
  A.~Kapustin, B.~Willett and I.~Yaakov,
  ``Exact Results for Wilson Loops in Superconformal Chern-Simons Theories with Matter,''
JHEP {\bf 1003}, 089 (2010).
[arXiv:0909.4559 [hep-th]].
}

\lref\KapustinSim{
A.~Kapustin,  2010 Simons Workshop talk, a video of this talk can be found at
{\tt
http://media.scgp.stonybrook.edu/video/video.php?f=20110810\_1\_qtp.mp4}
}

\lref\KapustinXQ{
  A.~Kapustin, B.~Willett and I.~Yaakov,
  ``Nonperturbative Tests of Three-Dimensional Dualities,''
JHEP {\bf 1010}, 013 (2010).
[arXiv:1003.5694 [hep-th]].
}

\lref\KapustinGH{
  A.~Kapustin,
  ``Seiberg-like duality in three dimensions for orthogonal gauge groups,''
arXiv:1104.0466 [hep-th].
}

\lref\KapustinJM{
  A.~Kapustin and B.~Willett,
  ``Generalized Superconformal Index for Three Dimensional Field Theories,''
[arXiv:1106.2484 [hep-th]].
}

\lref\KapustinVZ{
  A.~Kapustin, H.~Kim and J.~Park,
  ``Dualities for 3d Theories with Tensor Matter,''
JHEP {\bf 1112}, 087 (2011).
[arXiv:1110.2547 [hep-th]].
}

\lref\KapustinGUA{
  A.~Kapustin and N.~Seiberg,
  ``Coupling a QFT to a TQFT and Duality,''
JHEP {\bf 1404}, 001 (2014).
[arXiv:1401.0740 [hep-th]].
}

\lref\KarchUX{
  A.~Karch,
  ``Seiberg duality in three-dimensions,''
Phys.\ Lett.\ B {\bf 405}, 79 (1997).
[hep-th/9703172].
}

\lref\KarchSXI{
  A.~Karch and D.~Tong,
  ``Particle-Vortex Duality from 3d Bosonization,''
Phys.\ Rev.\ X {\bf 6}, 031043 (2016). 
[arXiv:1606.01893 [hep-th]].
}

\lref\KarchAUX{
  A.~Karch, B.~Robinson and D.~Tong,
  ``More Abelian Dualities in 2+1 Dimensions,''
JHEP {\bf 1701}, 017 (2017).
[arXiv:1609.04012 [hep-th]].
}

\lref\KimWB{
  S.~Kim,
  ``The Complete superconformal index for N=6 Chern-Simons theory,''
Nucl.\ Phys.\ B {\bf 821}, 241 (2009), [Erratum-ibid.\ B {\bf 864}, 884 (2012)].
[arXiv:0903.4172 [hep-th]].
}

\lref\KimCMA{
  H.~Kim and J.~Park,
  ``Aharony Dualities for 3d Theories with Adjoint Matter,''
[arXiv:1302.3645 [hep-th]].
}

\lref\KinneyEJ{
  J.~Kinney, J.~M.~Maldacena, S.~Minwalla and S.~Raju,
  ``An Index for 4 dimensional super conformal theories,''
Commun.\ Math.\ Phys.\  {\bf 275}, 209 (2007).
[hep-th/0510251].
}

\lref\KlebanovJA{
  I.~R.~Klebanov and A.~M.~Polyakov,
  ``AdS dual of the critical $O(N)$ vector model,''
Phys.\ Lett.\ B {\bf 550}, 213 (2002).
[hep-th/0210114].
}

\lref\KrattenthalerDA{
  C.~Krattenthaler, V.~P.~Spiridonov, G.~S.~Vartanov,
  ``Superconformal indices of three-dimensional theories related by mirror symmetry,''
JHEP {\bf 1106}, 008 (2011).
[arXiv:1103.4075 [hep-th]].
}

\lref\McG{
  S. M. Kravec, J. McGreevy, and B. Swingle,
  ``All-Fermion Electrodynamics And Fermion Number Anomaly Inflow,''
arXiv:1409.8339.
}

\lref\KutasovVE{
  D.~Kutasov,
  ``A Comment on duality in N=1 supersymmetric nonAbelian gauge theories,''
Phys.\ Lett.\ B {\bf 351}, 230 (1995).
[hep-th/9503086].
}

\lref\KutasovNP{
  D.~Kutasov and A.~Schwimmer,
  ``On duality in supersymmetric Yang-Mills theory,''
Phys.\ Lett.\ B {\bf 354}, 315 (1995).
[hep-th/9505004].
}

\lref\KutasovSS{
  D.~Kutasov, A.~Schwimmer and N.~Seiberg,
  ``Chiral rings, singularity theory and electric - magnetic duality,''
Nucl.\ Phys.\ B {\bf 459}, 455 (1996).
[hep-th/9510222].
}

\lref\LeeVP{
  K.~-M.~Lee and P.~Yi,
  ``Monopoles and instantons on partially compactified D-branes,''
Phys.\ Rev.\ D {\bf 56}, 3711 (1997).
[hep-th/9702107].
}

\lref\LeeVU{
  K.~-M.~Lee,
  ``Instantons and magnetic monopoles on R**3 x S**1 with arbitrary simple gauge groups,''
Phys.\ Lett.\ B {\bf 426}, 323 (1998).
[hep-th/9802012].
}

\lref\MaldacenaSS{
  J.~M.~Maldacena, G.~W.~Moore and N.~Seiberg,
  ``D-brane charges in five-brane backgrounds,''
JHEP {\bf 0110}, 005 (2001).
[hep-th/0108152].
}

\lref\MaldacenaJN{
  J.~Maldacena and A.~Zhiboedov,
  ``Constraining Conformal Field Theories with A Higher Spin Symmetry,''
J.\ Phys.\ A {\bf 46}, 214011 (2013).
[arXiv:1112.1016 [hep-th]].
}

\lref\MetlitskiEKA{
  M.~A.~Metlitski and A.~Vishwanath,
  ``Particle-vortex duality of 2d Dirac fermion from electric-magnetic duality of 3d topological insulators,''
[arXiv:1505.05142 [cond-mat.str-el]].
}

\lref\MetlitskiBPA{
  M.~A.~Metlitski, C.~L.~Kane and M.~P.~A.~Fisher,
  ``Symmetry-respecting topologically ordered surface phase of three-dimensional electron topological insulators,''
Phys.\ Rev.\ B {\bf 92}, no. 12, 125111 (2015).
}

\lref\MetlitskiYQA{
  M.~A.~Metlitski,
  ``$S$-duality of $u(1)$ gauge theory with $\theta =\pi$ on non-orientable manifolds: Applications to topological insulators and superconductors,''
[arXiv:1510.05663 [hep-th]].
}

\lref\MetlitskiDHT{
  M.~A.~Metlitski, A.~Vishwanath and C.~Xu,
  ``Duality and bosonization of (2+1)d Majorana fermions,''
  arXiv:1611.05049 [cond-mat.str-el].
}

\lref\MinwallaSCA{
  S.~Minwalla and S.~Yokoyama,
  ``Chern Simons Bosonization along RG Flows,''
JHEP {\bf 1602}, 103 (2016).
[arXiv:1507.04546 [hep-th]].
}

\lref\MlawerUV{
  E.~J.~Mlawer, S.~G.~Naculich, H.~A.~Riggs and H.~J.~Schnitzer,
  ``Group level duality of WZW fusion coefficients and Chern-Simons link observables,''
Nucl.\ Phys.\ B {\bf 352}, 863 (1991).
}

\lref\MSN{
G.~W.~Moore and N.~Seiberg,
  ``Naturality in Conformal Field Theory,''
  Nucl.\ Phys.\ B {\bf 313}, 16 (1989).}

\lref\MooreYH{
  G.~W.~Moore and N.~Seiberg,
  ``Taming the Conformal Zoo,''
Phys.\ Lett.\ B {\bf 220}, 422 (1989).
}

\lref\MoritaCS{
  T.~Morita and V.~Niarchos,
  ``F-theorem, duality and SUSY breaking in one-adjoint Chern-Simons-Matter theories,''
Nucl.\ Phys.\ B {\bf 858}, 84 (2012).
[arXiv:1108.4963 [hep-th]].
}

\lref\MrossIDY{
  D.~F.~Mross, J.~Alicea and O.~I.~Motrunich,
  ``Explicit derivation of duality between a free Dirac cone and quantum electrodynamics in (2+1) dimensions,''
[arXiv:1510.08455 [cond-mat.str-el]].
}

\lref\MulliganGLM{
  M.~Mulligan, S.~Raghu and M.~P.~A.~Fisher,
  ``Emergent particle-hole symmetry in the half-filled Landau level,''
[arXiv:1603.05656 [cond-mat.str-el]].
}

\lref\MuruganZAL{
  J.~Murugan and H.~Nastase,
  ``Particle-vortex duality in topological insulators and superconductors,''
[arXiv:1606.01912 [hep-th]].
}

\lref\NaculichPA{
  S.~G.~Naculich, H.~A.~Riggs and H.~J.~Schnitzer,
  ``Group Level Duality in {WZW} Models and {Chern-Simons} Theory,''
Phys.\ Lett.\ B {\bf 246}, 417 (1990).
}

\lref\NaculichNC{
  S.~G.~Naculich and H.~J.~Schnitzer,
  ``Level-rank duality of the U(N) WZW model, Chern-Simons theory, and 2-D qYM theory,''
JHEP {\bf 0706}, 023 (2007).
[hep-th/0703089 [HEP-TH]].
}

\lref\NakaharaNW{
  M.~Nakahara,
  ``Geometry, topology and physics,''
Boca Raton, USA: Taylor and Francis (2003) 573 p.
}

\lref\NakanishiHJ{
  T.~Nakanishi and A.~Tsuchiya,
  ``Level rank duality of WZW models in conformal field theory,''
Commun.\ Math.\ Phys.\  {\bf 144}, 351 (1992).
}

\lref\NguyenZN{
  A.~K.~Nguyen and A.~Sudbo,
  ``Topological phase fluctuations, amplitude fluctuations, and criticality in extreme type II superconductors,''
Phys.\ Rev.\ B {\bf 60}, 15307 (1999).
[cond-mat/9907385].
}

\lref\NiarchosJB{
  V.~Niarchos,
  ``Seiberg Duality in Chern-Simons Theories with Fundamental and Adjoint Matter,''
JHEP {\bf 0811}, 001 (2008).
[arXiv:0808.2771 [hep-th]].
}

\lref\NiarchosAA{
  V.~Niarchos,
  ``R-charges, Chiral Rings and RG Flows in Supersymmetric Chern-Simons-Matter Theories,''
JHEP {\bf 0905}, 054 (2009).=
[arXiv:0903.0435 [hep-th]].
}

\lref\NiarchosAH{
  V.~Niarchos,
  ``Seiberg dualities and the 3d/4d connection,''
JHEP {\bf 1207}, 075 (2012).
[arXiv:1205.2086 [hep-th]].
}

\lref\NiemiRQ{
  A.~J.~Niemi and G.~W.~Semenoff,
  ``Axial Anomaly Induced Fermion Fractionization and Effective Gauge Theory Actions in Odd Dimensional Space-Times,''
Phys.\ Rev.\ Lett.\  {\bf 51}, 2077 (1983).
}

\lref\VOstrik{
  V.~Ostrik and M.~Sun,
  ``Level-Rank Duality Via Tensor Categories,''
Comm. Math. Phys. 326 (2014) 49-61.
[arXiv:1208.5131 [math-ph]].
}

\lref\ParkWTA{
  J.~Park and K.~J.~Park,
  ``Seiberg-like Dualities for 3d ${\cal N}{=}2$ Theories with $SU(N)$ gauge group,''
JHEP {\bf 1310}, 198 (2013).
[arXiv:1305.6280 [hep-th]].
}

\lref\PaulyAMA{
  C.~Pauly,
  ``Strange duality revisited,''
Math.\ Res.\ Lett.\  {\bf 21}, 1353 (2014).
}

\lref\PeskinKP{
  M.~E.~Peskin,
  ``Mandelstam 't Hooft Duality in Abelian Lattice Models,''
Annals Phys.\  {\bf 113}, 122 (1978).
}

\lref\PolyakovFU{
  A.~M.~Polyakov,
  ``Quark Confinement and Topology of Gauge Groups,''
Nucl.\ Phys.\ B {\bf 120}, 429 (1977).
}

\lref\PolyakovMD{
  A.~M.~Polyakov,
  ``Fermi-Bose Transmutations Induced by Gauge Fields,''
Mod.\ Phys.\ Lett.\ A {\bf 3}, 325 (1988).
}

\lref\PotterCDN{
  A.~C.~Potter, M.~Serbyn and A.~Vishwanath,
  ``Thermoelectric transport signatures of Dirac composite fermions in the half-filled Landau level,''
Phys.\ Rev.\ X {\bf 6}, 031026 (2016).
[arXiv:1512.06852 [cond-mat.str-el]].
}

\lref\QiEW{
  X.~L.~Qi, T.~Hughes and S.~C.~Zhang,
  ``Topological Field Theory of Time-Reversal Invariant Insulators,''
Phys.\ Rev.\ B {\bf 78}, 195424 (2008).
[arXiv:0802.3537 [cond-mat.mes-hall]].
}

\lref\RabinoviciMJ{
  E.~Rabinovici, A.~Schwimmer and S.~Yankielowicz,
  ``Quantization in the Presence of {Wess-Zumino} Terms,''
Nucl.\ Phys.\ B {\bf 248}, 523 (1984).
}

\lref\RadicevicYLA{
  D.~Radicevic,
  ``Disorder Operators in Chern-Simons-Fermion Theories,''
JHEP {\bf 1603}, 131 (2016).
[arXiv:1511.01902 [hep-th]].
}

\lref\RadicevicWQN{
  D.~Radicevic, D.~Tong and C.~Turner,
  ``Non-Abelian 3d Bosonization and Quantum Hall States,''
JHEP {\bf 1612}, 067 (2016).
[arXiv:1608.04732 [hep-th]].
}

\lref\RazamatUV{
  S.~S.~Razamat,
  ``On a modular property of N=2 superconformal theories in four dimensions,''
JHEP {\bf 1210}, 191 (2012).
[arXiv:1208.5056 [hep-th]].
}

\lref\RedlichDV{
  A.~N.~Redlich,
  ``Parity Violation and Gauge Noninvariance of the Effective Gauge Field Action in Three-Dimensions,''
Phys.\ Rev.\ D {\bf 29}, 2366 (1984).
}

\lref\Rehren{
  K.-H.~Rehren,
  ``Algebraic Conformal QFT'',
3rd Meeting of the French-Italian Research Team on Noncommutative Geometry and Quantum Physics Vietri sul Mare, 2009.
}

\lref\RomelsbergerEG{
  C.~Romelsberger,
  ``Counting chiral primaries in N = 1, d=4 superconformal field theories,''
Nucl.\ Phys.\ B {\bf 747}, 329 (2006).
[hep-th/0510060].
}

\lref\RoscherWOX{
  D.~Roscher, E.~Torres and P.~Strack,
  ``Dual QED$_3$ at "$N_F = 1/2$" is an interacting CFT in the infrared,''
[arXiv:1605.05347 [cond-mat.str-el]].
}

\lref\SafdiRE{
  B.~R.~Safdi, I.~R.~Klebanov and J.~Lee,
  ``A Crack in the Conformal Window,''
[arXiv:1212.4502 [hep-th]].
}

\lref\SeibergBZ{
  N.~Seiberg,
  ``Exact results on the space of vacua of four-dimensional SUSY gauge theories,''
Phys.\ Rev.\ D {\bf 49}, 6857 (1994).
[hep-th/9402044].
}

\lref\SeibergPQ{
  N.~Seiberg,
  ``Electric - magnetic duality in supersymmetric nonAbelian gauge theories,''
Nucl.\ Phys.\ B {\bf 435}, 129 (1995).
[hep-th/9411149].
}

\lref\SulejmanpasicUWQ{
  T.~Sulejmanpasic, H.~Shao, A.~Sandvik and M.~Unsal,
  ``Confinement in the bulk, deconfinement on the wall: infrared equivalence between compactified QCD and quantum magnets,''
[arXiv:1608.09011 [hep-th]].
}

\lref\SeibergNZ{
  N.~Seiberg and E.~Witten,
  ``Gauge dynamics and compactification to three-dimensions,''
In *Saclay 1996, The mathematical beauty of physics* 333-366.
[hep-th/9607163].
}

\lref\SeibergQD{
  N.~Seiberg,
  ``Modifying the Sum Over Topological Sectors and Constraints on Supergravity,''
JHEP {\bf 1007}, 070 (2010).
[arXiv:1005.0002 [hep-th]].
}

\lref\SeibergRSG{
  N.~Seiberg and E.~Witten,
  ``Gapped Boundary Phases of Topological Insulators via Weak Coupling,''
PTEP {\bf 2016}, 12C101 (2016). 
[arXiv:1602.04251 [cond-mat.str-el]].
}

\lref\SeibergGMD{
  N.~Seiberg, T.~Senthil, C.~Wang and E.~Witten,
  ``A Duality Web in 2+1 Dimensions and Condensed Matter Physics,''
Annals Phys.\  {\bf 374}, 395 (2016).
[arXiv:1606.01989 [hep-th]].
}

\lref\SenthilJK{
  T.~Senthil and M.~P.~A.~Fisher,
  ``Competing orders, non-linear sigma models, and topological terms in quantum magnets,''
Phys.\ Rev.\ B {\bf 74}, 064405 (2006).
[cond-mat/0510459].
}

\lref\SezginRT{
  E.~Sezgin and P.~Sundell,
  ``Massless higher spins and holography,''
Nucl.\ Phys.\ B {\bf 644}, 303 (2002), Erratum: [Nucl.\ Phys.\ B {\bf 660}, 403 (2003)].
[hep-th/0205131].
}

\lref\SezginPT{
  E.~Sezgin and P.~Sundell,
  ``Holography in 4D (super) higher spin theories and a test via cubic scalar couplings,''
JHEP {\bf 0507}, 044 (2005).
[hep-th/0305040].
}

\lref\ShajiIS{
  N.~Shaji, R.~Shankar and M.~Sivakumar,
  ``On Bose-fermi Equivalence in a U(1) Gauge Theory With {Chern-Simons} Action,''
Mod.\ Phys.\ Lett.\ A {\bf 5}, 593 (1990).
}

\lref\Shamirthesis{
  I.~Shamir,
  ``Aspects of three dimensional Seiberg duality,''
  M.~Sc. thesis submitted to the Weizmann Institute of Science, April 2010.
}

\lref\ShenkerZF{
  S.~H.~Shenker and X.~Yin,
  ``Vector Models in the Singlet Sector at Finite Temperature,''
[arXiv:1109.3519 [hep-th]].
}

\lref\SonXQA{
  D.~T.~Son,
  ``Is the Composite Fermion a Dirac Particle?,''
Phys.\ Rev.\ X {\bf 5}, 031027 (2015).
[arXiv:1502.03446 [cond-mat.mes-hall]].
}

\lref\SpiridonovZR{
  V.~P.~Spiridonov and G.~S.~Vartanov,
  ``Superconformal indices for N = 1 theories with multiple duals,''
Nucl.\ Phys.\ B {\bf 824}, 192 (2010).
[arXiv:0811.1909 [hep-th]].
}

\lref\SpiridonovZA{
  V.~P.~Spiridonov and G.~S.~Vartanov,
  ``Elliptic Hypergeometry of Supersymmetric Dualities,''
Commun.\ Math.\ Phys.\  {\bf 304}, 797 (2011).
[arXiv:0910.5944 [hep-th]].
}

\lref\SpiridonovHF{
  V.~P.~Spiridonov and G.~S.~Vartanov,
  ``Elliptic hypergeometry of supersymmetric dualities II. Orthogonal groups, knots, and vortices,''
[arXiv:1107.5788 [hep-th]].
}

\lref\SpiridonovWW{
  V.~P.~Spiridonov and G.~S.~Vartanov,
  ``Elliptic hypergeometric integrals and 't Hooft anomaly matching conditions,''
JHEP {\bf 1206}, 016 (2012).
[arXiv:1203.5677 [hep-th]].
}

\lref\StrasslerFE{
  M.~J.~Strassler,
  ``Duality, phases, spinors and monopoles in $SO(N)$ and $spin(N)$ gauge theories,''
JHEP {\bf 9809}, 017 (1998).
[hep-th/9709081].
}

\lref\VasilievVF{
  M.~A.~Vasiliev,
  ``Holography, Unfolding and Higher-Spin Theory,''
J.\ Phys.\ A {\bf 46}, 214013 (2013).
[arXiv:1203.5554 [hep-th]].
}

\lref\VerstegenAT{
  D.~Verstegen,
  ``Conformal embeddings, rank-level duality and exceptional modular invariants,''
Commun.\ Math.\ Phys.\  {\bf 137}, 567 (1991).
}

\lref\WangUKY{
  C.~Wang, A.~C.~Potter and T.~Senthil,
  ``Gapped symmetry preserving surface state for the electron topological insulator,''
Phys.\ Rev.\ B {\bf 88}, no. 11, 115137 (2013).
[arXiv:1306.3223 [cond-mat.str-el]].
}

\lref\MPS{
  C. Wang, A. C. Potter, and T. Senthil,
  ``Classification Of Interacting Electronic Topological Insulators In Three Dimensions,''
Science {\bf 343} (2014) 629,
[arXiv:1306.3238].
}

\lref\WangLCA{
  C.~Wang and T.~Senthil,
  ``Interacting fermionic topological insulators/superconductors in three dimensions,''
Phys.\ Rev.\ B {\bf 89}, no. 19, 195124 (2014), Erratum: [Phys.\ Rev.\ B {\bf 91}, no. 23, 239902 (2015)].
[arXiv:1401.1142 [cond-mat.str-el]].
}

\lref\WangQMT{
  C.~Wang and T.~Senthil,
  ``Dual Dirac Liquid on the Surface of the Electron Topological Insulator,''
Phys.\ Rev.\ X {\bf 5}, no. 4, 041031 (2015). [arXiv:1505.05141 [cond-mat.str-el]].
}

\lref\WangFQL{
  C.~Wang and T.~Senthil,
  ``Half-filled Landau level, topological insulator surfaces, and three-dimensional quantum spin liquids,''
Phys.\ Rev.\ B {\bf 93}, no. 8, 085110 (2016). [arXiv:1507.08290 [cond-mat.st-el]].
}

\lref\WangGQJ{
  C.~Wang and T.~Senthil,
  ``Composite fermi liquids in the lowest Landau level,''
Phys.\ Rev.\ B {\bf 94}, 245107 (2016). 
[arXiv:1604.06807 [cond-mat.str-el]].
}

\lref\ArmoniVV{
  A.~Armoni, A.~Giveon, D.~Israel and V.~Niarchos,
  ``Brane Dynamics and 3D Seiberg Duality on the Domain Walls of 4D N=1 SYM,''
JHEP {\bf 0907}, 061 (2009).
[arXiv:0905.3195 [hep-th]].
}

\lref\WangCTO{
  C.~Wang and T.~Senthil,
  ``Time-Reversal Symmetric $U(1)$ Quantum Spin Liquids,''
Phys.\ Rev.\ X {\bf 6}, no. 1, 011034 (2016).
}

\lref\Whitehead{
  J.~H.~C.~Whitehead,
  ``On simply connected, 4-dimensional polyhedra,''
Comm.\ Math.\ Helv.\ {\bf 22} (1949) 48.
}

\lref\WilczekDU{
  F.~Wilczek,
  ``Magnetic Flux, Angular Momentum, and Statistics,''
Phys.\ Rev.\ Lett.\  {\bf 48}, 1144 (1982).
}

\lref\WilczekCY{
  F.~Wilczek and A.~Zee,
  ``Linking Numbers, Spin, and Statistics of Solitons,''
Phys.\ Rev.\ Lett.\  {\bf 51}, 2250 (1983).
}

\lref\GasserYG{
  J.~Gasser and H.~Leutwyler,
  ``Chiral Perturbation Theory to One Loop,''
Annals Phys.\  {\bf 158}, 142 (1984).
}

\lref\WillettGP{
  B.~Willett and I.~Yaakov,
  ``${\cal N}{=}2$ Dualities and $Z$-extremization in Three Dimensions,''
arXiv:1104.0487 [hep-th].
}

\lref\WittenHF{
  E.~Witten,
  ``Quantum Field Theory and the Jones Polynomial,''
Commun.\ Math.\ Phys.\  {\bf 121}, 351 (1989).
}

\lref\ClossetVP{
  C.~Closset, T.~T.~Dumitrescu, G.~Festuccia, Z.~Komargodski and N.~Seiberg,
  ``Comments on Chern-Simons Contact Terms in Three Dimensions,''
JHEP {\bf 1209}, 091 (2012).
[arXiv:1206.5218 [hep-th]].
}

\lref\CohenCC{
  E.~Cohen and C.~Gomez,
  ``Confinement And Chiral Symmetry Breaking With Twisted Gauge Configurations."
}

\lref\WittenXI{
  E.~Witten,
  ``The Verlinde algebra and the cohomology of the Grassmannian,''
In *Cambridge 1993, Geometry, topology, and physics* 357-422.
[hep-th/9312104].
}

\lref\WittenGF{
  E.~Witten,
  ``On S duality in Abelian gauge theory,''
Selecta Math.\  {\bf 1}, 383 (1995).
[hep-th/9505186].
}

\lref\WittenDS{
  E.~Witten,
  ``Supersymmetric index of three-dimensional gauge theory,''
In *Shifman, M.A. (ed.): The many faces of the superworld* 156-184.
[hep-th/9903005].
}

\lref\WittenYA{
  E.~Witten,
  ``SL(2,Z) action on three-dimensional conformal field theories with Abelian symmetry,''
In *Shifman, M. (ed.) et al.: From fields to strings, vol. 2* 1173-1200.
[hep-th/0307041].
}

\lref\WittenABA{
  E.~Witten,
  ``Fermion Path Integrals And Topological Phases,''
[arXiv:1508.04715 [cond-mat.mes-hall]].
}

\lref\WuGE{
  T.~T.~Wu and C.~N.~Yang,
  ``Dirac Monopole Without Strings: Monopole Harmonics,''
Nucl.\ Phys.\ B {\bf 107}, 365 (1976).
}

\lref\XuNXA{
  F.~Xu,
  ``Algebraic coset conformal field theories,''
Commun.\ Math.\ Phys.\  {\bf 211}, 1 (2000).
[math/9810035].
}

\lref\VafaTF{
  C.~Vafa and E.~Witten,
  ``Restrictions on Symmetry Breaking in Vector-Like Gauge Theories,''
Nucl.\ Phys.\ B {\bf 234}, 173 (1984).
}

\lref\BeniniDUS{
  F.~Benini, P.~S.~Hsin and N.~Seiberg,
  ``Comments on global symmetries, anomalies, and duality in (2 + 1)d,''
JHEP {\bf 1704}, 135 (2017).
[arXiv:1702.07035 [cond-mat.str-el]].
}

\lref\RedlichDV{
  A.~N.~Redlich,
  ``Parity Violation and Gauge Noninvariance of the Effective Gauge Field Action in Three-Dimensions,''
Phys.\ Rev.\ D {\bf 29}, 2366 (1984).
}

\lref\XuLXA{
  C.~Xu and Y.~Z.~You,
  ``Self-dual Quantum Electrodynamics as Boundary State of the three dimensional Bosonic Topological Insulator,''
Phys.\ Rev.\ B {\bf 92}, 220416 (2015). 
[arXiv:1510.06032 [cond-mat.str-el]].
}

\lref\RadicevicYLA{
  D.~Radicevic,
  ``Disorder Operators in Chern-Simons-Fermion Theories,''
JHEP {\bf 1603}, 131 (2016).
[arXiv:1511.01902 [hep-th]].
}

\lref\AharonyMJS{
  O.~Aharony,
  ``Baryons, monopoles and dualities in Chern-Simons-matter theories,''
JHEP {\bf 1602}, 093 (2016).
[arXiv:1512.00161 [hep-th]].
}

\lref\ZupnikRY{
   B.~M.~Zupnik and D.~G.~Pak,
   ``Topologically Massive Gauge Theories In Superspace,''
Sov.\ Phys.\ J.\  {\bf 31}, 962 (1988).
}

\lref\AppelquistTC{
  T.~Appelquist and D.~Nash,
  ``Critical Behavior in (2+1)-dimensional {QCD},''
Phys.\ Rev.\ Lett.\  {\bf 64}, 721 (1990).
}

\lref\FreedMX{
  D.~S.~Freed,
  ``Pions and Generalized Cohomology,''
J.\ Diff.\ Geom.\  {\bf 80}, no. 1, 45 (2008).
[hep-th/0607134].
}

\lref\ZwiebelWA{
  B.~I.~Zwiebel,
  ``Charging the Superconformal Index,''
JHEP {\bf 1201}, 116 (2012).
[arXiv:1111.1773 [hep-th]].
}

\lref\KomargodskiDMC{
  Z.~Komargodski, A.~Sharon, R.~Thorngren and X.~Zhou,
  ``Comments on Abelian Higgs Models and Persistent Order,''
[arXiv:1705.04786 [hep-th]].
}

\lref\CamanhoAPA{
  X.~O.~Camanho, J.~D.~Edelstein, J.~Maldacena and A.~Zhiboedov,
JHEP {\bf 1602}, 020 (2016).
[arXiv:1407.5597 [hep-th]].
}

\lref\LandsteinerCP{
  K.~Landsteiner, E.~Megias and F.~Pena-Benitez,
  ``Gravitational Anomaly and Transport,''
Phys.\ Rev.\ Lett.\  {\bf 107}, 021601 (2011).
[arXiv:1103.5006 [hep-ph]].
}

\lref\BanerjeeIZ{
  N.~Banerjee, J.~Bhattacharya, S.~Bhattacharyya, S.~Jain, S.~Minwalla and T.~Sharma,
  ``Constraints on Fluid Dynamics from Equilibrium Partition Functions,''
JHEP {\bf 1209}, 046 (2012).
[arXiv:1203.3544 [hep-th]].
}

\lref\JensenKJ{
  K.~Jensen, R.~Loganayagam and A.~Yarom,
  ``Thermodynamics, gravitational anomalies and cones,''
JHEP {\bf 1302}, 088 (2013).
[arXiv:1207.5824 [hep-th]].
}

\lref\BonettiELA{
  F.~Bonetti, T.~W.~Grimm and S.~Hohenegger,
  ``One-loop Chern-Simons terms in five dimensions,''
JHEP {\bf 1307}, 043 (2013).
[arXiv:1302.2918 [hep-th]].
}

\lref\DiPietroBCA{
  L.~Di Pietro and Z.~Komargodski,
  ``Cardy formulae for SUSY theories in $d =$ 4 and $d =$ 6,''
JHEP {\bf 1412}, 031 (2014).
[arXiv:1407.6061 [hep-th]].
}

\lref\ArdehaliHYA{
  A.~Arabi Ardehali, J.~T.~Liu and P.~Szepietowski,
  ``High-Temperature Expansion of Supersymmetric Partition Functions,''
JHEP {\bf 1507}, 113 (2015).
[arXiv:1502.07737 [hep-th]].
}

\lref\FeiOHA{
  L.~Fei, S.~Giombi, I.~R.~Klebanov and G.~Tarnopolsky,
  ``Generalized $F$-Theorem and the $\epsilon$ Expansion,''
JHEP {\bf 1512}, 155 (2015).
[arXiv:1507.01960 [hep-th]].
}

\lref\GiombiXXA{
  S.~Giombi and I.~R.~Klebanov,
  ``Interpolating between $a$ and $F$,''
JHEP {\bf 1503}, 117 (2015).
[arXiv:1409.1937 [hep-th]].
}

\lref\KapustinGUA{
  A.~Kapustin and N.~Seiberg,
  ``Coupling a QFT to a TQFT and Duality,''
JHEP {\bf 1404}, 001 (2014).
[arXiv:1401.0740 [hep-th]].
}

\lref\GaiottoKFA{
  D.~Gaiotto, A.~Kapustin, N.~Seiberg and B.~Willett,
  ``Generalized Global Symmetries,''
JHEP {\bf 1502}, 172 (2015).
[arXiv:1412.5148 [hep-th]].
}

\lref\Freed{
  D.~S.~Freed, Z.~Komargodski, and N.~Seiberg,
  to appear.
}

\lref\DijkgraafPZ{
  R.~Dijkgraaf and E.~Witten,
  ``Topological Gauge Theories and Group Cohomology,''
Commun.\ Math.\ Phys.\  {\bf 129}, 393 (1990).
}

\lref\KadanoffKZ{
  L.~P.~Kadanoff and H.~Ceva,
  ``Determination of an opeator algebra for the two-dimensional Ising model,''
Phys.\ Rev.\ B {\bf 3}, 3918 (1971).
}

\lref\GrossKZA{
  D.~J.~Gross and P.~F.~Mende,
Phys.\ Lett.\ B {\bf 197}, 129 (1987).
}

\lref\KarlinerHD{
  M.~Karliner, I.~R.~Klebanov and L.~Susskind,
Int.\ J.\ Mod.\ Phys.\ A {\bf 3}, 1981 (1988).
}

\lref\Shankar{
  R.~Shankar and N.~Read,
  ``The $\theta = \pi$ Nonlinear $\sigma$ Model Is Massless,''
Nucl.\ Phys.\ B {\bf 336}, 457 (1990).
}

\lref\WeissRJ{
  N.~Weiss,
  ``The Effective Potential for the Order Parameter of Gauge Theories at Finite Temperature,''
Phys.\ Rev.\ D {\bf 24}, 475 (1981).
}

\lref\DashenET{
  R.~F.~Dashen,
  ``Some features of chiral symmetry breaking,''
Phys.\ Rev.\ D {\bf 3}, 1879 (1971).
}

\lref\Hitoshi{
http://hitoshi.berkeley.edu/221b/scattering3.pdf
}

\lref\RosenzweigAY{
  C.~Rosenzweig, J.~Schechter and C.~G.~Trahern,
  ``Is the Effective Lagrangian for QCD a Sigma Model?,''
Phys.\ Rev.\ D {\bf 21}, 3388 (1980).
}

\lref\NathIK{
  P.~Nath and R.~L.~Arnowitt,
  ``The U(1) Problem: Current Algebra and the Theta Vacuum,''
Phys.\ Rev.\ D {\bf 23}, 473 (1981).
}

\lref\GreensiteZZ{
  J.~Greensite,
  ``An introduction to the confinement problem,''
Lect.\ Notes Phys.\  {\bf 821}, 1 (2011).
}

\lref\SeibergRS{
  N.~Seiberg and E.~Witten,
  ``Electric - magnetic duality, monopole condensation, and confinement in N=2 supersymmetric Yang-Mills theory,''
Nucl.\ Phys.\ B {\bf 426}, 19 (1994), Erratum: [Nucl.\ Phys.\ B {\bf 430}, 485 (1994)].
[hep-th/9407087].
}

\lref\ColemanUZ{
  S.~R.~Coleman,
  ``More About the Massive Schwinger Model,''
Annals Phys.\  {\bf 101}, 239 (1976).
}

\lref\SeibergAJ{
  N.~Seiberg and E.~Witten,
  ``Monopoles, duality and chiral symmetry breaking in N=2 supersymmetric QCD,''
Nucl.\ Phys.\ B {\bf 431}, 484 (1994).
[hep-th/9408099].
}

\lref\Deepak{
D.~Naidu, ``Categorical Morita equivalence for group-theoretical categories, " Communications in Algebra 35.11 (2007): 3544-3565.
APA	
}

\lref\SusskindAA{
  L.~Susskind,
Phys.\ Rev.\ D {\bf 49}, 6606 (1994).
[hep-th/9308139].
}

\lref\DubovskySH{
  S.~Dubovsky, R.~Flauger and V.~Gorbenko,
  ``Effective String Theory Revisited,''
JHEP {\bf 1209}, 044 (2012).
[arXiv:1203.1054 [hep-th]].
}

\lref\AharonyIPA{
  O.~Aharony and Z.~Komargodski,
  ``The Effective Theory of Long Strings,''
JHEP {\bf 1305}, 118 (2013).
[arXiv:1302.6257 [hep-th]].
}

\lref\HellermanCBA{
  S.~Hellerman, S.~Maeda, J.~Maltz and I.~Swanson,
  ``Effective String Theory Simplified,''
JHEP {\bf 1409}, 183 (2014).
[arXiv:1405.6197 [hep-th]].
}

\lref\AthenodorouCS{
  A.~Athenodorou, B.~Bringoltz and M.~Teper,
  ``Closed flux tubes and their string description in D=3+1 SU(N) gauge theories,''
JHEP {\bf 1102}, 030 (2011).
[arXiv:1007.4720 [hep-lat]].
}

\lref\AharonyHDA{
  O.~Aharony, N.~Seiberg and Y.~Tachikawa,
  ``Reading between the lines of four-dimensional gauge theories,''
JHEP {\bf 1308}, 115 (2013).
[arXiv:1305.0318 [hep-th]].
}

\lref\CallanSA{
  C.~G.~Callan, Jr. and J.~A.~Harvey,
  ``Anomalies and Fermion Zero Modes on Strings and Domain Walls,''
Nucl.\ Phys.\ B {\bf 250}, 427 (1985).
}

\lref\tHooftXSS{
  G.~'t Hooft et al.,
    ``Recent Developments in Gauge Theories. Proceedings, Nato Advanced Study Institute, Cargese, France, August 26 - September 8, 1979,''
NATO Sci.\ Ser.\ B {\bf 59}, pp.1 (1980).
}

\lref\AppelquistVG{
  T.~Appelquist and R.~D.~Pisarski,
  ``High-Temperature Yang-Mills Theories and Three-Dimensional Quantum Chromodynamics,''
Phys.\ Rev.\ D {\bf 23}, 2305 (1981).
}

\lref\HarveyIT{
  J.~A.~Harvey,
  ``TASI 2003 lectures on anomalies,''
[hep-th/0509097].
}

\lref\tHooftUJ{
  G.~'t Hooft,
  ``A Property of Electric and Magnetic Flux in Nonabelian Gauge Theories,''
Nucl.\ Phys.\ B {\bf 153}, 141 (1979).
}

\lref\Brower{
R.~C.~Brower and J.~Harte.
Physical Review 164.5 (1967): 1841.
}

\lref\HenningsonHP{
  M.~Henningson,
  ``Wilson-'t Hooft operators and the theta angle,''
JHEP {\bf 0605}, 065 (2006).
[hep-th/0603188].
}

\lref\CreutzXU{
  M.~Creutz,
  ``Spontaneous violation of CP symmetry in the strong interactions,''
Phys.\ Rev.\ Lett.\  {\bf 92}, 201601 (2004).
[hep-lat/0312018].
}

\lref\ChenPG{
  X.~Chen, Z.~C.~Gu, Z.~X.~Liu and X.~G.~Wen,
  ``Symmetry protected topological orders and the group cohomology of their symmetry group,''
Phys.\ Rev.\ B {\bf 87}, no. 15, 155114 (2013).
[arXiv:1106.4772 [cond-mat.str-el]].
}

\lref\PappadopuloJK{
  D.~Pappadopulo, S.~Rychkov, J.~Espin and R.~Rattazzi,
Phys.\ Rev.\ D {\bf 86}, 105043 (2012).
[arXiv:1208.6449 [hep-th]].
}

\lref\LegRed{
Askey, Richard. "Orthogonal expansions with positive coefficients." Proceedings of the American Mathematical Society 16.6 (1965): 1191-1194.
}

\lref\ClossetVP{
  C.~Closset, T.~T.~Dumitrescu, G.~Festuccia, Z.~Komargodski and N.~Seiberg,
  ``Comments on Chern-Simons Contact Terms in Three Dimensions,''
JHEP {\bf 1209}, 091 (2012).
[arXiv:1206.5218 [hep-th]].
}

\lref\AnberKEA{
  M.~M.~Anber, E.~Poppitz and T.~Sulejmanpasic,
  ``Strings from domain walls in supersymmetric Yang-Mills theory and adjoint QCD,''
Phys.\ Rev.\ D {\bf 92}, no. 2, 021701 (2015).
[arXiv:1501.06773 [hep-th]].
}

\lref\JainGZA{
  S.~Jain, S.~Minwalla and S.~Yokoyama,
  ``Chern Simons duality with a fundamental boson and fermion,''
JHEP {\bf 1311}, 037 (2013).
[arXiv:1305.7235 [hep-th]].
}

\lref\WittenDF{
  E.~Witten,
  ``Constraints on Supersymmetry Breaking,''
Nucl.\ Phys.\ B {\bf 202}, 253 (1982).
}

\lref\SeibergRSG{
  N.~Seiberg and E.~Witten,
  ``Gapped Boundary Phases of Topological Insulators via Weak Coupling,''
PTEP {\bf 2016}, no. 12, 12C101 (2016).
[arXiv:1602.04251 [cond-mat.str-el]].
}

\lref\AmatiWQ{
  D.~Amati, M.~Ciafaloni and G.~Veneziano,
Phys.\ Lett.\ B {\bf 197}, 81 (1987).
}

\lref\KomargodskiKEH{
  Z.~Komargodski and N.~Seiberg,
  ``A Symmetry Breaking Scenario for QCD$_3$,''
[arXiv:1706.08755 [hep-th]].
}

\lref\TachikawaCHA{
  Y.~Tachikawa and K.~Yonekura,
  ``On time-reversal anomaly of 2+1d topological phases,''
PTEP {\bf 2017}, no. 3, 033B04 (2017).
[arXiv:1610.07010 [hep-th]].
}

\lref\TachikawaNMO{
  Y.~Tachikawa and K.~Yonekura,
  ``More on time-reversal anomaly of 2+1d topological phases,''
[arXiv:1611.01601 [hep-th]].
}

\lref\forthc{
  J.~Gomis, Z.~Komargodski, and N.~Seiberg, 
  To Appear.
}

\lref\FrishmanDQ{
  Y.~Frishman, A.~Schwimmer, T.~Banks and S.~Yankielowicz,
  ``The Axial Anomaly and the Bound State Spectrum in Confining Theories,''
Nucl.\ Phys.\ B {\bf 177}, 157 (1981).
}

\lref\ShimizuASF{
  H.~Shimizu and K.~Yonekura,
  ``Anomaly constraints on deconfinement and chiral phase transition,''
[arXiv:1706.06104 [hep-th]].
}

\lref\ChermanTEY{
  A.~Cherman, S.~Sen, M.~Unsal, M.~L.~Wagman and L.~G.~Yaffe,
  ``Order parameters and color-flavor center symmetry in QCD,''
[arXiv:1706.05385 [hep-th]].
}

\lref\KapustinLWA{
  A.~Kapustin and R.~Thorngren,
  ``Anomalies of discrete symmetries in three dimensions and group cohomology,''
Phys.\ Rev.\ Lett.\  {\bf 112}, no. 23, 231602 (2014).
[arXiv:1403.0617 [hep-th]].
}

\lref\ElitzurXJ{
  S.~Elitzur, Y.~Frishman, E.~Rabinovici and A.~Schwimmer,
  ``Origins of Global Anomalies in Quantum Mechanics,''
Nucl.\ Phys.\ B {\bf 273}, 93 (1986).
}

\lref\WangTXT{
  C.~Wang, A.~Nahum, M.~A.~Metlitski, C.~Xu and T.~Senthil,
  ``Deconfined quantum critical points: symmetries and dualities,''
[arXiv:1703.02426 [cond-mat.str-el]].
}

\lref\GattringerBAA{
  C.~Gattringer, T.~Kloiber and M.~Müller-Preussker,
  ``Dual simulation of the two-dimensional lattice U(1) gauge-Higgs model with a topological term,''
Phys.\ Rev.\ D {\bf 92}, no. 11, 114508 (2015).
[arXiv:1508.00681 [hep-lat]].
}

\lref\KapustinZVA{
  A.~Kapustin and R.~Thorngren,
  ``Anomalies of discrete symmetries in various dimensions and group cohomology,''
[arXiv:1404.3230 [hep-th]].
}

\lref\CreutzWF{
  M.~Creutz,
  ``Quark masses and chiral symmetry,''
Phys.\ Rev.\ D {\bf 52}, 2951 (1995).
[hep-th/9505112].
}

\lref\CreutzXU{
  M.~Creutz,
  ``Spontaneous violation of CP symmetry in the strong interactions,''
Phys.\ Rev.\ Lett.\  {\bf 92}, 201601 (2004).
[hep-lat/0312018].
}

\lref\KomargodskiDMC{
  Z.~Komargodski, A.~Sharon, R.~Thorngren and X.~Zhou,
  ``Comments on Abelian Higgs Models and Persistent Order,''
[arXiv:1705.04786 [hep-th]].
}

\lref\KomargodskiSMK{
  Z.~Komargodski, T.~Sulejmanpasic and M.~\"Unsal,
  ``Walls, Anomalies, and (De)Confinement in Quantum Anti-Ferromagnets,''
[arXiv:1706.05731 [cond-mat.str-el]].
}

\lref\ShankarDU{
  R.~Shankar and G.~Murthy,
  ``Deconfinement in d=1: A Closer look,''
Phys.\ Rev.\ B {\bf 72}, 224414 (2005).
[cond-mat/0508242].
}

\lref\WittenSP{
  E.~Witten,
  ``Large N Chiral Dynamics,''
Annals Phys.\  {\bf 128}, 363 (1980).
}

\lref\RabinoviciMJ{
  E.~Rabinovici, A.~Schwimmer and S.~Yankielowicz,
  ``Quantization in the Presence of {Wess-Zumino} Terms,''
Nucl.\ Phys.\ B {\bf 248}, 523 (1984).
}

\lref\KomargodskiKEH{
  Z.~Komargodski and N.~Seiberg,
  ``A Symmetry Breaking Scenario for QCD$_3$,''
[arXiv:1706.08755 [hep-th]].
}

\lref\BeniniDUS{
  F.~Benini, P.~S.~Hsin and N.~Seiberg,
  ``Comments on global symmetries, anomalies, and duality in (2 + 1)d,''
JHEP {\bf 1704}, 135 (2017).
[arXiv:1702.07035 [cond-mat.str-el]].
}

\lref\TanizakiBAM{
  Y.~Tanizaki and Y.~Kikuchi,
  ``Vacuum structure of bifundamental gauge theories at finite topological angles,''
JHEP {\bf 1706}, 102 (2017).
[arXiv:1705.01949 [hep-th]].
}

\lref\Hatcher{
A.~Hatcher, ``Algebraic topology,'' Cambridge University Press, 2002.
}

\lref\ZamolodchikovZR{
  A.~B.~Zamolodchikov and A.~B.~Zamolodchikov,
  ``Massless factorized scattering and sigma models with topological terms,''
Nucl.\ Phys.\ B {\bf 379}, 602 (1992).
}


\vskip-60pt
\Title{} {\vbox{\centerline{Time-Reversal Breaking in QCD$_4$, Walls, }
\centerline{}
\centerline{and Dualities in 2+1 Dimensions
}}  }

\vskip-15pt

\centerline{Davide Gaiotto,${}^{1}$ Zohar Komargodski,${}^{2,3}$ and Nathan Seiberg${}^4$}
\vskip15pt
\centerline{\it ${}^1$  Perimeter Institute for Theoretical Physics,
Waterloo, Ontario, N2L 2Y5, Canada }
\centerline{\it ${}^2$ Department of Particle Physics and Astrophysics, Weizmann Institute of Science, Israel }
\centerline{\it ${}^3$   Simons Center for Geometry and Physics, Stony Brook University, Stony Brook, NY}
\centerline{\it ${}^4$ School of Natural Sciences, Institute for Advanced Study, Princeton, NJ 08540, USA}

\bigskip

\noindent
We study $SU(N)$ Quantum Chromodynamics (QCD) in 3+1 dimensions with $N_f$ degenerate fundamental quarks with mass $m$ and a $\theta$-parameter.  For generic $m$ and $\theta$ the theory has a single gapped vacuum. However, as $\theta$ is varied through $\theta=\pi$ for large $m$ there is a first order transition. For $N_f=1$ the first order transition line ends at a point with a massless $\eta'$ particle (for all $N$) and for $N_f>1$ the first order transition ends at $m=0$, where, depending on the value of $N_f$, the IR theory has free Nambu-Goldstone bosons, an interacting conformal field theory, or a free gauge theory.  Even when the $4d$ bulk is smooth, domain walls and interfaces can have interesting phase transitions separating different $3d$ phases.  These turn out to be the phases of the recently studied $3d$ Chern-Simons matter theories, thus relating the dynamics of QCD$_4$ and QCD$_3$, and, in particular, making contact with the recently discussed dualities in 2+1 dimensions. For example, when the massless $4d$ theory has an $SU(N_f)$ sigma model, the domain wall theory at low (nonzero) mass supports a $3d$ massless $\C\P^{N_f-1}$ nonlinear $\sigma$-model with a Wess-Zumino term, in agreement with the conjectured dynamics in 2+1 dimensions.

\bigskip
\Date{August 2017}

\newsec{Introduction}

As is well known, the dynamics of $4d$ QCD with massless quarks depends on the number of flavors $N_f$ and colors $N$.  For $N_f\ge {11\over 2} N$ it is IR free.  It is meaningful only as an effective IR theory and its IR dynamics is rather simple.  For $N_{CFT}\le N_f<{11\over 2}N$ with some yet unknown $N_{CFT}(N)$ the theory flows to a nontrivial fixed point.  For $1<N_f<N_{CFT}$ it breaks its chiral symmetry leading at low energies to a nonlinear sigma model with target space $SU(N_f)$.  For $N_f=1$ it is gapped with a unique vacuum.  And for $N_f=0$ the theory has an additional parameter, $\theta$.  For generic $\theta$ it is gapped with a unique vacuum, but as we vary $\theta$ through $\pi$ we cross a first order phase associated with the spontaneously broken time-reversal symmetry at that point. The vacuum is therefore doubly degenerate at $\theta=\pi$.

This picture has not been rigorously derived.  In fact, although the phase transition at $\theta=\pi$ can be derived at large $N$ \refs{\WittenBC\WittenVV\WittenSP\DiVecchiaYFW-\WittenUKA} and was argued to exist at finite $N$ using anomalies~\GaiottoYUP, it could be that at small values of $N$ the picture is different~\GaiottoYUP.  Also, it might be that for some $N$ and $N_f$ there are additional more exotic phases.  Here we will ignore these possibilities.

One purpose of this note is to make some comments about the extension of this picture to the massive theory.  For simplicity, we turn on equal masses $m$ for all the quarks.  When $m\not=0$ the theory depends on $\theta$.  However, because of the chiral anomaly, the bulk physics depends only on $m^{N_f}e^{i\theta}$.  So, without loss of generality, throughout this note we will take $m$ real and non-negative and we will explore the dependence of the theory on the complex number $m^{N_f}e^{i\theta}$.

Using a combination of arguments based on the expected behavior at $m=0$, matching the large $m$ theory with the $N_f=0$ theory, large $N$ results, and anomalies, we will present a coherent picture of the phase diagram.  For generic $m$ and $\theta$ the theory has a unique gapped vacuum.  It is typically the case that at $\theta=\pi$ the theory has a first order transition associated with the spontaneous breaking of its time-reversal symmetry there.  For $N_f>1$ this transition persists all the way to $m=0$, i.e.\ it is on the entire negative real axis in the complex $m^{N_f}e^{i\theta}$ plane.  It ends at $m=0$, where we find the IR behavior we mentioned above (which depends on $N_f$, i.e.\ a chiral Lagrangian for $1<N_f<N_{CFT}$ and a free gauge theory or a conformal field theory for $N_f\geq N_{CFT}$).  For $N_f=1$ the situation is different, essentially because the $m=0$ theory does not have an enhanced symmetry compared to the  $m\neq 0$ theory. For $N_f=1$ the first order transition runs along the negative real axis of $me^{i\theta}$ and ends at a negative point $me^{i\theta}=-m_0$.  (Large $N$ arguments imply~\refs{\WittenSP, \DiVecchiaYFW} that the theory with $m=0$ has a non-degenerate gapped vacuum. We will assume that this remains the case for lower values of $N$.  Then, $m_0$  is a positive number.)  The low energy theory at this point includes a single massless particle, which we will identify with the $\eta'$ particle.  The reason for that identification is that for large $N$ the endpoint $m_0$ goes to zero and this particle is the Nambu-Goldstone boson of the axial $U(1)$ symmetry.

We will study in detail domain walls and interfaces in these systems.  We should distinguish between different notions.

First, whenever the theory has more than one vacuum, e.g.\ in the time-reversal symmetry broken situations at $\theta=\pi$, there can be dynamical domain walls separating between the two vacua.  These are dynamical excitations of the system.  In all our examples the domain wall separates between two gapped ground states -- the lowest excitation in the bulk has nonzero energy $M$.\foot{Sometime, as in the pure gauge $N_f=0$ theory, these two gapped phases are in different SPT phases, that is, the two gapped phases have different local-terms for background fields \GaiottoYUP.} It is often the case that there are nontrivial excitations with energy much lower than $M$ living on the domain wall.  They are described by a $3d$ quantum field theory. It may also be that the domain wall does not have excitations with energy much smaller than $M$, but it supports a $3d$ Topological QFT (TQFT).  One of the goals of this paper is to identify these $3d$ QFTs.  These $3d$ QFTs are valid only up to energies of order $M$.  At higher energies the bulk cannot be ignored and the theory is no longer a purely $3d$ QFT.

Second, we will consider the system with space-dependent coupling constants.  Specifically, we will let $\theta$ be position dependent and we will let it interpolate smoothly between, say $\theta=0$ and $\theta=2\pi k$ for some $k$.  Since $\theta$ is position dependent, this is not an excitation of the system.  Yet, it is an interesting system to study.  It is important that the physics of this interface follows from the UV $4d$ Lagrangian without additional data.\foot{Here we assume that the UV theory does not include additional couplings of the dynamical fields to derivatives of $\theta$.}  As we will see, the physics of the interface depends on how fast $\theta$ varies, i.e.\ it depends on $|\grad \theta|$.

Third, we can consider sharp interfaces.  Here the parameters change abruptly.  For example, we can have a sharp interface between $\theta_1$ and $\theta_2$.  This situation can be viewed as a limit of the previous case, as the gradient becomes larger.  However, unlike the smooth interface, here the physics is not universal.  There is an ambiguity in adding degrees of freedom supported only on the interface.  Anomaly considerations can constrain the dynamics, but they do not uniquely determine it.

Now let us summarize the results of this paper.

\medskip

\item{1.} {\it Pure Yang-Mills theory.} At $\theta=\pi$ time-reversal symmetry is spontaneously broken. The domain wall theory is comprised of the usual center of mass mode but in addition there is an $SU(N)_1$ Chern-Simons theory. This is an example where the domain wall supports a nontrivial $3d$ TQFT \GaiottoYUP.

\item{2.} {\it QCD with $N_f=1$}. For $m\gg m_0$ and $\theta=\pi$ the theory is well described by pure Yang-Mills theory at $\theta=\pi$ and therefore has two degenerate ground states related by time-reversal symmetry. The domain wall theory contains $SU(N)_1$ Chern-Simons theory. As we lower the mass $m$ this TQFT eventually disappears and we obtain a domain wall theory that contains only the center of mass mode. There is therefore a phase transition on the domain wall while the bulk is entirely gapped and smooth.\foot{A similar phenomenon was predicted to occur in quantum anti-ferromagnets, i.e.~in the 2+1-dimensional Abelian Higgs model with monopole operators~\refs{\SulejmanpasicUWQ,\KomargodskiDMC,\KomargodskiSMK}.}

\item{3.}{\it QCD with $N_{CFT}>N_f>1$}. Here again at large $m$ and $\theta=\pi$ we have $SU(N)_1$ Chern-Simons theory. As we lower the mass, there is a phase transition and we obtain a $\C\P^{N_f-1}$ non-linear $\sigma$-model with some Wess-Zumino term. The $\C\P^{N_f-1}$ non-linear $\sigma$-model is clearly visible in the chiral Lagrangian. There is therefore again a phase transition on the domain wall while the bulk is gapped and smooth.

\item{4.} {\it QCD with $N_{CFT}\leq N_f$}. We do not study this case in detail in this paper. But we note that a natural conjecture is that here the domain wall theory is given by an $SU(N)_1$ Chern-Simons theory for all $m$ (which should be below the Landau pole scale when $N_f\geq {11\over 2} N$).

\medskip

The phases in all the four cases above are captured by the conjectured phases of the Chern-Simons matter theory
\eqn\threed{SU(N)_{1-N_f/2}+N_f\ {\rm fermions}~,}
where the fermions are in the fundamental representation of $SU(N)$. This model has been studied in detail in the literature on $3d$ Chern-Simons matter theories.

\medskip

\item{1.} $N_f=0$. This model~\threed\ is just a pure Chern-Simons TQFT.

\item{2.}$N_f=1$. The three-dimensional model is $SU(N)_{1/2}+\psi$. This model has $N_f=2|k|$ where $k$ is the Chern-Simons level (which is equal to 1/2 in our case). Therefore, this model is within the regime where the dualities of~\refs{\JainGZA\RadicevicYLA\AharonyMJS\KarchSXI\MuruganZAL\SeibergGMD\HsinBLU\RadicevicWQN\KachruRUI
\KachruAON\KarchAUX\MetlitskiDHT\AharonyJVV-\BeniniDUS} apply. There are conjecturally two phases separated by a transition. One phase is $SU(N)_0$, i.e.\ an empty, trivial theory on one side of the transition and $SU(N)_1$ on the other side of the transition. These are exactly the phases of the domain wall in QCD with one flavor, as reviewed above. Furthermore, there is a bosonic dual theory describing the transition, $U(1)_{-N}+\phi$, with $\phi$ having charge $1$.

\item{3.} $N_f>1$. Here $N_f>|2k|$ and hence the above-mentioned dualities cannot be used in their simplest form \HsinBLU.  However, these dualities are still useful.  According to \KomargodskiKEH\ for $N_*>N_f>1$ (with some unknown $N_*$) this Chern-Simons matter theory~\threed\ has two transitions, of which one is between $SU(N)_1$ Chern-Simons theory and a massless $\C\P^{N_f-1}$  sigma model. These are precisely the phases of the domain wall theory when the number of flavors in four dimensions satisfies $1<N_f<N_{CFT}$.  A dual bosonic description is given by $U(1)_{-N}+N_f\ \phi$. For $N_f\geq N_*$ it was conjectured in~\KomargodskiKEH\ that the Grassmannian phase disappears, as does the symmetry breaking phase on the domain wall in four dimensions when $N_f\geq N_{CFT}$. This suggests that $N_{CFT}$ and $N_*$ are possibly related.

\medskip

The dynamics of QCD in four dimensions therefore leads to intricate dynamics on the domain wall, reproducing the phases of nontrivial Chern-Simons matter theories.\foot{Connections between dynamics in 3+1 dimensions and 2+1 dimensions via domain wall constructions were also studied, for instance, in~\ArmoniVV.}
We will also see that QCD in four dimensions can reproduce the phases of more general 2+1 dimensional Chern-Simons matter theories by considering interfaces in addition to domain walls. However, here we do not explore the subject of interfaces in QCD exhaustively and leave it for the future.

The techniques used here can be applied in a wide variety of other examples such as theories with orthogonal or symplectic gauge groups, softly deformed supersymmetric theories, quiver theories, theories with adjoint matter fields (about which there are already a few intriguing observations~\refs{\AcharyaDZ\TachikawaCHA-\TachikawaNMO}, and see also~\forthc), etc. It would be nice to study the bulk phases, domain walls, interfaces, and corresponding dynamics and dualities in 2+1 dimensions.

In section 2 we consider the dynamics, domain walls, and interfaces of pure Yang-Mills theory. This section is mostly a review, but we make a few new observations there. In section 3 we consider one flavor QCD and again study both the bulk and domain walls phases. In section 4 we study the case of $N_{CFT}>N_f>1$ degenerate quarks both at finite $N$ and in the large $N$ limit (following~\refs{\RosenzweigAY,\NathIK,\WittenSP,\DiVecchiaYFW}). In section 5 we make some comments about anomaly matching on domain walls and interfaces. Some additional results about the chiral Lagrangian are collected in an appendix.

\newsec{Pure Yang-Mills Theory }

Consider the Lagrangian of pure Yang-Mills theory with gauge group $SU(N)$
\eqn\Lagintroi{{\cal L}=-{1\over 4g^2} Tr(F\wedge\star F)+{i\theta\over 8\pi^2}Tr(F\wedge F)~,}
where we used Euclidean signature (as we will throughout this note) and hence the factor of $i$ in front of the $\theta$-term.
We will study this theory on a four-dimensional manifold $M_4$.  The instanton number is quantized as
\eqn\topc{{1\over 8\pi^2}\int_{M_4} Tr(F\wedge F) \in  \Z~. }
As a result, the theory with $\theta$ is equivalent to the theory with $\theta+2\pi$. More precisely, the Hamiltonians with $\theta$ and $\theta+2\pi$ are similar and the similarity transformation is implemented by the unitary operator
\eqn\Unitt{U=e^{{i\over 4\pi} \int_{M_3} CS(A) }~,}
with $CS(A)$ being the standard Chern-Simons action for an $SU(N)$ gauge field and $M_3$ is a space-like slice.

The symmetries of the theory include the one-form global symmetry $\Z_N$ associated with the center of the gauge group \refs{\KapustinGUA,\GaiottoKFA} (and see references therein).  It acts on the fundamental Wilson loop $W_F=Tr_F P e^{i \int A}$ by
\eqn\Wtran{W_F\to e^{2\pi i \over N} W_F~,}
and at $\theta=0,\pi$ there is in addition also a time-reversal symmetry (equivalently, $CP$ symmetry). The theory at $\theta=0$ has no anomalies and it is believed to have a trivial, unique, gapped ground state. At $\theta=\pi$ there is a mixed 't Hooft anomaly between the time-reversal symmetry and the $\Z_N$ one-form global symmetry \GaiottoYUP. As a result, the theory cannot have a trivial ground state. One way to saturate this anomaly in the infrared is to break time-reversal symmetry spontaneously. In fact, this occurs in the planar limit $N=\infty$ \refs{\WittenBC}. We will assume that this is how the anomaly is saturated also at finite $N$. (For low values of $N$, especially for $N=2$, there are also other plausible scenarios~\GaiottoYUP.)

As a result of the spontaneous breaking of time-reversal symmetry at $\theta=\pi$, the theory admits a domain wall. The domain wall theory cannot be trivial in the infrared because of anomaly inflow. Indeed, the bulk has a mixed 't Hooft anomaly involving time-reversal symmetry and the $\Z_N$ one-form symmetry. Since time-reversal symmetry is broken in the bulk, the domain wall theory must have an 't Hooft anomaly for its one-form $\Z_N$ symmetry.  In more detail, let $B$ be the two-form $\Z_N$ gauge field. Then, the $3d$ theory on the wall can be coupled to $B$, but the partition function is not gauge invariant. The non-gauge invariance can be canceled by  a bulk term 
\eqn\bulktwo{{2\pi i(1-N) \over 2N}\int_{{\cal M}_4} B\cup B~.}
(More precisely, the Pontryagin square has to be used. Furthermore, this expression is valid only for even $N$; for odd $N$ see~\GaiottoYUP\ and some comments below. See also~\DieriglXTA\
 for an earlier discussion of similar ideas.)

From the $\Z_N$ anomaly~\bulktwo\  one can infer that the domain wall theory is an $SU(N)_1$ Chern-Simons theory~\GaiottoYUP.\foot{Note that the $SU(N)_1$ theory is related by level-rank duality to $U(1)_{-N}$, so one might try to claim that the theory on the wall can also be thought of as $U(1)_{-N}$.  However, this level-rank duality is valid only when these two theories are viewed as spin-Chern-Simons theories~\HsinBLU.  (In fact, for odd $N$ the $U(1)_{-N}$ theory exists only as a spin-Chern-Simons theory. For even $N$, $U(1)_N$ is a non-spin theory but its anomaly is ${2\pi i  \over 2N}\int_{{\cal M}_4} B\cup B$ rather than~\bulktwo. For more details, see~\forthc.)  Since our microscopic theory does not need a spin manifold, this level-rank duality cannot be used.  As a result, the theory on the wall is $SU(N)_1$ and should not be thought of as $U(1)_{-N}$. } Hence, the $\Z_N$ symmetry is spontaneously broken on the domain wall and probe quarks are deconfined near the wall (see also~\refs{\AnberKEA,\KomargodskiSMK}).

We can estimate the tension of the domain wall. Clearly it scales like $\Lambda^3$. But for large $N$ it would be important later that the tension in fact scales like \eqn\tens{T\sim N \Lambda^3~.}
This factor of $N$ follows simply from the fact that at large $N$ the action scales like $g^{-2}\sim N$.

\subsec{Digression about Anomalies in Chern-Simons Theory}

In preparation for our discussion of interfaces we would like to recall a few facts concerning Chern-Simons theories. $SU(N)_k$ Chern-Simons theory is well-defined for integer $k$ and does not require a spin structure. The theory has a $\Z_N$ one-form symmetry acting on its line operators \refs{\KapustinGUA,\GaiottoKFA}. We can think of this one-form symmetry as being associated with the center of the gauge group. This symmetry is spontaneously broken since the Wilson lines are deconfined. Suppose we couple the $SU(N)_k$ theory to a background two-form gauge field $B$ valued in $\Z_N$. The partition function may not be gauge invariant, i.e.\ the $\Z_N$ symmetry could have an 't Hooft anomaly.

Let us classify the possible anomalies. The possible  anomalies correspond to local functionals of $B$ in four dimensions. More precisely, the functionals of $B$ in four dimensions are required to be local, well defined, and gauge invariant on closed four-manifolds.
For even $N$ they are given by
\eqn\Bloc{{2\pi i K\over 2N} \int_{{\cal M}_4} B\cup B~,}
where the distinct local terms are labeled by $K=0, 1, \cdots 2N-1$.  (Actually, for spin manifolds with a choice of spin structure only $K\mod N$ matters.)
For odd $N$ the distinct local terms are labeled by $K=0,2,\cdots 2N-2$ \refs{\KapustinGUA,\GaiottoKFA}.

For even $N$,  the $SU(N)_k$ theory leads to the anomaly~\Bloc\ with 
\eqn\Kkmod{K=k-kN \ {\rm mod} \ 2N~,}
and as a result $SU(N)_{2N}$ has no anomaly, while $SU(N)_N$ has no anomaly on spin manifolds. (Here we neglect framing and other gravitational anomalies.) Equivalently, we could say that $SU(N)_k$ and $SU(N)_{k\pm 2N}$ have the same anomalies. On spin manifolds, $SU(N)_k$ and $SU(N)_{k\pm N}$ have the same anomalies.
For odd $N$ the situation is somewhat simpler and $SU(N)_k$ has exactly the same anomalies as $SU(N)_{k\pm N}$ on both spin and non-spin manifolds.\foot{Another way to state the same facts is to note that for odd $N$ the $PSU(N)_N$ Chern-Simons theory is a nonspin TQFT, while for even $N$ it is a spin-TQFT.}

\subsec{Interfaces}

Now let us discuss interfaces in the theory.  Unlike the domain walls, these are not excitations in the original system~\Lagintroi, but rather, they are obtained by letting some of the parameters be space-time dependent.

We will let $\theta$ depend on one of the coordinates $x$ in $\R^4$ interpolating between $\theta=0$ at $x\to -\infty$ to $\theta=2\pi k$ with integer (positive) $k$ as $x\to +\infty$. The two vacua at the two ends are exactly the same as long as we do not couple the theory to external background fields.
They are both non-degenerate, gapped, and confining. In fact, they are related by a similarity transformation using $U^k$ with $U$ of~\Unitt.

However, if we couple the theory to a $\Z_N$ two-form background gauge field $B$ for the one-form global symmetry, the vacua on the two sides are different.  The theory labeled by $(\theta,K)$ is the same as the theory labeled by  $(\theta + 2\pi ,K-1+N)$~\refs{\KapustinGUA,\GaiottoKFA,\GaiottoYUP}.  Therefore, starting at $(\theta=0, K=0)$ and interpolating as above by changing only $\theta$ we end up with $(\theta=2\pi k, K=0) \sim (\theta=0, K=k-kN\mod 2N)$, i.e.\ a state equivalent to $\theta=0$, but with the additional local term~\Bloc.

What is the $2+1$ dimensional theory on the interface?  We argued in the introduction that as long as $\theta$ varies in a smooth way, this question has a unique universal answer, which depends only on the UV Lagrangian and the gradient of $\theta$.  Anomaly considerations alone do not lead to a unique answer.  For example, one possibility is an $SU(N)_k$ Chern-Simons theory and another is $[SU(N)_1]^k$. And there are many other possibilities. For even $N$ we could have, for example $SU(N)_{k+2Np}$ for any integer $p$, and for odd $N$ the anomaly is consistent with $SU(N)_{k+Np}$ for  any integer $p$.

Let us now study various possibilities for varying $\theta$  between $0$ and $2\pi k$.

\medskip

\item{1.} Let us first analyze the case of adiabatically varying $\theta(x)$ with $|\grad \theta| \ll \Lambda$, with $\Lambda$ the dynamical scale of the theory.  In this case at any point in space we are approximately in the vacuum of the theory at that value of $\theta(x)$. The function of the vacuum energy is continuous as a function of $\theta$ but non-differentiable when $\theta\in\pi+2\pi\Z$. This is the first order transition that we described above, where the time-reversal symmetry is broken spontaneously due to an anomaly. The energy density is therefore peaked at these points and the interface in fact breaks up to $k$ copies of the $SU(N)_1$ theory that we discussed above. The interface theory is therefore given by
\eqn\slow{[SU(N)_1]^k~.}
This clearly matches the anomaly~\Bloc.

 \item{2.} Now let us assume that $\theta$ interpolates from $0$ to $2\pi k$ rapidly, $|\grad \theta| \gg \Lambda$, but with finite $\grad\theta$. In that case the interface theory is naively
      \eqn\abrupt{SU(N)_k~.}
This is certainly the correct answer for sufficiently small $k$ compared to $N$.\foot{ On both sides of the interface we have a time-reversal invariant bulk theory. Yet, the theory on the interface is not time-reversal invariant since the profile of $\theta$ breaks time-reversal symmetry explicitly.  Furthermore, for generic $K$, the local term~\Bloc\ breaks time-reversal symmetry explicitly.} When $k$ becomes of the order of $N$, we may be able to decrease $k$ in absolute value without changing the anomaly by subtracting $N$ or $2N$ in the cases that $N$ is odd or even, respectively.
At the moment it is therefore unclear if~\abrupt\ is correct for all $k$ or only for $k$ sufficiently small. We leave this interesting question for the future.

\medskip

\ifig\Quiver{A quiver theory that reproduces the phases of the $\theta$-angle interface.  }%
{\epsfxsize4in\epsfbox{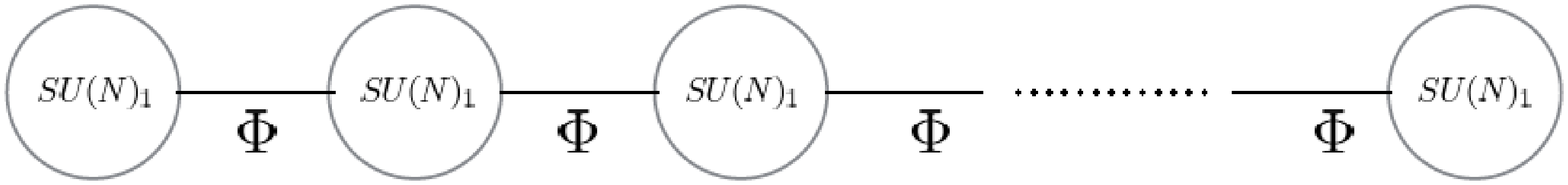}}

As we change the gradient from small to large, there must be a transition on the interface between~\slow\ and~\abrupt. The transition may be first or second order. In case it is second order it is especially useful to write a 2+1 dimensional model that would reproduce the dynamics of the interface. A natural guess is
$[SU(N)_1]^k+$bifundamentals, i.e.\ a linear quiver as in~\Quiver. When the mass squared is large and positive the phase is an
$[SU(N)_1]^k$ Chern-Simons theory and when the scalars condense it is an $SU(N)_k$ Chern-Simons theory.\foot{This picture is not yet entirely precise since this model with only quadratic and quartic terms for the bifundamental scalars has a $U(1)^{k-1}$ global symmetry, which is spontaneously broken in the phase where the scalars condense. To remove these Nambu-Goldstone bosons it is important to add to the Lagrangian terms like $\det\Phi$, where $\Phi$ is a bifundamental field. These terms break this accidental symmetry explicitly.}

The bifundamental scalars  are reminiscent of the open string modes that connect D-branes in string theory. Here the D2-branes support $SU(N)_1$ theories in 2+1 dimensions.
In terms of Yang-Mills theory, the bi-fundamental fields $\Phi$ are simply the confining strings. Intuitively, when the domain walls are far apart these strings are very long and their mass scales like their length (multiplied by the tension, which is of order $\Lambda$). Therefore, the bi-fundamental mass in $3d$ scales linearly with the inverse gradient of $\theta$ for small gradients. As the domain walls approach each other, the strings become shorter and shorter until at some point we can intuitively imagine that they become tachyonic (which is a standard phenomenon for short open strings). Then, they condense and the theory changes to $SU(N)_k$. As we remarked above, this picture may break down for $k$ of order $N$.

\newsec{Quantum Chromodynamics with One Flavor}

Here we study the theory \Lagintroi\ with an additional quark
\eqn\Lagintroi{{\cal L}=-{1\over 4g^2} Tr(F\wedge\star F)+{i\theta\over 8\pi^2}Tr(F\wedge F)+i\bar\psi\slash{D}\psi+i\bar{\tilde \psi}{\slash D}\tilde\psi+\left(m\tilde\psi\psi+c.c.\right)~.}
This theory depends on the complex parameter $m$ and on $\theta$ only through the combination $me^{i\theta}$.  So we will take $m$ real and non-negative and view the theory as a function of this complex parameter.

The global symmetry of the theory includes a continuous $U(1)_B$ acting on the quarks as
\eqn\UoneB{U(1)_B:\qquad \psi\to e^{i\alpha}\psi,~\qquad \tilde\psi\to e^{-i\alpha}\tilde \psi~.}
For $\alpha=2\pi k /N$ with integer $k$ this is a gauge transformation and so it is more precise to think about the global symmetry as $U(1)_B/\Z_{N}$. This simply means that the gauge invariant operators of the theory must carry $U(1)_B$ charge in multiples of $N$, i.e.\ these are the baryons. For $\theta = 0,\ \pi$ the theory also has a $CP$ symmetry (equivalently, time-reversal symmetry).

A well-known observation about this theory is that at $m=0$ there is no new symmetry. It therefore appears that $m=0$ is not a special point (although it is a well-defined point\foot{One might think that since for $N_f=1$ the quark mass suffers from an additive renormalization, the point $m=0$ is not well defined.  However, at short distances this additive renormalization is softer than the bare mass $m$ and therefore they can be easily separated.  Specifically, consider the chirality violating two point function $\langle \psi_\alpha\tilde \psi_\beta\rangle$ at high momentum $p$.  The bare mass $m$ contributes ${m\over p^2}+\cdots$.  The additive instanton contribution includes a factor of $\exp\left({-{8\pi^2\over g^2(\mu)}}\right) \sim \left({\Lambda\over \mu}\right)^{{11\over 3}N-{2\over 3}N_f}$ and therefore, up to logarithmic corrections, its contribution is $\sim {1\over |p|}\left({\Lambda\over |p|}\right)^{{11\over 3}N-{2\over 3}}$, which is softer than the contribution due to the bare mass $m$.}).  However, the real $me^{i\theta}$ line is a special subspace of the full complex plane because it preserves $CP$.

At large $m$ we can integrate out the quark and we end up with pure Yang-Mills theory at $\theta$. For generic $\theta$ it has a unique ground state and at $\theta=\pi$ it has two vacua.  (As we said above, this could be different for small values of $N$.) This means that in the complex $me^{i\theta}$ plane there is a first order line coming from infinity along the negative real axis and it must end at some point with $m=m_0$.  This point must be along the negative real axis ($\theta=\pi$) because, as we said in the introduction, the theory for $m=0$ has a trivial vacuum (this can be explicitly demonstrated at large $N$ and we will assume that it continuous to be true at finite $N$ even though this assumption is not crucial for us).

\ifig\Nfone{The phases of QCD with one flavor. At $me^{i\theta}=-m_0$  there is a massless $\eta'$ particle.  Along the real $me^{i\theta}$ axis the theory is time reversal invariant.  The $\eta'$ field condenses along that line to the left of $-m_0$, but not to the right of $-m_0$.  The expression for the effective Lagrangian is as in the large $N$ limit, expanded around $m\approx m_0 $ and $|\eta'|\ll 1$. }%
{\epsfxsize3.5in\epsfbox{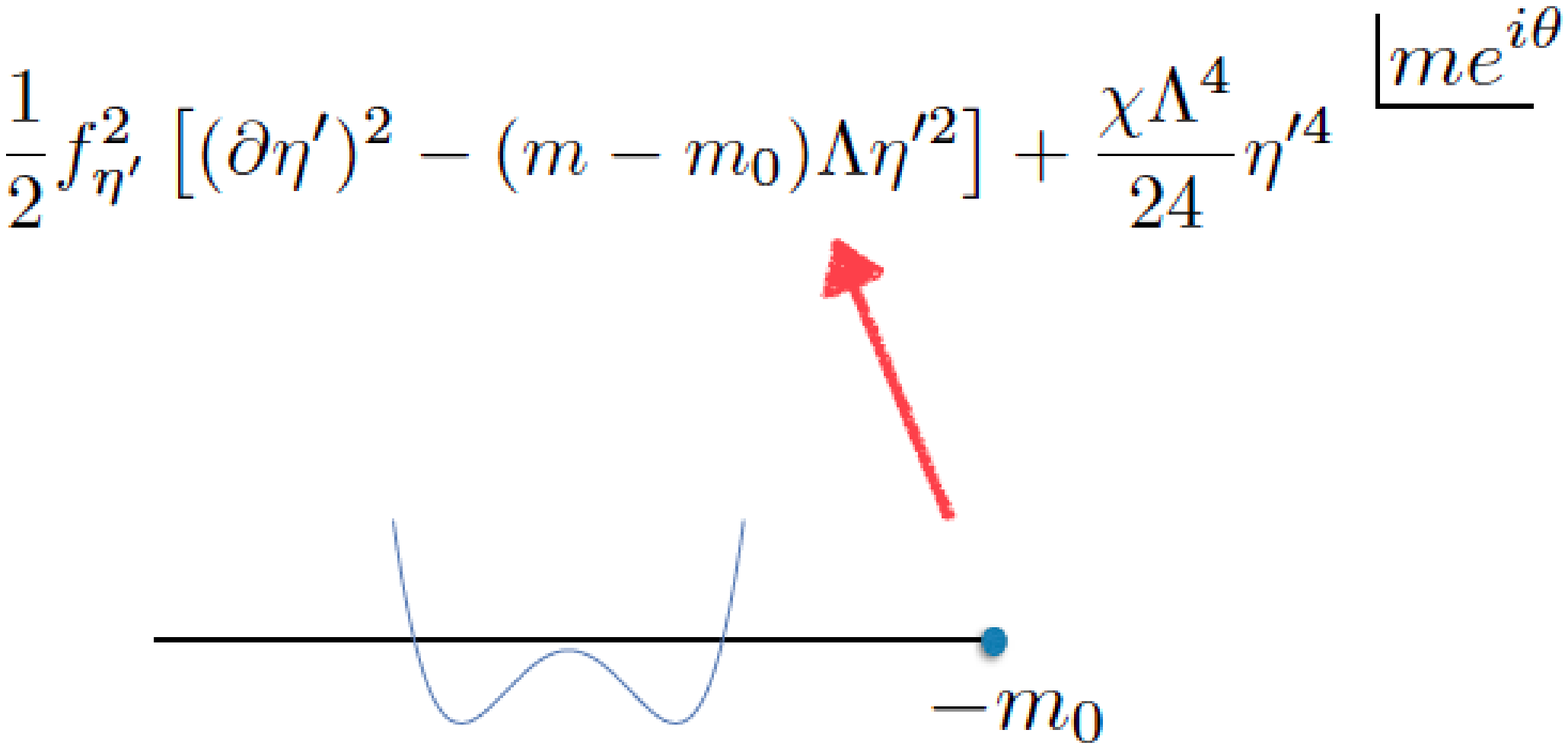}}

Therefore, as we change $m$ with $\theta=\pi$, $CP$ should be restored at $m=m_0$.  The transition at this point must be second order because it is the end point of a first order line. This is depicted in \Nfone.

It is natural to assume that this transition is described by a real pseudoscalar field $\eta'$. Its low energy dynamics is essentially free.  It can loosely be called the ``$4d$ Ising theory.''
For $\theta=\pi $ and $m$ close to $m_0$ its dynamics can be described by the effective low energy Euclidean theory
\eqn\effetap{\half f_{\eta'}^2 (\partial \eta')^2 + \half\mu^2 (\eta')^2 +\lambda(\eta')^4 + \cdots~.}
Since $\eta'$ is massless at $m=m_0$,
\eqn\muetap{\mu^2 \sim (m_0-m) + \cdots~.}
It is important that $\eta'$ is a pseudoscalar because its condensation needs to break the $CP$ symmetry spontaneously. We identify the $\eta'$ field with the phase of fermion condensate
\eqn\etapd{\langle \psi\tilde\psi\rangle \sim e^{i\eta'}~, }
and hence with the $\eta'$ particle.

We conclude that the $SU(N)$ gauge theory with $N_f=1$ for any $N$ has a value of the bare mass where the $\eta'$ particle is massless!

Our description of the emergence of the $\eta'$ particle can be understood particularly easily in the large $N$ limit. Indeed, in the planar limit the axial anomaly disappears and the axial symmetry is restored at $m=0$.  Then the Goldstone boson of its spontaneous breaking $\eta'$ is exactly massless.  $1/N$ corrections give the $\eta'$ a small mass~\WittenVV.  For $m=0$ the low energy description of the planar theory with $N_f=1$ is given by \refs{\RosenzweigAY,\NathIK,\WittenSP,\DiVecchiaYFW}
\eqn\LargeN{{\cal L}= \frac12 f_{\eta'}^2(\del\eta')^2+\half\Lambda^4\chi \eta'^2+\cdots~,}
where the corrections to this Lagrangian are suppressed by powers of $N$. In the large $N$ limit $f_{\eta'}^2\sim N\Lambda^2$ and $\Lambda$ as well as $\chi$ are fixed and do not scale with $N$. $\chi$ is known as the topological susceptibility.  It is determined in the pure gauge $SU(N)$ theory and it is important that it is positive \WittenVV.
The physical mass of $\eta'$ is therefore $m^2_{\eta'}={\Lambda^4\chi \over f_{\eta'}^2} \sim {\Lambda^2\over N}$. The potential is proportional to $\eta'^2$ for $\eta'\in [-\pi,\pi]$ and beyond that the potential is defined in a periodic fashion. There is therefore a non-differentiable singularity at $\eta'=\pi \mod 2\pi$.

Now imagine adding to the Lagrangian a small mass term $m\psi\tilde\psi+c.c.$. This is reflected by modifying the Lagrangian~\LargeN\ to \refs{\RosenzweigAY,\NathIK,\WittenSP,\DiVecchiaYFW}
\eqn\LargeNi{{\cal L}= \frac12 f_{\eta'}^2(\del\eta')^2-m\Lambda f_{\eta'}^2\cos(\eta'+\theta) +\half \Lambda^4\chi \eta'^2~+\cdots.}
where as always $m$ is real and positive and $\theta$ is $2\pi$ periodic.  We define the QCD scale $\Lambda$ such that the second term in \LargeNi\ is as we wrote it.  The sign of that term is determined such that for $\theta=0$ the theory has a unique ground state at $\eta'=0$ for all $m$.  Of course, the Lagrangian \LargeNi\ is valid only to leading order in $m\over \Lambda$ and to the order we wrote in the $ {1\over N}$ expansion.

Let us analyze the potential of \LargeNi\
\eqn\mini{V=-m\Lambda f_{\eta'}^2\cos(\eta'+\theta)+\half\Lambda^4\chi\eta'^2~.}
For $\theta=0$ it has a single minimum at $\eta'=0$.  For generic $\theta$ there is also a unique minimum at some value of $\eta'$. For $\theta=\pi$ the situation is more interesting.  The behavior changes at
\eqn\mstar{ m_0={\chi\Lambda^3\over f_{\eta'}^2}\sim {\Lambda\over N}~,}
where $\eta'$ is massless.
For $m<m_0$ there is a unique minimum at $\eta'=0$.  This is the analytic continuation of the situation at $\theta=0$.  For $m>m_0$ there are two vacua at $\eta'=\pm \eta'_0$, with $\eta'_0$ interpolating from $0$ at $m\roughly>m_0$ to $\pi$ at $m\to \infty$ (the chiral Lagrangian approximation breaks down beforehand).

Note that expanding around $m \approx m_0$ with $\eta'\ll 1$ the effective Lagrangian \LargeNi\ goes over to \effetap. It is therefore consistent with~\Nfone.

Let us discuss the case $\theta=\pi$ and ${1\over N} \ll {m \over \Lambda}\ll 1$ in more detail.  Here the minima are at $\eta' =\pm \eta'_0$ with $\eta'_0$ approaching $\pi$
\eqn\calcran{ \eta'_0=\pi-{\pi\chi\Lambda^3\over mf_{\eta'}^2}+\cdots~.}
It is important that these two minima are not physically close to each other in field space, although they seem to be close to each other in the  $\eta'$ coordinate. Recall that at $\eta'=\pi$ the potential is not differentiable. Such singularities of the effective potential typically mean that some heavy degrees of freedom are in fact important. And indeed, we expect that as we cross the singular point at $\eta'=\pi$ we need to rearrange the heavy fields significantly. As a result, despite appearances,  the two ground states do not get physically close to each other as we increase the mass.

When $CP$ is spontaneously broken, i.e.\ for $\theta=\pi$ with $m>m_0$, the theory has domain walls.  For sufficiently large $m$ the domain wall is the same as in pure Yang-Mills theory at $\theta=\pi$, i.e.\ it supports an $SU(N)_{1}$ Chern-Simons theory. However, as $m$ is reduced toward $ m_0$, this can no longer be true.  Here the bulk is described by the $\eta'$ field, which is simply a pseudoscalar field theory with two minima, and hence its domain wall theory is trivial.  We see that for some finite value of the bulk mass $m_{transition}>m_0$ there must be a transition on the domain wall from $SU(N)_{1}$ to a trivial domain wall (which we can think about as $SU(N)_0$).  It is interesting that throughout this process of reducing $m$ and going through a phase transition on the wall the bulk remains gapped and is essentially unchanged.  Yet, the domain wall undergoes a phase transition.

Let us estimate the tension of the domain wall in various limits (we always take $m>m_0$ and $\theta=\pi$ so that $CP$ is spontaneously broken and the domain wall exists).

\medskip

\item{1.} $0<m-m_0\ll {\Lambda\over N}$. Here the domain wall can be understood using the $\eta'$ Lagrangian~\LargeNi, which can be further simplified to
    \eqn\simpLag{\half f_{\eta'}^2\left[(\partial\eta')^2-(m-m_0)\Lambda \eta'^2\right]+{\chi \Lambda^4\over 24}\eta'^4~.}
    The domain wall interpolates between $\pm \eta'_0 = \pm \left({6(m-m_0) f_{\eta'}^2 \over \chi\Lambda^3}\right)^\half$.  The tension of the domain wall is therefore given by solving the equation of motion for $\eta'$ that interpolates between $\pm \eta'_0$ and then substituting in
\eqn\Tenslon{T= \int d x \left (\half f_{\eta'}^2\left[(\del\eta')^2-(m-m_0)\Lambda \eta'^2\right]+{\chi \Lambda^4\over 24}\eta'^4 \right) ~.}
We can rescale the coordinate $x$ as well as the $\eta'$ field such that it interpolates between $\pm 1$ and convert the formula for the tension to
\eqn\regionone{T= {f_{\eta'}^4(m-m_0)^{3/2}\over \chi\Lambda^{5/2}}   \int d x \left ((\partial\eta')^2-\eta'^2+\half\eta'^4 \right) \sim {f_{\eta'}^4(m-m_0)^{3/2}\over \chi\Lambda^{5/2}}  ~. }
This domain wall is trivial in the sense that it has no nontrivial degrees of freedom other than the obvious center of mass.
\item{2.} ${\Lambda\over N} \ll m-m_0\approx m\ll\Lambda$. Also in this region we can trust the chiral Lagrangian except that it is now better approximated by
    \eqn\seconappL{{\cal L}= f_{\eta'}^2\left[\half (\partial\eta')^2+ m \Lambda \cos(\eta')\right]~.}
    In this region, as we explained around~\calcran, the two minima appear to be close to each other, but they are in fact far away. It is therefore still beneficial for the domain wall to interpolate between the two vacua by having the $\eta'$ field go around the circle, avoiding $\eta=\pi$.  After rescaling the coordinates, the tension of this domain wall is given by
    \eqn\regiontwo{T=f_{\eta'}^2m^{1/2}\Lambda^{1/2} \int dx \left [\half(\partial\eta')^2+\cos(\eta') \right]\sim f_{\eta'}^2m^{1/2}\Lambda^{1/2}\sim Nm^\half\Lambda^{5/2}~.}

\medskip

These two regions $0<m-m_0\ll {\Lambda\over N}$ and ${\Lambda\over N} \ll m-m_0\approx m\ll\Lambda$ are smoothly related and give a coherent picture for $m_0<m \ll \Lambda$ and ${1\over N}\ll 1$.  For these values of the parameters the $\eta'$ field performs a continuously increasing field excursion on the circle as $m$ is increased.  Hence, there is no phase transition between them.  Throughout this region the domain wall is trivial in the sense that it has no nontrivial degrees of freedom other than the obvious center of mass.

Finally, we discuss the limit $m\gg\Lambda$. Here we can simply integrate out the quark and we remain with a domain wall in pure Yang-Mills theory at $\theta=\pi$. The tension scales like
\eqn\pureagain{T\sim N\Lambda^3~.}
Here $\Lambda$ should be the strong coupling scale of the pure YM theory that remains after integrating out the quark. The difference between it and the original strong coupling scale is subleading in $N$.  In terms of the large $N$ picture with $\eta'$ we can say that at some point it becomes favourable for the $\eta'$ field to jump over the singular point $\eta'=\pi$ and not go around the whole circle. This happens when $m$ becomes of order $\Lambda$, where the expansion in $m\over \Lambda$ above is no longer valid.

\medskip

It is important to know whether the transition on the domain wall is first order or second order.  From the large $N$ discussion above it seems that for small $m$ the $\eta'$ coordinate interpolates through $\eta'=0$ and for large $m$ it interpolates through $\eta'=\pi$.  This might indicate that the trajectory of $\eta'$ jumps and hence the transition is first order.  If this is the case, the physics of the transition involves massive modes on the wall.  It could also involve massive modes in the $4d$ bulk of the system.  If this is the case, we cannot describe the transition in terms of a $3d$ quantum field theory.

If, on the other hand, the transition is second order, since the bulk is gapped, it must have a purely $3d$ quantum field theory description.

Regardless of which of these two options materializes, we now present a $3d$ quantum field theory that has exactly the two phases of the domain wall.
Consider the theory\foot{We follow the notations, the conventions, and the results of \KomargodskiKEH.}
\eqn\CSpro{SU(N)_{1/2}+\psi~,}
i.e.\ $SU(N)_{1/2}$ Chern-Simons theory coupled to a fundamental fermion $\psi$.
If we give $\psi$ a large mass, depending on the sign of the mass, we end up with either $SU(N)_{1}$ pure Chern-Simons theory or the trivial theory. These are the expected phases of the domain wall theory.
It is further conjectured (see~\refs{\HsinBLU,\KomargodskiKEH} and references within) that these are the only two phases of the theory, with a single phase transition.
We can also describe this theory in terms of a bosonic dual \refs{\AharonyMJS,\HsinBLU}
\eqn\CSproi{U(1)_{-N}+\phi~,}
i.e.\ $U(1)_{-N}$ Chern-Simons theory with a single scalar of charge 1. When the scalar Higgses the gauge group we end up with a trivial theory and when it has a positive mass squared we end up with $U(1)_{-N}$ Chern-Simons theory, which is dual to $SU(N)_1$ Chern-Simons theory as a spin TQFT.\foot{Here we can use level-rank duality because both theories need a choice of a spin structure (more generally a spin$^c$ structure) \HsinBLU.} The monopole operator in the theory~\CSproi\ has half-integer spin if $N$ is odd and integer spin if $N$ is even. In the fermionic language~\CSpro\ this is the baryon operator \RadicevicYLA. It extends into the bulk as the worldline of a heavy baryon. Indeed, the spin of the baryon in $SU(N)$ gauge theory with one flavor is half-integer if $N$ is odd and integer if $N$ is even.

\newsec{Quantum Chromodynamics with $N_f>1$ Quarks}

Consider QCD with $N_f$ quarks
\eqn\Lagintro{{\cal L}=-{1\over 4g^2} Tr(F\wedge\star F)+{i\theta\over 8\pi^2}Tr(F\wedge F)+i\sum_{i=1}^{N_f}\bar\psi_i {\cal D}\psi^i+i\sum_{i=1}^{N_f}\bar{\tilde \psi}^i {\cal D}\tilde\psi_i+\left(m_i\tilde\psi_i\psi^i+c.c.\right)~.}
We will first study in detail the bulk phases of the theory and then we will consider the domain walls (and in appendix A we also briefly consider some interfaces).

For generic masses this theory has a $U(1)^{N_f}/\Z_{N}$ global symmetry, where the quotient is by a $\Z_N$ subgroup, which is part of the gauge group. This quotient will be important in section 5. For simplicity we will consider the situation with equal masses $m$ and then the global symmetry is $U(N_f)/\Z_{N}$.  As for $N_f=1$, without loss of generality we can let the common mass $m$ be real and non-negative and parameterize the theories by the complex number $m^{N_f}e^{i\theta}$.  The generalization to arbitrary masses is straightforward.\foot{Also, the coupling to background gauge fields for the global symmetry leads to additional discrete parameters.  This will be discussed in section 5.}
For generic $\theta$ the system is not $CP$ invariant, but for $\theta = 0\mod \pi$ it is $CP$ invariant. (If we rotate $\theta$ into the mass matrix, the system is $CP$ invariant when $\det m$ is real.)  As we will see, the physics at $\theta=0$ and at $\theta=\pi$ are different.

\ifig\NfLarge{The phases of QCD with $N_f>1$ degenerate flavors. At $m=0$ the global $SU(N_f)\times SU(N_f)$ symmetry is spontaneously broken to the diagonal $SU(N_f)$ and the low energy theory is gapless. For positive $m$ and $\theta=\pi$ there are two vacua and for generic $\theta$ there is a single vacuum. Unlike $N_f=1$, here the $\eta'$ particle is never massless.}%
{\epsfxsize3.3in\epsfbox{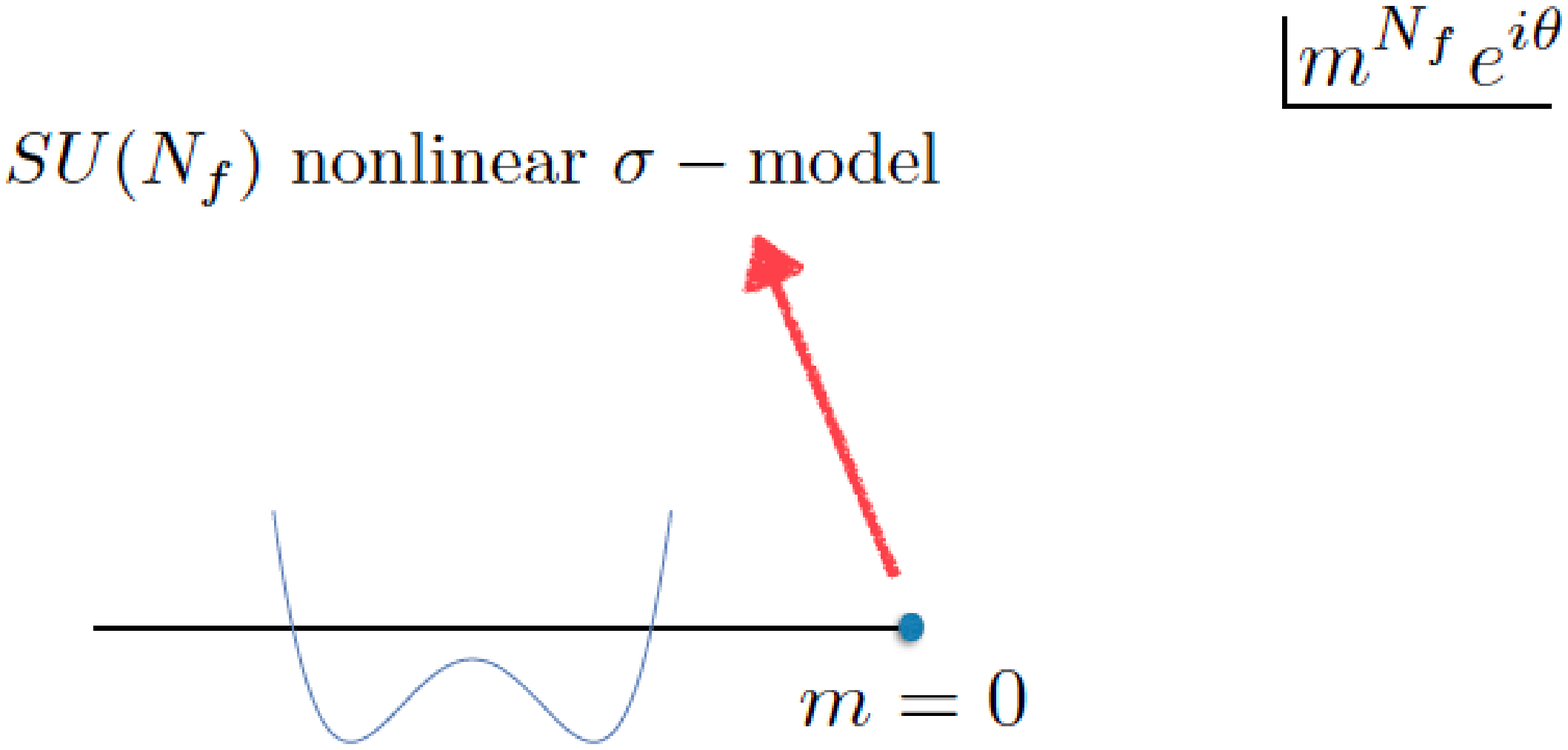}}

Consider $m\gg \Lambda$, where $\Lambda$ is the dynamical scale of the system.  Then we can integrate out the quarks.
For $\theta=0$ there is a confining trivial vacuum.  And for $\theta=\pi$ we obtain two ground states related by $CP$.
In \NfLarge\ we summarize the phases of the theory in the $m^{N_f}e^{i\theta}$ plane.

Now consider the theory around $m=0$. We restrict to $N_{CFT}>N_f>1$, where the massless theory is described by a chiral Lagrangian.  We will first study the finite $N$ theory, ignoring the $\eta'$ particle. The Lagrangian at $m=0$, which is the second order transition point of~\NfLarge, is given by
\eqn\kineticcal{{\cal L}={f_\pi^2\over 2}\Tr\left(\del U \del U^\dagger\right)+\cdots~,}
where the higher order terms are suppressed by additional derivatives and $U$ is an $SU(N_f)$ matrix, expressed in terms of the pions $\pi^a$ as $U=e^{i\pi^a T^a}$  and $a$ is an index in the adjoint of the unbroken group.

Now suppose that we add a mass term $m$ for the $N_f$ quarks.
\eqn\masstermNf{ m\sum \tilde\psi_i\psi^i+c.c.~.}
As usual, as long as $m\ll \Lambda$ we can analyze the effect of the mass using the chiral Lagrangian, where to leading order we simply need to add a potential in terms of the fermion condensate
$\langle \psi\tilde\psi\rangle \sim f_\pi^2 \Lambda$. The distinction between $f_\pi$ and $\Lambda$ will be important later when we study the phases of the theory at large $N$. The potential in the chiral Lagrangian is therefore
\eqn\pot{V= -\half f_\pi^2\Lambda  me^{i\theta/N_f}\Tr U+c.c.~.}
In this expression we absorbed a dimensionless constant into a definition of $\Lambda$.
Clearly the theory~\kineticcal\ with the potential~\pot\ is $SU(N_f)$ preserving. We can also verify that $\theta$ is $2\pi$ periodic. Indeed, the shift $\theta\to \theta+2\pi$ can be undone with the change of variables  $U\to e^{-{2\pi i \over N_f}}U$.
For $\theta=0$ the $CP$ symmetry acts as $U\to U^\dagger$.  And for $\theta=\pi$ it acts as
$U\to e^{-{2\pi i \over N_f}}U^\dagger$.

Let us consider the minimization of the potential for $\theta=0$ and $\theta=\pi$.
Starting with $\theta=0$, assuming that $SU(N_f)$ is not spontaneously broken, the vacua are potentially at $U=e^{2\pi i k /N_f}\unit$ with $k\in \Z$. Minimizing~\pot\ over $k$ we find that the minimum is at $U=\unit$, which is $CP$ invariant. Therefore, $CP$ is unbroken. One can check that this vacuum is gapped and the assumption that $SU(N_f)$ is not broken is self-consistent.

Next, we turn to $\theta=\pi$.  Again, we examine the $SU(N_f)$-invariant vacua
\eqn\SUNfi{U=e^{2\pi i k /N_f}\unit ~.}
Substituting this into~\pot\ we find that the potential of the configurations above is
\eqn\potthet{V=-m N_ff_\pi^2\Lambda \cos\left({\pi(2  k+1)\over N_f}\right)~. }
This is minimized for $k=0$ and $k=-1$, i.e.\
\eqn\vacschiral{U=e^{-{2\pi i\over N_f}}\unit~,\qquad U=\unit~.}
These two solutions are related to each other by $U\to e^{-{2\pi i\over N_f}}U^\dagger$ and hence $CP$ is spontaneously broken. ($CP$ breaking in the chiral Lagrangian was first discussed in \DashenET\ and later by many others including \refs{\WittenSP,\DiVecchiaYFW} and \refs{\CreutzXU,\CreutzWF}.)

Let us check whether the assumption that $SU(N_f)$ is preserved is self consistent by studying the small fluctuations around these two minima. A general $SU(N_f)$ matrix with entries on the diagonal is given by
\eqn\diagcon{U={\rm diag}(e^{i\alpha_1},e^{i\alpha_2},...,e^{i\alpha_{N_f}})~,\qquad \sum_{k=1}^{N_f} \alpha_k =0 \ {\rm mod} \ 2\pi~.}
The energy of this configuration is
\eqn\energy{V=-m f_\pi^2\Lambda \sum_{k=1}^{N_f} \cos(\alpha_k+\pi/N_f)~.}
For $N_f=2$ the potential $V$ vanishes identically and hence it costs no energy to interpolate between the two $SU(2)$-preserving minima~\vacschiral. (For a detailed discussion of this point see~\SmilgaDH.) For $N_f>2$ we can expand around \vacschiral\ and verify that they are true local minima.  In the special case of $N_f=2$ whether $SU(2)$ is broken or $CP$ is broken is determined by the higher-order terms in the chiral Lagrangian.

\subsec{Bulk Phases at Large $N$}

Let us now  repeat the analysis above in the large $N$ limit, where we have to include the $\eta'$ field.
At the massless point, we now have instead of~\kineticcal\ a Lagrangian written in terms of a $U(N_f)$ matrix $U$ \refs{\RosenzweigAY,\NathIK,\WittenSP,\DiVecchiaYFW}
\eqn\kineticLargeN{{\cal L} = {f_\pi^2\over 2} \Tr(\partial U\partial U^\dagger)+\half\chi\Lambda^4|\log \det U|^2~,}
with $\chi$  a positive constant.  Note that in the large $N$ limit there is no separate kinetic term of the form $\partial(\Tr U )\partial (\Tr U^\dagger)$; i.e.\ $f_\pi = f_{\eta'}$ to leading order. In~\kineticLargeN\ the logarithm is defined to be between $-i\pi$ and $+i\pi$. This guarantees the invariance under $U\to e^{2\pi i\over N_f}U$.
The ground states of the large $N$ theory at zero mass are therefore obtained when $\log \det U = 0 \ {\rm mod}\ 2\pi$; i.e.\ when the matrix $U$ is an $SU(N)$ matrix.

As in \pot, adding a mass term for the fermions corresponds to adding another term to the chiral Lagrangian
\eqn\kineticLargeN{- \half me^{i\theta/N_f}\Lambda f_\pi^2\Tr U+c.c.~.}
So we should study the potential \refs{\RosenzweigAY,\NathIK,\WittenSP,\DiVecchiaYFW}
\eqn\potlargeNf{- \half me^{i\theta/N_f}\Lambda f_\pi^2\Tr U+c.c. + \half\chi\Lambda^4|\log \det U|^2~.}

We parameterise the matrix $U$ as
\eqn\Uparmf{U=e^{i\eta' \over N_f}\tilde U~,}
where $\tilde U$ is an $SU(N_f)$ matrix.  This parametrization is subject to the identification
\eqn\parami{(\tilde U,\eta') \sim (e^{-{2\pi i\over N_f}}\tilde U,\eta'+2\pi)~,}
so we can limit $\eta'$ to the range $[-\pi,\pi]$.

We expect the global $SU(N_f)$ symmetry to be unbroken and therefore we look for minima of the form $\tilde U = e^{2\pi i k \over N_f}\unit $ for some integer $k$. The potential as a function of $\eta'$ and $k$ is
\eqn\potketa{V(k,\eta')= -mN_f \Lambda f_\pi^2 \cos\left({\eta' +\theta +2\pi  k\over N_f}\right) + \half \chi \Lambda^4 \eta'^2~.}
As a check, for $N_f=1$ this coincides with \mini.

For $\theta=0$ we obtain $-mN_f \Lambda f_\pi^2 \cos({\eta' +2\pi  k \over N_f}) + \half \chi \Lambda^4 \eta'^2$, which is clearly minimized at $\eta'=k=0$.  Hence, there is a unique minimum at $U=\unit$, which preserve the global $SU(N_f)$ symmetry as well as time-reversal symmetry.

For $\theta=\pi$ the potential is  $-mN_f \Lambda f_\pi^2 \cos({\eta'+ \pi(2  k+1) \over N_f}) + \half \chi \Lambda^4 \eta'^2$. Its minima are at $(k=0,\eta'=- \eta'_0)$ and $(k=-1, \eta'=+\eta'_0)$ for some $\eta'_0$, which depends on the parameters.  These two minima are related by $CP$ ($\eta'\to -\eta'$ and $k \to -k-1$) and hence $CP$ is spontaneously broken.

It is instructive to examine various limits.  Recall that our analysis is valid in the limits ${1\over N},\ {m\over \Lambda} \ll 1$.  We will analyze separately the cases ${m\over \Lambda} \ll{1\over N}\ll 1$ and ${1\over N}\ll {m\over \Lambda} \ll 1$.

For ${m\over \Lambda} \ll{1\over N}\ll 1$ the physics is essentially identical to the analysis for finite $N$, where we ignored the $\eta'$ field.  More precisely, $\eta'_0={mf_\pi^2 \over \chi \Lambda^3}\sin({\pi\over N_f})+\cdots\sim {N m\over \Lambda} \ll 1$ and the two minima are those of~\vacschiral.

For ${1\over N}\ll {m\over \Lambda} \ll 1$ the dynamics of $\eta'$ is important. The solution in this limit is $\eta'_0= \pi(1-{N_f\chi \Lambda^3\over mf_\pi^2}+\cdots )$, namely, the matrices that minimize the potential are
\eqn\Umifpi{U=e^{-{i\pi\over N_f}\pm i{\pi\chi \Lambda^3\over  mf_\pi^2}+\cdots} \unit  ~. }
As in~\calcran, while the two minimal seem close to each other, $\eta'=-\pi$ is a singular point and the minima are in fact far apart.

We see that the $\eta'$ particle is always massive and for $\theta=\pi$ the time-reversal symmetry is broken for every positive $m$.

\subsec{Domain Walls and 2+1-Dimensional Dynamics}

In our study here we will ignore the $\eta'$ field and will study the domain wall (and briefly some interfaces) of QCD with $N_f>1$ flavors.
We fix $\theta=\pi$ and vary $m$.  As we said, the theory breaks its time-reversal symmetry spontaneously for every nonzero $m$.  Hence, there is a domain wall that connects the two ground states.

For $m\gg\Lambda$ the quarks are very heavy, we can integrate them out, and we remain with pure Yang-Mills theory with $\theta=\pi$. The domain wall theory in the deep infrared supports a TQFT
\eqn\CSM{ m\gg\Lambda : \qquad\qquad SU(N)_1 \quad \leftrightarrow \quad U(1)_{-N}~,}
where we used the duality between them as spin TQFTs \HsinBLU.

Now let us consider the domain wall theory for $m\ll\Lambda$. We can study it in the chiral Lagrangian. The two vacua are given by~\vacschiral. Clearly, each of these vacua preserves the $SU(N_f)$  symmetry. We would like to consider a smooth configuration that interpolates between them. Since any such smooth configuration must break $SU(N_f)$ away from the vacua, once there is one such configuration, there must be a manifold of such configurations. We will argue in appendix A that the lowest energy domain wall theory preserves the symmetry $S[U(1)\times U(N_f-1)]$. Therefore the domain wall theory is gapless, supporting degrees of freedom parameterizing the coset
\eqn\lowmassdw{m\ll\Lambda: \qquad\qquad {SU(N_f)\over S[U(1)\times U(N_f-1)]}=\C\P^{N_f-1}~.}
There is therefore a massless nonlinear $\sigma$-model on the domain wall at small $m$. More precisely, for $N_f>2$, the chiral Lagrangian also includes a Wess-Zumino term and as a result the 2+1-dimensional domain wall theory also has an induced Wess-Zumino term.  The special case $N_f=2$ will be discussed below.

We see that there is a phase transition on the domain wall separating the two phases~\CSM\ and \lowmassdw.  As we said above, we do not know whether this transition is first or second order.  And if it is a first order transition it is not clear that there should be a $3d$ QFT describing it.  Nevertheless, we would like to present a natural $3d$ QFT, which has these phases.
One can think about the domain wall theory as
\eqn\threeduni{SU(N)_{1-N_f/2} +N_f\ {\rm fermions}~.}

According to the conjecture of~\KomargodskiKEH\ this theory has two phase transitions. The domain wall theory accounts for the phases around one of these transitions, which is also described by the dual bosonic theory
\eqn\threedunii{U(1)_{-N} +N_f\ {\rm scalars}~.}
It is manifest that for positive mass squared in this theory we are in the phase~\CSM\ and for negative mass squared we have the $\C\P^{N_f-1}$ manifold; i.e.\ the phase~\lowmassdw. The Chern-Simons term in~\threedunii\ gives rise to a Wess-Zumino term in the nonlinear $\sigma$-model with target space $\C\P^{N_f-1}$ \KomargodskiKEH.

We see that the domain wall theory can be naively thought of as~\threeduni.  And then, using the conjectured dynamics in~\KomargodskiKEH\ it produces the topological phase \CSM\ and the symmetry breaking phase~\lowmassdw.  Conversely, this picture provides some more evidence to that conjectured dynamics.

Let us now discuss two slightly atypical cases. For $N_f=1$ the coset~\lowmassdw\ is trivial and indeed the domain wall theory has a trivial phase. This is the phase where the $\eta'$ field condenses, as explained in section 3. For $N_f=2$ the bulk $3+1$-dimensional theory has no continuous Wess-Zumino term, but instead it has a $\Z_2$-valued $\theta$-parameter associated with $\pi_4(SU(2))=\Z_2$ \refs{\WittenTW,\WittenTX}.  It is nontrivial for odd $N$ and makes the Skyrmions/baryons fermions for odd $N$. Correspondingly, the $\C\P^1$  2+1 dimensional nonlinear $\sigma$-model has no continuous Wess-Zumino term and instead it has a $\Z_2$ valued $\theta$-parameter~\FreedRLK\ (and see references therein). This term is closely related to the Hopf term at $\theta=\pi$ \WilczekCY.  As in $3+1$ dimensions, this $\theta$-term transforms the Skyrmions into fermions for odd $N$. These $2+1$ dimensional Skyrmions originate from baryons in the bulk, which subsequently become baryons of the domain wall theory~\threeduni.

\newsec{Anomalies in QCD and Anomaly Inflow to 2+1 Dimensions}

In this section we show that when the greatest common divisor $\gcd(N,N_f)\not=1$ the theory at $\theta=\pi$ has a mixed 't Hooft anomaly between the continuous global symmetry and time-reversal symmetry, forcing nontrivial physics at long distances.

Our discussion here will be very similar to the analysis in \BeniniDUS\ of the global symmetries and their 't Hooft anomalies in the analogous $3d$ theory.  In fact, the anomaly in the $4d$ theory means that there must be a nontrivial $3d$ theory on domain walls, and that $3d$ theory can be one of the theories discussed in \BeniniDUS. We therefore obtain the anomaly of~\BeniniDUS\ by anomaly inflow from an anomaly in $4d$, which also includes time-reversal symmetry.

First, we should determine the global symmetry of our system.  For $\theta=0\mod\pi$ the system has time-reversal symmetry.  The symmetry that commutes with the Lorentz group includes charge conjugation symmetry (which we will ignore) and a continuous internal symmetry $G$.  Let us discuss the latter.

$SU(N_f)\times U(1)$ is a global symmetry of the system under which the fermions transform in the fundamental representation $({\bf N_f},1)$.  However, this symmetry does not act faithfully.  First, only $U(N_f)=(SU(N_f)\times U(1))/\Z_{N_f}$ acts faithfully on the fundamental fields.  Second, a $\Z_N$ subgroup of this group acts in the same way as the center of the $SU(N)$ gauge group and hence it is a gauge symmetry.  It is convenient to consider two groups that act.  The  group that acts on the fields in the Lagrangian is
\eqn\Kdef{K={SU(N)\times SU(N_f)\times U(1)\over \Z_N\times \Z_{N_f}} = {SU(N)\times U(N_f)\over \Z_N}~,}
where the $SU(N)$ factor is the gauge group.
The global symmetry group is
\eqn\Gdef{G={K\over SU(N)} =  {SU(N_f)\times U(1)\over \Z_N\times \Z_{N_f}}={U(N_f)\over \Z_N}~.}
The gauge invariant local operators are in representations of $G$ rather than of $K$.  They are therefore in representations of $U(N_f)$, but that group is not represented faithfully.

We can represent $K$ in terms of three group elements $\Big(u\in SU(N), v\in SU(N_f), \rho \in U(1)\Big)$ with the identifications
\eqn\Kident{(u,v,\rho) \sim (e^{2\pi i /N} u, v, e^{-2\pi i/N} \rho) \sim  ( u, e^{2\pi i /N_f}v, e^{-2\pi i/N_f} \rho)~,}
which follow from the two factors in the denominator of \Kdef.  Similarly, we can describe $G$ in terms of two group elements $\Big( v\in SU(N_f), \rho \in U(1)\Big)$ subject to the identifications that follow from \Gdef\
\eqn\Gident{(v,\rho) \sim (v, e^{-2\pi i/N} \rho) \sim  ( e^{2\pi i /N_f}v, e^{-2\pi i/N_f} \rho)~.}
Clearly, \Gident\ follows from performing the $SU(N)$ quotient on \Kident.

$G$ has two interesting subgroups.  First, $U(1) / \Z_N$, which is isomorphic to $U(1)$, is the baryon number symmetry. Under the original $U(1)$ factor the quarks have charge one and under this $U(1)$ group the baryons have charge 1.  Second, $SU(N_f)/ \Z_d$ with $d=\gcd(N,N_f)$ is the flavor symmetry that acts faithfully.  It means that the gauge invariant local operators of the theory are not in all $SU(N_f)$ representations.  The number of boxes in their Young tableaux should be a multiple of $d$.\foot{This is familiar from QCD with three light quarks.  The mesons and baryons are in ${\bf 1}$, ${\bf 8}$, and ${\bf 10}$ of the $SU(3)$ flavor symmetry.  There can be exotic particles with larger $SU(3)$ representations, but the number of boxes in their Young tableaux must be a multiple of three. All of them are in representations of $PSU(3)=SU(3)/ \Z_3$.}

The global symmetry of the system allows us to couple it background classical gauge fields.  We can easily couple the system to $SU(N_f)\times U(1)$ gauge fields.  But the quotients in \Gdef\ lead to additional options.  We can couple the system to background gauge fields of $G$, which are not $SU(N_f)\times U(1)$ gauge fields.  The quotient by $\Z_{N_f}$ is relatively simple -- it is straightforward to couple the system to $U(N_f)$ gauge fields.  But the quotient by $\Z_N$ is more interesting, because it acts also on the $SU(N)$ dynamical gauge fields. (See~\CohenCC\ for an early work on combined twists for the flavor symmetry  and gauge bundle.)  This means that if the background is a $G=U(N_f)/\Z_N$ gauge field, which is not a $U(N_f)$ gauge field, the dynamical gauge fields in the problem are not in an $SU(N)$ bundle, but in a $PSU(N)=SU(N)/\Z_N$ bundle.

We see that by turning on nontrivial $G$ classical gauge fields for the global symmetry, we can force the system to be in a nontrivial $PSU(N)$ bundle.  It was emphasized in \refs{\KapustinGUA,\GaiottoKFA} that this is not a new field theory, but instead, this is an observable in the original $SU(N)$ gauge theory.

Note that since our matter fields are in the fundamental representation of $SU(N)$, our system does not have a one-form global symmetry.  Yet, with appropriate twists in the flavor symmetry we can place it in $PSU(N)$ bundles.  In \refs{\GaiottoKFA,\GaiottoYUP} the presence of such bundles led to anomalies involving the one-form global symmetry.  Here, following \BeniniDUS, these twisted bundles will lead to anomalies involving ordinary global symmetries. (A similar analysis can be found in~\refs{\KomargodskiDMC,\KomargodskiSMK}.)

Another way to approach the problem is to turn the $SU(N_f)$ gauge fields to dynamical gauge fields. Then there is a one-form global symmetry, $\Z_{\gcd(N,N_f)}$.
Equivalently, we can say that the $SU(N_f)$ gauge fields are spurions for the explicitly broken one-form symmetry in QCD. Now there could be a mixed anomaly between the one-form symmetry and time-reversal symmetry. This point of view was used in~\ShimizuASF\ (see also~\refs{\TanizakiBAM,\ChermanTEY}).

Next, we look for anomalies.  Clearly, there is nothing wrong with coupling the system to $G$ gauge fields and therefore there is no pure flavor anomaly.  The situation with the time-reversal symmetry at $\theta=\pi$ is different.  Here the symmetry depends on the $2\pi$ periodicity of $\theta$.  However, nontrivial $G$ bundles can force the gauge fields to be in nontrivial $PSU(N)$ bundles, which are not $SU(N)$ bundles.  For these, (on spin manifolds) the periodicity of $\theta$ is $2\pi N$ rather than $2\pi$ and therefore the $\theta=\pi$ system might not be time-reversal invariant.  This is essentially the same phenomenon observed in \GaiottoYUP, except that here we use flavor symmetries rather than one-form symmetries to create fractional $PSU(N)$ instantons.

In order to analyze this problem in detail we should try to add to the $\theta=\pi$ Lagrangian counterterms in the background fields such that it is time-reversal invariant even for the non-trivial bundles. Since all our bundles will be $PSU(N)$, $PSU(N_f)$, and $U(1)$ bundles, all the relevant $\theta$-terms can be expressed in terms of ordinary instanton numbers, except that because of the twists they can be fractional. We have denoted the field strength of the original $SU(N)$ gauge field by $F$.  Now, we add an $SU(N_f)$ gauge field with field strength $F_f$ and a $U(1)$ gauge field\foot{Our system has fermions and therefore it involves a spin manifold with a choice of spin structure. We can generalize it by choosing a spin$^c$ structure and letting the $U(1)$ field that couples to the fundamental quarks be a spin$^c$ connection.  For simplicity, we will not do it here.} with field strength $F_B$.  And we generalize them to be gauge fields of $K$ \Kdef\ rather than simply $SU(N)\times SU(N_f)\times U(1)_B$ gauge fields.

We started with ${\theta\over 8\pi^2}\Tr (F\wedge F)$ at $\theta=\pi$, and we add to it two local counterterms in the background fields
\eqn\lagwith{{\pi\over 8\pi^2}\Tr (F\wedge F) + {\theta_f\over 8\pi^2} \Tr( F_f\wedge F_f)  +{ \theta_B\over 8\pi^2} F_B\wedge F_B ~.}
If the background fields are simply $SU(N_f)\times U(1)$ fields, then the coefficients of the counterterms $\theta_f$ and $\theta_B$ are $2\pi$-periodic and the system is time-reversal invariant when they are integer multiples of $\pi$.  However, because of the $\Z_N\times \Z_{N_f}$ quotient this is not necessarily true.

We would like to know the conditions on the integers $r$, $s$, and $t$ in
\eqn\lagwiths{{\pi r\over 8\pi^2}\Tr (F\wedge F) + {\pi s\over 8\pi^2} \Tr( F_f\wedge F_f)  +{ \pi t\over 8\pi^2} F_B\wedge F_B }
such that this Lagrangian is time-reversal invariant.  This would be the case if twice that Lagrangian is trivial.  Then, we have a consistent Chern-Simons theory
\eqn\CScond{{SU(N)_r\times SU(N_f)_s\times U(1)_t \over \Z_N\times \Z_{N_f} }~.}
Precisely this problem was analyzed in \BeniniDUS\ (see equations (2.2), (2.3) there), where it was found that we need\foot{The first condition follows from the consistency of the $\Z_N$ quotient.  The second conditions follows from the consistency of the $\Z_{N_f}$ quotient. Together they imply that $t$ is divisible both by $N$ and by $N_f$ (but not necessarily by $NN_f$). The third condition follows from the mutual consistency of the two quotients.  We need to make a Wilson line in some $SU(N)_r$ representation times a charge $t/N$ Wilson line of $U(1)_t$ have trivial braiding with a Wilson line in some $SU(N_f)_s$ representation times a charge $t/N_f$ Wilson line of $U(1)_t$.  This is determined by $e^{2\pi i t \over N N_f }$, which should be set to one~\BeniniDUS. Equivalently, we can choose a configuration of the fluxes in four dimensions as follows. Take the four-manifold to be $\T^4$ and choose a minimal 't Hooft flux \tHooftXSS\ in the
$PSU(N)$ bundle over the two cycle in the directions $12$  (namely the integral of $u_2(PSU(N))$ over the $12$ cycle is $1$, where $u_2$ is the $\Z_N$ valued two-form obstruction of $PSU(N)$ bundles to be $SU(N)$ bundles).  Similarly, take a minimal 't Hooft flux in the $34$ direction for $PSU(N_f)$. All the other fluxes are taken to vanish.  In this case the first and second terms in~\lagwiths\ vanish (modulo the usual integer instantons), while the third term is given by ${1\over 8\pi^2} \int_{\T^4} F_B\wedge F_B = {1\over NN_f}\mod 1~.$
Therefore one obtains that $t$ should be a multiple of $NN_f$, as we found above.}
\eqn\CScondr{t-Nr\in N^2\Z \qquad, \qquad  t -N_fs \in N_f^2\Z \qquad,\qquad t\in NN_f\Z~.}

We are interested in the case $r=1$.  Then, these conditions can be satisfied if and only if the greatest common divisor $\gcd(N,N_f)=1$. In that case there are always integer $p$ and $q$ such that $pN_f=1+qN$.  Then we can set $s=pN$ and $t= N(1+qN)=pNN_f$.  This means that that the terms in the Lagrangian \lagwith, \lagwiths\ are
\eqn\Tint{\eqalign{
&{\pi\over 8\pi^2}\Tr (F\wedge F)
+{\pi p N \over 8\pi^2}\Tr (F_f\wedge F_f)+ {\pi pNN_f \over 8\pi^2} F_B \wedge F_B =\cr
&{\pi\over 8\pi^2}\Tr (F\wedge F)
+{\pi p N \over 8\pi^2}\Tr (\CF_f\wedge \CF_f)~,}}
where $\CF_f=F_f+\unit F_B$ is a $U(N_f)$ field.  Clearly, there are also other choices.

If $\gcd(N,N_f)\not=1$ we cannot satisfy the conditions \CScondr\ with $r=1$ and therefore we cannot find a time-reversal invariant expression of the form \lagwith.  This means that even though the Lagrangian of the dynamical fields is time-reversal invariant, when we couple the system to background gauge fields for the symmetry $G$ there is no way to preserve the time-reversal symmetry.  This means that the system has a mixed 't Hooft anomaly between the global symmetry $G$ and time-reversal symmetry. There should be a way to write this anomaly as coming from a five-dimensional bulk term. We do not attempt to do it here.

This anomaly constrains the bulk physics.  It can be matched by either a symmetric vacuum with a nontrivial infrared theory (which could be topological) or by breaking either of the symmetries spontaneously.  Indeed, as we discussed in section~4, we expect the time-reversal symmetry to be spontaneously broken.  Then the domain wall theory that we proposed
\eqn\anomaly{SU(N)_{1-N_f/2}+N_f\ {\rm fermions}~}
should have an anomaly matching the expected anomaly inflow from the bulk; i.e.\ it should
have a pure $U(N_f)/\Z_N$ anomaly when $\gcd(N,N_f)\neq 1$. This is in precise agreement with the result of~\BeniniDUS, where the anomaly of~\anomaly\ was computed directly. This shows that our proposal for the dynamics of the domain wall theory is consistent with anomaly inflow.

It would be interesting to investigate further the consequences of this mixed time-reversal/flavor symmetry anomaly in QCD. For instance, if the gauge group is $SU(3)$, the anomaly exists only when the number of flavors is a multiple of three.

\bigskip
\noindent{\bf Acknowledgments}

We would like to thank O.~Aharony, F.~Benini, C.~Cordova, P.-S.~Hsin, A.~Kapustin, J.~Maldacena, G.~Perez, T. Sulejmanpasic, M.~Unsal,  and E.~Witten for useful discussions.
The work of D.G. was supported by the Perimeter Institute for Theoretical Physics. Research at the Perimeter Institute
is supported by the Government of Canada through Industry Canada and by the Province of
Ontario through the Ministry of Economic Development and Innovation.
 Z.K. is supported in part by an Israel Science Foundation
center for excellence grant and by the I-CORE program of the Planning and Budgeting Committee
and the Israel Science Foundation (grant number 1937/12). Z.K. is also supported by the ERC
STG grant 335182 and by the Simons Foundation grant 488657 (Simons Collaboration on the Non-Perturbative
Bootstrap).
NS was supported in part by DOE grant DE-SC0009988.

\appendix{A}{Wall and Interfaces in the Chiral Lagrangian}

\subsec{Walls in the chiral Lagrangian}

We consider the Lagrangian
\eqn\Chiral{{\cal L} = {f_\pi^2 \over 2}  \left[Tr (\del U \del U^\dagger) - m\Lambda e^{i\theta/N_f} Tr U+c.c.\right]~}
with $\theta=\pi$. Let us use the $SU(N_f)$ symmetry to bring $U$ to a diagonal form
\eqn\diagform{U=\left(\matrix{e^{i\alpha_1}  &0\cdot\cdot & \cdots \cr 0 &e^{i\alpha_2}\cdots  & \cdots   \cr
 \cdots &\cdots  & \cdots \cr
\cdots&\cdots  & e^{i\alpha_{N_f}}}\right)~,\qquad \sum \alpha_i = 0 \ {\rm mod }  \ 2\pi ~.}
Substituting this into~\Chiral\ we obtain
\eqn\Chirali{ {\cal L}={f_\pi^2 \over 2} \left[  \sum_i (\del \alpha_i)^2- 2m\Lambda\sum_i\cos\left(\alpha_i+{\pi\over N_f}\right)  \right]~.  }
The ground states are clearly at $\alpha_i=-{2\pi\over N_f} $ and $\alpha_i=0$. These two ground states are related by time-reversal and they are both manifestly gapped (as worked out in section 4) for $N_f>2$.

Here we study a configuration that interpolates between the two matrices, $U=\unit$ and $U=e^{-{2\pi i \over N_f} }\unit$. The $SU(N_f)$ symmetry must be spontaneously broken since there is no way to interpolate between these two vacua with $SU(N_f)$-preserving matrices. Let us divide the phases $\alpha_i$ into two groups, $\alpha_1=...=\alpha_k$ and $\alpha_{k+1}=...=\alpha_{N_f}$. Suppose without loss of generality that the first group goes continuously from $0$ to $-{2\pi\over N_f}$. Then the second group would end up at $\alpha_{k+1}=...=\alpha_{N_f}=-{2\pi\over N_f} +{2\pi \over N_f-k}$.  Therefore we end up at the required vacuum $U=e^{-{2\pi i \over N_f} }\unit$ only if $N_f-k=1$.

In order to study the domain wall we denote $\alpha_1=...\alpha_{N_f-1} \equiv \alpha $ and $\alpha_{N_f}=-(N_f-1)\alpha$. Substituting this into~\Chirali\ we obtain
\eqn\Chiralii{ {\cal L}={f_\pi^2 \over 2} \left[ N_f(N_f-1)(\del \alpha)^2- 2m\Lambda(N_f-1)\cos\left(\alpha+{\pi\over N_f}\right)- 2m\Lambda\cos\left(-(N_f-1)\alpha+{\pi\over N_f}\right)\right]~. }
We can estimate the tension of the resulting domain wall by rescaling the coordinate orthogonal to the wall as
$x=x'/\sqrt{m\Lambda}$. Then the the tension is
\eqn\Tensiaa{\eqalign{
T={f_\pi^2 \over 2} \sqrt{m\Lambda} \int dx' &\Bigg[ N_f(N_f-1)(\del \alpha)^2\cr
&- 2(N_f-1)\cos\left(\alpha+{\pi\over N_f}\right)- 2\cos\left(-(N_f-1)\alpha+{\pi\over N_f}\right)\Bigg]~, }}
with $\alpha$ changing from 0 to $-{2\pi\over N_f}$. Therefore the tension scales like
\eqn\tenscaling{T\sim f_\pi^2\sqrt{m\Lambda}~.}
The width of the domain wall is also fixed by dimensional analysis to be of order $1\over \sqrt{m\Lambda}$.

In the large $N$ limit the tension \tenscaling\ is $\sim Nm^{1/2}\Lambda^{5/2}$.
As we increase the mass of the quarks, the domain wall's tension rises and its width $1\over \sqrt{m\Lambda}$ decreases.  The expressions above are valid for $m\ll \Lambda$.  For $m\sim\Lambda$ and larger the width becomes $\Lambda^{-1}$ and the tension scales like $N\Lambda^3$.

\subsec{Interfaces in the Chiral Lagrangian}

Now we discuss the problem of interfaces in the chiral Lagrangian. Here we imagine that $\theta$ varies in space along one direction such that at one end we have $\theta=0$ and on the other side we have $\theta=2\pi  k$ with integer $k$. (For simplicity of the discussion we limit ourselves to $0<k\le{N_f\over 2}$.) These two bulk vacua are identical and they are trivial and gapped. In order to minimize the bulk energy on one side we therefore have $U=\unit$ and on the other side $U=e^{-{2\pi i k\over N_f}} \unit$.

The physics of the interface depends on how fast $\theta$ changes.  We start with the case that $\theta$ changes very slowly (we will soon specify slowly relative to what). Then the vacuum changes adiabatically with a transition whenever $\theta$ crosses $\theta=(2m+1)\pi $ for integer $m$. At every such $m$ we switch from $U=e^{-{2m\pi i\over N_f}}\unit$ to $e^{-{(2m+2)\pi i\over N_f}}\unit$. In order to do so, the system produces a domain wall with tension~\tenscaling\ and width $1\over \sqrt{m\Lambda}$.  Since the width of the domain wall is $1\over \sqrt{m\Lambda} $, this is the preferred way for the system to minimize the energy as long as $|\nabla \theta| \ll \sqrt {m\Lambda}$.  We conclude that in this limit the theory on the interface is given by
\eqn\interfaceslow{|\nabla \theta| \ll \sqrt {m\Lambda}:\qquad [\C\P^1]^k~.}

As we increase the gradient of  $\theta$, it eventually becomes preferable to make a rapid transition from $U=\unit$ to $U=e^{-{2\pi i k\over N_f}} \unit$. Repeating the analysis of the previous subsection we find that the the eigenvalues of~\diagform\ should be divided into two groups such that $\alpha_1=...=\alpha_{N_f-k}$ and $\alpha_{N_f-k+1}=...=\alpha_{N_f}$. The first group interpolates from $0$ to $-{2\pi k \over N_f}$ and the second group interpolates between $0$ and $-{2\pi k \over N_f}+{2\pi}$. We expect that the width of this configuration is again $1\over\sqrt{m\Lambda}$. Here the symmetry is spontaneously broken as
\eqn\interfacefast{|\nabla \theta| \gg \sqrt {m\Lambda}:\qquad {U(N_f)\over U(k)\times U(N_f-k)}~.}
For a related discussion see \TachikawaXVS.

We conclude that as we change the gradient of $\theta$ through $|\grad\theta|\sim {1\over\sqrt{m\Lambda}}$ the theory on the interface undergoes a phase transition from~\interfaceslow\ to \interfacefast. For $m\ll \Lambda$ this transition can be described in the chiral Lagrangian and it appears to be first order. The model~\interfacefast\ is closely related to the Grassmannian phase that was suggested in~\KomargodskiKEH, however, we leave the details for the future.

\listrefs

\bye